\documentclass[12pt,reprint,aps,groupedaddress,nofootinbib,prd,twocolumn]{revtex4-2}

\usepackage[utf8]{inputenc} 
\usepackage[T1]{fontenc}    
\usepackage{url}            
\usepackage{booktabs}       
\usepackage{amsfonts}       
\usepackage{nicefrac}       
\usepackage{microtype}      
\usepackage{lipsum}
\usepackage{xcolor}
\usepackage{amsmath}
\usepackage{amssymb}
\usepackage{empheq} 
\usepackage{ptdr-definitions}
\usepackage{bbm}
\usepackage{xspace}
\usepackage{relsize}
\usepackage{multirow}
\usepackage{subfloat}
\usepackage{graphicx}
\usepackage{dcolumn}
\usepackage[most]{tcolorbox} 

\graphicspath{ {./images/} }

\newcommand{\bn}{\ensuremath{{\boldsymbol{\nu}}}\xspace}

\newcommand{\bt}{{\ensuremath{{\boldsymbol{\theta}}}}\xspace}

\newcommand{\bzero}{{\ensuremath{\mathbf{0}}}\xspace}
\newcommand{\bx}{\ensuremath{\boldsymbol{x}}\xspace}

\newcommand{\ddd}{\ensuremath{{\textrm{d}}}\xspace}
\newcommand{\PW}{\ensuremath{\cmsSymbolFace{W}}\xspace}
\newcommand{\PZ}{\ensuremath{\cmsSymbolFace{Z}}\xspace}
\newcommand{\ggH}{\ensuremath{\cmsSymbolFace{ggH}}\xspace}

\newcommand{\pTmiss}{\ensuremath{\vec{p}_\text{T}^{\,\text{miss}}}\xspace}
\newcommand{\mT}{\ensuremath{m_\text{T}}\xspace}
\newcommand{\ptH}{\ensuremath{\pt^\PH}\xspace}
\newcommand{\VV}{\ensuremath{\cmsSymbolFace{VV}}\xspace}

\makeatletter
\renewcommand*{\@fnsymbol}[1]{%
  \ensuremath{%
    \ifcase#1\or
      1\or 2\or 3\or 4\or 5\or 6\or 7\or 8\or 9%
    \else
      \@ctrerr
    \fi
  }%
}
\makeatother

\usepackage{hyperref}

\begin{document}

\preprint{HEPHY-ML-25-01}

\title{Unbinned inclusive cross-section measurements with machine-learned systematic uncertainties}

\author{Lisa Benato}
\author{Cristina Giordano}
\author{Claudius Krause}
\author{Ang Li}
\author{Robert~Sch{\"o}fbeck}
\author{Dennis Schwarz}
\author{Maryam Shooshtari}
\author{Daohan Wang}
\affiliation{\vspace{.1cm}
  Institute for High Energy Physics, Austrian Academy of Sciences
  Dominikanerbastei 16, 1010 Vienna, Austria \vspace{.1cm}
}
\date{\today}

\begin{abstract}
We introduce a novel methodology for addressing systematic uncertainties in unbinned inclusive cross-section measurements and related collider-based inference problems. Our approach incorporates known analytic dependencies on parameters of interest, including signal strengths and nuisance parameters. When these dependencies are unknown, as is frequently the case for systematic uncertainties,  dedicated neural network parametrizations provide an approximation that is trained on simulated data. The resulting machine-learned surrogate captures the complete parameter dependence of the likelihood ratio, providing a near-optimal test statistic.
As a case study, we perform a first-principles inclusive cross-section measurement of $\textrm{H}\rightarrow\tau\tau$ in the single-lepton channel, utilizing simulated data from the FAIR Universe Higgs Uncertainty Challenge. Results in Asimov data, from large-scale toy studies, and using the Fisher information demonstrate significant improvements over traditional binned methods. 
Our computer code ``Guaranteed Optimal Log-Likelihood-based Unbinned Method'' (GOLLUM) for machine-learning and inference is publicly available.

\end{abstract}

\maketitle

\section{Introduction}
In recent years, the Large Hadron Collider (LHC)~\cite{Evans:2008zzb}  has shifted from a discovery instrument to a precision measurement machine, allowing unprecedented accuracy in the measurements of parameters of the Standard Model (SM) and the processes and phenomena it predicts.
A toolkit of statistical methods employs the likelihood function to extract parameter values, perform hypothesis tests, and set exclusion limits~\cite{Cowan:2010js,ATLAS:2011tau,Cranmer:2014lly,CMS:2024onh}. The exact likelihood for high-dimensional experimental data is, however, usually computationally intractable. Traditionally, LHC analyses eschew this issue by reducing high-dimensional measurements to low-dimensional summary statistics via binning and with the help of physically motivated observables. The likelihood is then tractable, but the dimensional reduction inevitably causes information loss, resulting in suboptimal performance.

To exploit the LHC's full potential, we must avoid information loss and work with a high-dimensional, unbinned likelihood function. While that function remains intractable in experimental settings, it can be inferred from detailed simulations with modern machine learning~(ML) tools~\cite{Brehmer:2018kdj,Brehmer:2018eca,Brehmer:2018hga,Brehmer:2019xox,Kong:2022rnd}, a technique known as (neural) simulation-based inference (SBI)~\cite{Cranmer:2019eaq}. SBI eliminates the need for binning and summary statistics, while allowing unbinned approximations related to the effects of the parton shower, the hadronization, and the detector response. 
With this, it opens new avenues for precision analyses~\cite{Bahl:2024meb}. Additionally, once the initial training phase is complete, likelihood-based inference is computationally fast.

Although phenomenological studies have explored SBI~\cite{Brehmer:2019gmn,Barman:2021yfh,Bahl:2021dnc,GomezAmbrosio:2022mpm,Barrue:2023ysk,Mastandrea:2024irf,Cheng:2024yig,Cheng:2025ewj}, large-scale deployment in experimental analyses remains limited by the relative simplicity of tools for addressing systematic uncertainties. In Ref.~\cite{Schofbeck:2024zjo}, the various methods for estimating uncertainties in binned analyses are promoted to the unbinned case. The ML-based parameterizations of systematic uncertainties, either in groups of correlated effects or one by one, allow for a flexible and refinable approach that largely follows the workflow of the binned case. Analysis development begins with the most important processes and systematic effects and can later be improved step-by-step without invalidating earlier partial results.  
Building on this work, we demonstrate here how the methodology of Ref.~\cite{Schofbeck:2024zjo} can be applied to unbinned cross-section measurements.

We showcase the developments in the FAIR Universe Higgs Uncertainty Challenge~\cite{Bhimji:2024bcd} (FAIR-HUC), achieving highly competitive results and tying \emph{ex aequo} for first place.
Our code ``Guaranteed Optimal Log-Likelihood-based Unbinned Method'' (\textsc{GOLLUM}) is publicly available~\cite{GOLLUM}. 

The ATLAS Collaboration~\cite{ATLAS:2008xda} recently explored unbinned cross-section measurements with classifier-based modeling of systematic uncertainties~\cite{ATLAS:2024ynn, ATLAS:2024jry}. In contrast, we learn, in principle, arbitrarily accurate parametric models of systematic effects and also include the cases where they do not factorize.

The structure of this paper is as follows. In Sec.~\ref{sec:samples}, we introduce the $\PH \rightarrow \tau \tau$ process and describe the FAIR-HUC simulated datasets, including the available event-level features, the event selections, and the systematic uncertainties. We develop the profiled likelihood ratio test statistic, both in a binned reference case and in the fully unbinned setting, in Sec.~\ref{sec:modeling}. Using the FAIR-HUC dataset, we demonstrate how refinable modeling incorporates known analytic dependencies into an ansatz for the likelihood function. Unknown dependencies are captured via a neural network-based parametrization of several systematic effects whose details we describe in Sec.~\ref{sec:ML}. Section~\ref{sec:limit-setting} presents results obtained from applying these techniques to the FAIR-HUC datasets, including studies on Asimov data~\cite{Cowan:2010js}, large-scale toy studies, and the Fisher information matrix. We conclude in Sec.~\ref{sec:outlook}.

\section{Processes and datasets}\label{sec:samples}

\subsection{The \texorpdfstring{$\PH \rightarrow \tau \tau$}{H tau tau} process}\label{sec:dataset}

The study of the $\PH\to\tau\tau$ process at the LHC has been systematically pursued by the ATLAS~\cite{ATLAS:2008xda} and CMS~\cite{CMS:2008xjf} Collaborations across multiple datasets and center-of-mass energies. Before the discovery of the Higgs~(\PH) boson, searches in the 7~\TeV dataset obtained during Run~I operations by CMS~\cite{CMS:2012bkm} and ATLAS~\cite{ATLAS:2012gfw} provided upper limits as a function of the \PH boson mass. After the Higgs boson discovery, the first evidence of the $\PH\to\tau\tau$ decay was reported using the combined 7 and 8~TeV datasets~\cite{CMS:2014wdm,ATLAS:2015xst}. The observation of the process with a significance of $5\sigma$ was achieved by combining earlier results with Run~II data at 13~\TeV~\cite{CMS:2017zyp,ATLAS:2018ynr}. Subsequently, inclusive cross-section measurements were performed~\cite{CMS:2022kdi,ATLAS:2022yrq}. Both collaborations refined the differential measurements by targeting the boosted selections and the Simplified Template Cross Section (STXS)~\cite{CMS:2021gxc,CMS:2024jbe,ATLAS:2024wfv}. These results collectively establish the $\PH\to\tau\tau$ decay as a key probe of the Higgs boson's properties at the LHC.

The \(\PH\to\tau\tau\) final state is classified by the leptonic~(\(\tau_\textrm{lep}\)) or hadronic~(\(\tau_\textrm{had}\)) decays of the $\tau$ leptons. Leptonic decays follow \(\tau\to\ell\nu_{\ell}\nu_{\tau}\), where \(\ell=(e,\mu)\), while hadronic decays produce charged and neutral hadrons along with a neutrino. The fully leptonic final state (\(\tau_\textrm{lep}\tau_\textrm{lep}\)) has a clean signature but suffers from reduced mass resolution due to multiple neutrinos. 
In contrast, hadronic decays allow for better momentum reconstruction with a single neutrino but face larger backgrounds from misidentified jets. The fully hadronic channel (\(\tau_\textrm{had}\tau_\textrm{had}\)), in particular, has the highest branching fraction but is significantly affected by jet misidentification. 
The semi-leptonic channel (\(\tau_\textrm{lep}\tau_\textrm{had}\)) balances signal yield and background suppression, benefiting from efficient identification of the $\tau_\text{lep}$ while retaining good mass resolution from the $\tau_\text{had}$. Combining all channels optimizes sensitivity to \(\PH\to\tau\tau\), with the semi-leptonic final state typically contributing most. 

\subsection{Features, event samples, and selections}
As the signal process, the FAIR-HUC datasets contain the semi-leptonic $\PH\to\tau_\textrm{lep}\tau_\textrm{had}$ process generated with \textsc{Pythia}~8.2~\cite{Sjostrand:2014zea} and processed with \textsc{Delphes }~3.5.0~\cite{deFavereau:2013fsa}. 
Each event contains a lepton with transverse momentum~(\pt) above 20~\GeV and with a pseudo-rapidity~($\eta$) satisfying $|\eta^\ell|\leq 2.5$. The $\tau_\text{had}$ satisfies\footnote{These thresholds are similar to the choices in Ref.~\cite{ATLAS:2018ynr}.} $\pt^{\tau_\text{had}}>26$~\GeV and $|\eta^{\tau_\text{had}}|<2.69$.

The 28 available event features capture key kinematic properties of the $\tau_\text{had}$, the lepton, the jets, and the neutrino system, along with derived quantities constructed from these primary observables. 

\begin{table*}
    \caption{List of event features used in the analysis. Primary features correspond to direct kinematic observables, while derived features incorporate correlations and high-level event properties.}
    \label{tab:event_features}
    \begin{ruledtabular}
    \renewcommand{\arraystretch}{1.2}
    \begin{tabular}{l l l l}
        \textbf{Symbol} & \textbf{Description} & \textbf{Symbol} & \textbf{Description} \\
        \hline
        \(\pt^{\tau_\text{had}}\) & Transverse momentum of the $\tau_\text{had}$ &
        \(\pt^{\ell}\) & Transverse momentum of the lepton \\
        \(\eta^{\tau_\text{had}}\) & Pseudorapidity of the $\tau_\text{had}$ &
        \(\eta^{\ell}\) & Pseudorapidity of the lepton \\
        \(\phi^{\tau_\text{had}}\) & Azimuthal angle of the $\tau_\text{had}$ &
        \(\phi^{\ell}\) & Azimuthal angle of the lepton \\
        \(\pTmiss\) & Missing transverse momentum &
        \(\phi^\text{miss}\) & Azimuthal angle of missing transverse momentum \\
        \(\pt^{j_1}\) & Transverse momentum of the leading jet &
        \(\pt^{j_2}\) & Transverse momentum of the subleading jet \\
        \(\eta^{j_1}\) & Pseudorapidity of the leading jet &
        \(\eta^{j_2}\) & Pseudorapidity of the subleading jet \\
        \(\phi^{j_1}\) & Azimuthal angle of the leading jet &
        \(\phi^{j_2}\) & Azimuthal angle of the subleading jet \\
        \(N_j\) & Number of reconstructed jets &
        $\sum_{\text{jets}} p_{\text{T}}$ & Sum of transverse momenta of all jets \\
        \(m_\text{T}(\ell, \pTmiss)\) & Transverse mass of lepton and missing \(\pt\) &
        \(m_{\text{vis}}\) & Visible invariant mass of \(\tau_\text{had}\) and \(\ell\) \\
        \ptH & Modulus of vector sum of \(\tau_\text{had}, \ell, \pTmiss\) &
        \(m^{j_1j_2}\) & Invariant mass of the two leading jets \\
        \(\Delta\eta^{j_1j_2}\) & Pseudorapidity separation of leading jets &
        \(\eta^{j_1} \cdot \eta^{j_2}\) & Product of leading jet pseudorapidities \\
        \(\pt^\text{tot}\) & Modulus of vector sum of \(\pTmiss\), \(\pt^{\tau_\text{had}}\), \(\pt^{\ell}\), \(\pt^{j_1}\), \(\pt^{j_2}\) &  
\(\sum \pt\) & Scalar sum of \(\pTmiss\), \(\pt^{\tau_\text{had}}\), \(\pt^{\ell}\), and all jets\\
        \(C^\text{miss}_\phi\) & Azimuthal centrality of \(\pTmiss\) w.r.t. \(\tau_\text{had}, \ell\) &
        \(C^\ell_\eta\) & Pseudorapidity centrality of lepton w.r.t. jets \\
        \(\Delta R(\tau_\text{had}, \ell)\) & Angular separation between \(\tau_\text{had}\) and \(\ell\) &
        \(\pt^{\ell}/\pt^{\tau_\text{had}}\) & Ratio of transverse momenta of lepton and \(\tau_\text{had}\)  \\
    \end{tabular}
    \end{ruledtabular}
\end{table*}

The primary features include the transverse momentum (\(\pt^{\tau_\text{had}}\), \(\pt^{\ell}\), \(\pt^{j_1}\), \(\pt^{j_2}\)), pseudorapidity (\(\eta^{\tau_\text{had}}\), \(\eta^{\ell}\), \(\eta^{j_1}\), \(\eta^{j_2}\)), and azimuthal angle (\(\phi^{\tau_\text{had}}\), \(\phi^{\ell}\), \(\phi^{j_1}\), \(\phi^{j_2}\)) of the $\tau_\text{had}$, the lepton, and up to two leading jets. Additionally, the missing transverse momentum (\pTmiss) and its azimuthal angle (\(\phi^\text{miss}\)) are included.

Derived observables are computed from these primary features and incorporate information relevant for distinguishing \(\PH\to\tau\tau\) signal events from background processes. These include kinematic quantities such as the transverse mass of the lepton and missing transverse energy (\(m_\text{T}(\ell, \pTmiss)\)), the visible invariant mass of the $\tau_\text{had}$ and the lepton (\(m_{\text{vis}}\)), the modulus of the vector sum of the $\tau_\text{had}$, lepton, and missing transverse momentum (\(\pt^\text{H}\)), the invariant mass of the two leading jets (\(m^{j_1j_2}\)), the number of reconstructed jets (\(N_j\)), and the sum of transverse jet momenta~($\sum_\text{jets}\pt$).
Additional variables exploit correlations between objects, such as the pseudorapidity separation (\(\Delta\eta^{j_1j_2}\)) and product (\(\eta^{j_1} \cdot \eta^{j_2}\)) between the two leading jets, the angular separation \(\Delta R(\tau_\text{had}, \ell)\) between the $\tau_\text{had}$ and the lepton, and event-level features such as the modulus of the total transverse momentum sum (\(\pt^\text{tot}\)) and the sum of transverse momenta of the $\tau_\text{had}$, lepton, and all jets~($\sum \pt$). In traditional analyses, the observables $\Delta\eta^{j_1j_2}$, $\eta^{j_1} \cdot \eta^{j_2}$, and $m^{j_1j_2}$ are used to identify signal events origination from vector-boson fusion~(VBF) production.
Furthermore, two centrality observables are included: the azimuthal centrality of the missing transverse energy with respect to the $\tau_\text{had}$ and the lepton (\(C^\text{miss}_\phi\)) and the pseudorapidity centrality of the lepton with respect to the two jets (\(C^\ell_\eta\)). The ratio of the transverse momenta of the lepton and the $\tau_\text{had}$ is also used (\(\pt^{\ell}/\pt^{\tau_\text{had}}\)). 

These engineered features are designed to enhance sensitivity to the Higgs boson decay topology and improve background rejection. Participants in the FAIR-HUC are free to utilize, modify, or omit these derived features based on their optimization strategy.
All features are summarized in Table~\ref{tab:event_features} with further implementation details provided in Ref.~\cite{Bhimji:2024bcd}.

Besides the $\PH\to\tau\tau$ signal, the FAIR-HUC dataset contains three background components: \(\PZ\to\tau\tau\), top quark pair production~(\ttbar), and diboson (\(\VV\)) processes. The \(\PZ\to\tau\tau\) background is irreducible, as it shares the same final state but differs due to the absence of Higgs-mediated production. The \ttbar background arises from semi-leptonic top decays, where hadronic \(\tau\) decays and missing transverse momentum resemble the signal topology. The \VV processes, primarily \(\PW\PW\), contribute when both bosons decay leptonically or when one produces the $\tau_\text{had}$. These backgrounds contribute to distinct kinematic regions and must be separated from the signal and from each other for an accurate measurement of the signal strength.
Backgrounds from instrumental effects, non-prompt leptons, or fake leptons are not included.

We employ the features to define two signal-enriched regions~(SRs), targeting the $\PH\rightarrow\tau\tau$ process, and three signal-depleted control regions~(CRs), which help constrain uncertainties in the rates of background processes and the calibration of physics objects.
In three of these regions, we perform unbinned cross-section measurements that fully exploit the high-dimensional feature correlation. 

The SR \texttt{lowMT-VBFJet} aims to select the VBF Higgs boson production by requiring \(\pt^{j_1} > 50\)~\GeV and \(\pt^{j_2} > 30\)~\GeV, referred to as the \texttt{VBFJet} selection. No explicit requirement on \(\Delta\eta^{j_1j_2}\) is applied, as this variable is used in the training of ML surrogates. To suppress the dominant \ttbar background, we impose \(\mT \leq 70\)~\GeV. 
The second SR, \texttt{lowMT-noVBFJet-ptH100}, targets gluon fusion~(\ggH), which predominantly produces the \PH boson at high \pt. This region is defined by \(\ptH > 100\)~\GeV, \(\mT \leq 70\)~\GeV, and an overlap removal implemented by inverting the \texttt{VBFJet} selection.
The CR \texttt{highMT-VBFJet} combines the \texttt{VBFJet} selection with \(\mT > 70\)~\GeV and is strongly enriched in \ttbar production due to the absence of a \(\Delta\eta^{j_1j_2}\) requirement.
The \(\PZ\to\tau\tau\) production is enriched in the \texttt{highMT-noVBFJet} region by selecting events with \(\mT > 70\)~\GeV and inverting the \texttt{VBFJet} selection. This region can be further split to target \ttbar production and \VV production individually using multivariate discriminants explained in Sec.~\ref{sec:CR-binning}.
The \texttt{lowMT-noVBFJet-ptH0to100} region selects a small yield of low-\pt \ggH events with \(\ptH \leq 100\)~\GeV and \(\mT < 70\)~\GeV. Because it is background-dominated and contains 95\% of the total yield, its main purpose is to constrain the overall normalization.
The event selections are summarized in Table~\ref{tab:selections}.

\begin{table*}
    \caption{Summary of the event selections. 
    The ``UB'' label indicates whether the selection is included in the unbinned analysis. Signal regions~(SR) and control regions~(CR) are treated the same. The predicted Poisson yields correspond to $\mathcal{L}\sigma$ where $\mathcal{L}$ denotes the luminosity and are reported for each sub-process and region. The last column shows the signal-to-background ratio. The total yields in the ``inclusive'' row are slightly lower than the total yields in Ref.~\cite{Bhimji:2024bcd} because of selections on $\pt^\PH$ and $\pt^\ell$ mandated by FAIR-HUC. The regions marked with $\ast$ are subsets of the \texttt{highMT-noVBFJet} region and target the \ttbar and \VV normalization via requirements on multivariate discriminants explained in Sec.~\ref{sec:CR-binning}.}
    \label{tab:selections}
    \renewcommand{\arraystretch}{1.3}
    \begin{ruledtabular}
    \begin{tabular}{l l c r r r r r}
        \multirow{2}{*}{\textbf{Region}} & \multirow{2}{*}{\textbf{Requirements}} & \multirow{2}{*}{\textbf{Type}} & \multicolumn{4}{c}{\textbf{Poisson yield} $\mathcal{L}\sigma$} & \multicolumn{1}{c}{\multirow{2}{*}{$S/B$}} \\
         &  &  & \multicolumn{1}{c}{\(\PH\to\tau\tau\)} & \multicolumn{1}{c}{\(\PZ\to\tau\tau\)} & \multicolumn{1}{c}{\(\ttbar\)} & \multicolumn{1}{c}{\(\VV\)} &  \\
        \hline
        \multirow{3}{*}{\texttt{lowMT-VBFJet}} & \(\pt^{j_1} > 50\)~\GeV & \multirow{3}{*}{UB, SR}\\
         & \(\pt^{j_2} > 30\)~\GeV & & 225.8 &  41\,280.4 &  15\,425.9 &   313.4 & \(3.96 \cdot 10^{-3}\)\\
         & \(\mT \leq 70\)~\GeV \\
        \hline
        \multirow{3}{*}{\texttt{highMT-VBFJet}} & \(\pt^{j_1} > 50\)~\GeV & \multirow{3}{*}{UB, CR}\\
         & \(\pt^{j_2} > 30\)~\GeV & & 14.7 &     721.7 &  16\,768.6 &   193.2 & \(8.30 \cdot 10^{-4}\) \\
         & \(\mT > 70\)~\GeV \\       
        \hline
        \multirow{3}{*}{\texttt{lowMT-noVBFJet-ptH100}} & \(\ptH > 100\)~\GeV & \multirow{3}{*}{UB, SR} \\
         & \(\mT \leq 70\)~\GeV & &  57.0 &  17\,379.8 &     674.6 &    79.0 & \(3.14 \cdot 10^{-3}\) \\
         & veto on \texttt{VBFJet} \\
        \hline
        \multirow{3}{*}{\texttt{lowMT-noVBFJet-ptH0to100}} & \(\ptH \leq 100\)~\GeV  \\
         & \(\mT \leq 70\)~\GeV & CR &  642.5 & 837\,928.7 &   3\,360.1 & 1\,438.5 & \(7.62 \cdot 10^{-4}\)\\
         & veto on \texttt{VBFJet} \\
         \hline
        \multirow{2}{*}{\texttt{highMT-noVBFJet}} & \(\mT > 70\)~\GeV & \multirow{2}{*}{CR}&   \multirow{2}{*}{26.0} &   \multirow{2}{*}{3\,826.9} &   \multirow{2}{*}{5\,054.3} & \multirow{2}{*}{1\,409.4} & \multirow{2}{*}{\(2.53 \cdot 10^{-3}\)} \\
         & veto on \texttt{VBFJet} \\
         \hline
         \texttt{inclusive} & & &   966.0 & 901\,137.5 & 41\,283.4 &  3\,433.5  & \(1.02 \cdot 10^{-3}\)\\
        \hline
        \multirow{3}{*}{\texttt{highMT-noVBFJet-tt}$^\ast$} & \(\mT > 70\)~\GeV & &   \multirow{3}{*}{3.2} &   \multirow{3}{*}{203.7} &   \multirow{3}{*}{3\,821.8} & \multirow{3}{*}{258.5} & \multirow{3}{*}{\(7.36 \cdot 10^{-4}\)} \\
        & veto on \texttt{VBFJet} & CR \\
        & $\hat f_{\ttbar}>0.4$ \\
        \hline
        \multirow{4}{*}{\texttt{highMT-noVBFJet-VV}$^\ast$} & \(\mT > 70\)~\GeV & \multirow{4}{*}{CR} &   \multirow{4}{*}{2.8} &   \multirow{4}{*}{165.7} &   \multirow{4}{*}{292.2} & \multirow{4}{*}{724.5} & \multirow{4}{*}{\(2.3 \cdot 10^{-3}\)} \\
        & veto on \texttt{VBFJet} \\
        & $\hat f_{\ttbar}\leq 0.4$ \\
        & $\hat f_{\VV}> 0.5$ \\
    \end{tabular}
\end{ruledtabular}
\end{table*}

\begin{figure*}
    \centering
    \includegraphics[width=0.33\linewidth]{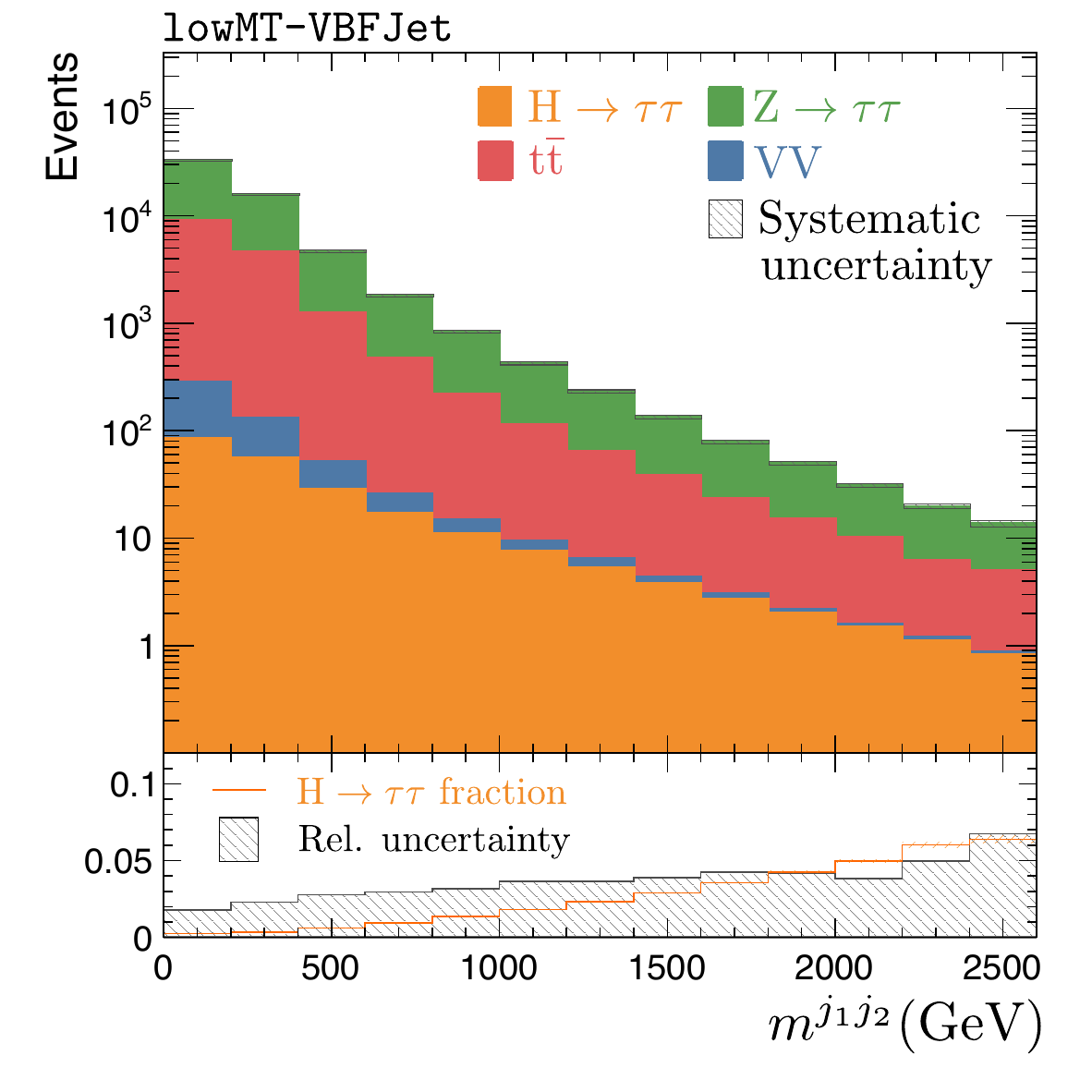}\hfill
    \includegraphics[width=0.33\linewidth]{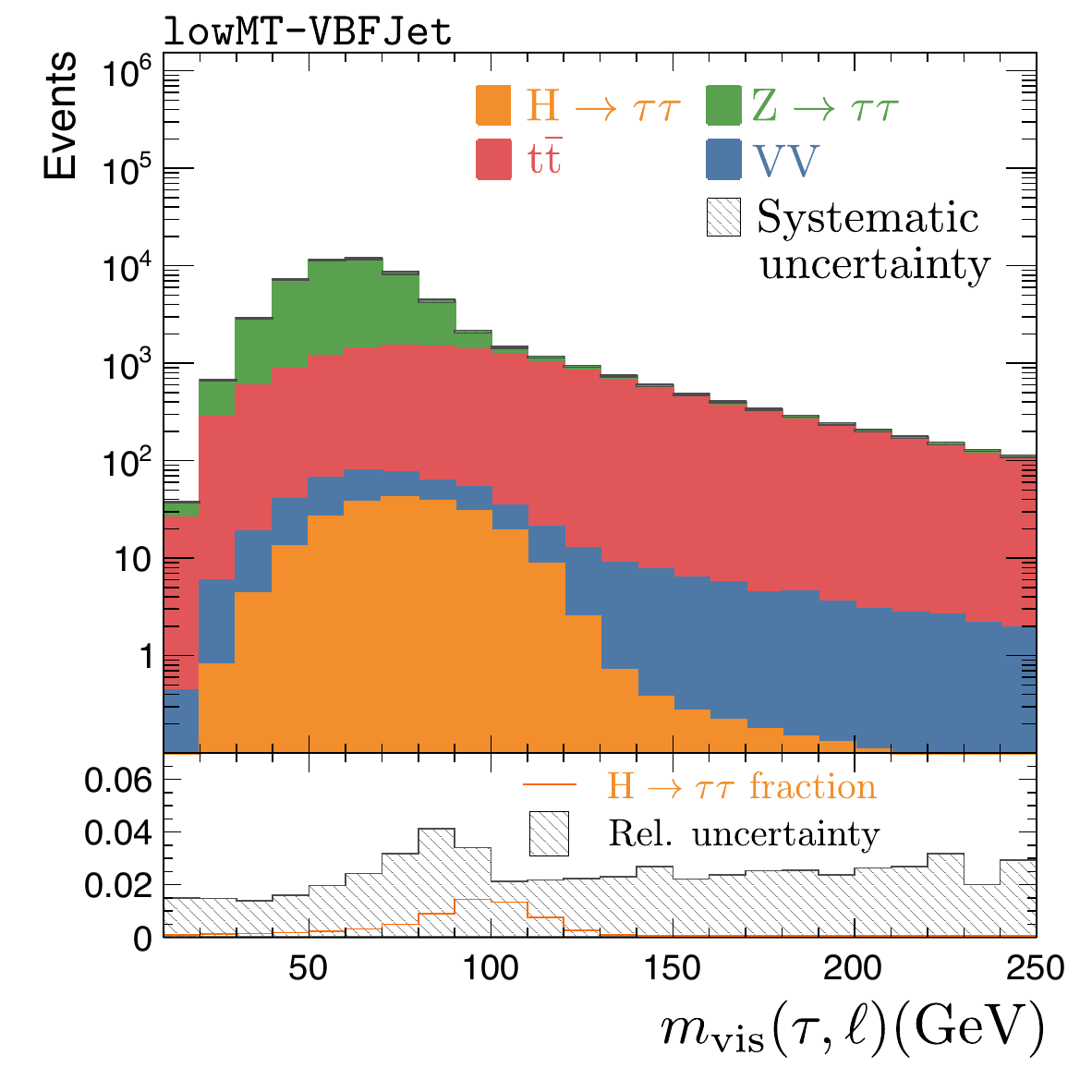}
    \includegraphics[width=0.33\linewidth]{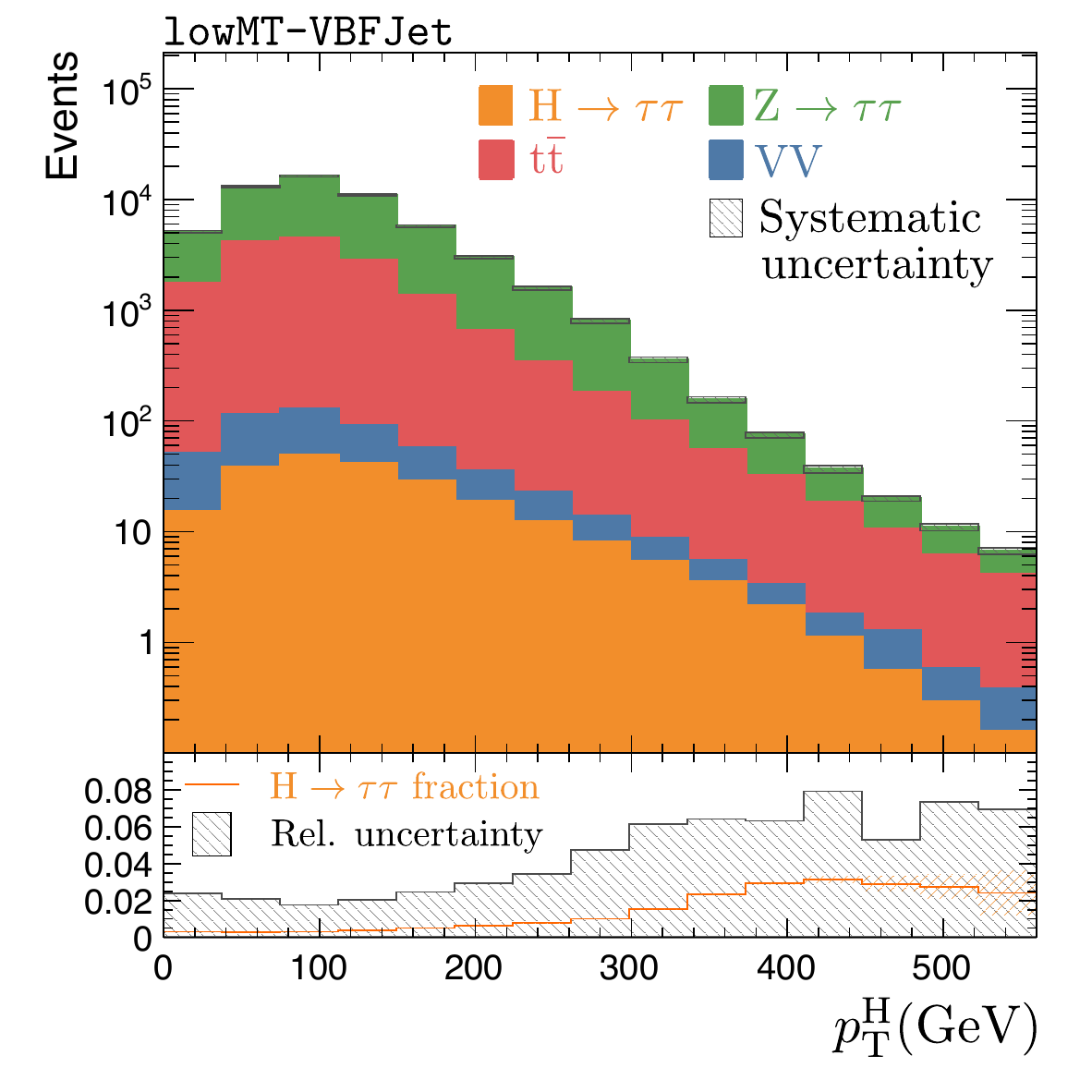}\hfill
    \caption{\label{fig:training-data} Training data distributions of $m^{j_1j_2}$~(left), $m_\text{vis}(\tau,\ell)$~(middle), and $p_\text{T}^\text{H}$~(right) in the main signal region \texttt{lowMT-VBFJet}. The total systematic uncertainty, obtained from uncorrelated $\pm1\sigma$ variations, is shown as a hashed band. In the bottom pad, the relative total uncertainty is compared to the fraction of the $\PH\to\tau\tau$ signal.}
\end{figure*}

\subsection{Systematic uncertainties}\label{sec:sys}

The systematic uncertainties of the FAIR-HUC can be categorized into ``calibration''-type and ``normalization''-type. The calibration-type uncertainties affect event reconstruction and selection, propagating from the primary features to the derived quantities. 
These include the jet energy scale (\texttt{jes}), associated with a nuisance parameter \(\nu_{\text{jes}}\), and the tau energy scale (\texttt{tes}), associated with \(\nu_{\text{tes}}\). When obtaining confidence intervals, these nuisance parameters contribute a penalty term to the likelihood from a hypothetical auxiliary measurement that is approximated with a standard normal distribution constrained to $\pm 10$ standard deviations. A variation of one standard deviation in \texttt{jes} and \texttt{tes} corresponds to 1\% shifts in the jet and tau energies, respectively. The ``soft \pTmiss'' energy scale, associated with \(\nu_{\text{met}}\), follows a log-normal distribution with mean zero and standard deviation \(\sigma_{\text{met}} = 1\) constrained to the interval~\([0,5]\). A computer code for event modification is provided~\cite{Bhimji:2024bcd} that recomputes primary and derived features as a function of $\nu_\text{jes}$,  $\nu_\text{tes}$, and $\nu_\text{met}$. The motivation for these choices and further details are provided in Ref.~\cite{Bhimji:2024bcd}. For toy studies, the nuisance parameter likelihood function is interpreted as a probability density function and sampled.

Log-normal uncertainties are used to vary the normalization of each process, scaling the yield by \((1+\alpha)^{\nu}\), where \(\alpha\) is a constant and \(\nu\) the corresponding nuisance parameter, associated with a standard normal distribution constrained to $\pm10$ standard deviations. Three normalization uncertainties are considered. The parameter \(\nu_{\text{bkg}}\) modifies the overall background normalization with a small impact of \(\alpha_{\text{bkg}} = 0.001\). The \ttbar background normalization uncertainty has \(\alpha_{\ttbar} = 0.02\), while the \VV normalization uncertainty is estimated at \(\alpha_{\VV} = 0.25\)~\cite{Bhimji:2024bcd}. These normalization uncertainties scale the background yields without altering their individual kinematic distributions but have indirect shape effects on the cumulative background. 

In total, the model includes seven parameters, abbreviated by \bn: the signal strength \(\mu\), three nuisance parameters related to calibration uncertainties (\(\nu_\text{jes}\), \(\nu_\text{tes}\), \(\nu_\text{met}\)), and three nuisance parameters  (\(\nu_\text{bkg}\), \(\nu_{\ttbar}\), \(\nu_\VV\)) controlling the background normalization. 

Figure~\ref{fig:training-data} shows examples of distributions of kinematic features in the \texttt{lowMT-VBFJet} region as provided in the training data. Consistent with results from ATLAS and CMS, this region will dominate the sensitivity to the signal strength. The effects of systematic uncertainties at the $1\sigma$ level exceed the small signal contribution almost everywhere. With signal-to-background ratios at most at the percent level, sophisticated methods are required to reliably isolate the small \(\PH\to\tau\tau\) contribution and to reduce the uncertainties.

Generating event samples for specific points in the model's parameter space is straightforward and allows, in principle, the evaluation of the likelihood.
However, for limit setting, we must evaluate the likelihood continuously and efficiently, and we must do so for unseen data. The following two sections describe how we address this challenge in the SBI context. The FAIR-HUC systematics represent the leading uncertainties for a $\PH\to\tau\tau$ cross-section measurement, but real-world analyses can include hundreds of nuisance parameters. Refinable modeling --- the ability to improve the description of individual backgrounds and/or systematic effects without affecting other parts of the analysis --- is, therefore, practically important. 

The statistical treatment of additional uncertainty sources for unbinned analyses, such as renormalization and factorization scale variations or \cPqb-tagging efficiencies, is discussed in Ref.~\cite{Schofbeck:2024zjo}.

\section{Modeling}\label{sec:modeling}

\subsection{The extended likelihood ratio test statistic}
In this section, we discuss the general statistical setup for unbinned modeling. It is suitable for FAIR-HUC, but more widely applicable to general unbinned LHC data analyses.

We aim to set confidence intervals on the signal strength parameter $\mu$ of the $\PH\to\tau\tau$ process. To this end, we compute the profiled likelihood ratio test statistic
\begin{align}    
q_{\mu}(\mathcal{D})&=-2\log\frac{\max_{\bn\phantom{,\bn}}L(\mathcal{D}|\mu,\bn)}{\max_{\mu,\bn}L(\mathcal{D}|\mu,\bn)}\label{eq:test-statistic}
\end{align}
of the observed unbinned dataset $\mathcal{D}$, comprising a variable-length list of high-dimensional feature vectors $\bx_i$, $\mathcal{D}=\{\bx_i\}_{i=1}^{N_\text{obs}}$.
In our case, \bx has 28 components, as listed in Table~\ref{tab:event_features}.
In the absence of \bn, the Neyman-Pearson lemma~\cite{Neyman:1933wgr} states that $q_\mu(\mathcal{D})$ is the optimal test statistic for all values of $\mu$. No such guarantee is known in the presence of nuisance parameters, but the profiled likelihood ratio is nevertheless widely adopted because of its asymptotic behavior: 
If $\mu$ is the true value, $q_{\mu}$ is asymptotically distributed as a $\chi^2$ distribution with one degree of freedom for any value of $\bn$ according to Wilks' theorem~\cite{Wilks:1938dza}. 
The approximate confidence interval $(\mu_{16}, \mu_{84})$ is given by the solutions to $q_{\mu}(\mathcal{D})=1$ which are obtained numerically by scanning $\mu$ and minimizing $\bn$ for each value of $\mu$. 

Next, we derive the profiled likelihood ratio, which is the main test statistic used in this work. We assume the likelihood is ``extended'' by a Poisson factor with observation $N_\text{obs}$~\cite{Barlow:1990vc}. The remaining factor, the probability to observe a number of $N_\text{obs}$ independently and identically distributed events $\bx_i\sim p(\bx|\mu,\bn)$, then leads to
\begin{align}
    L(\mathcal{D}|\mu,\bn)=\textrm{P}(N_\text{obs}|\mathcal{L}\sigma(\mu,\bn))\prod_{i=1}^{N_\text{obs}}p(\bx_i|\mu,\bn).\label{eq:extended-likelihood}
\end{align}
where $\mathcal{L}$ denotes the luminosity, $\sigma(\mu,\bn)$ the total cross-section, and $\textrm{P}$ the probability density function of the Poisson process.
We convert the normalized probability density function $p(\bx|\mu,\bn)$ to the differential cross sections via
\begin{align}
    \ddd\sigma(\bx|\mu,\bn)=\sigma(\mu,\bn)p(\bx|\mu,\bn)\ddd\bx\label{eq:normalization}
\end{align}
so we have $\sigma(\mu,\bn)=\int\ddd\sigma(\bx|\mu,\bn)$ for the inclusive cross section. 
To simplify the profiling, we divide both the numerator and denominator likelihoods in Eq.~\ref{eq:test-statistic} by a reference likelihood at $(\mu,\bn)=(1,\bzero)$~\cite{Schofbeck:2024zjo}, corresponding to a signal strength of $\mu=1$ and the nominal value for the nuisance parameters.
This yields
\begin{align}
    q_{\mu}(\mathcal{D}) = \min_{\bn} u(\mathcal{D}|\mu,\bn) - \min_{\mu,\bn} u(\mathcal{D}|\mu,\bn),\label{eq:test-stat}
\end{align}
where   
\begin{align}
-\frac{1}{2}u(\mathcal{D}|\mu,\bn) &= -\mathcal{L}(\sigma(\mu,\bn) - \sigma(1,\bzero)) \nonumber\\&+ \sum_{i=1}^{N_\text{obs}}\log\left(\frac{\ddd\sigma(\bx_i|\mu,\bn)}{\ddd\sigma(\bx_i|1,\bzero)}\right)\label{eq:likelihoodratio-normalized}
\end{align}
is the log-likelihood ratio to the reference. The first term in Eq.~\ref{eq:test-stat} is the best-fit value of $u$ for a given $\mu$ over the nuisance parameters, while the second term corresponds to the best-fit value, also including $\mu$ in the fit.

The term in the logarithm in Eq.~\ref{eq:likelihoodratio-normalized} is the differential cross-section ratio~(DCR). With the inclusive cross-section and the DCR, we can evaluate Eq.~\ref{eq:likelihoodratio-normalized}. These quantities, therefore, are the goal of unbinned SBI-based strategy. We compute the quantities with ML surrogates in Sec.~\ref{sec:ML}.

In a binned approach, where $\mathcal{D}$ is represented by a number of observed counts~ $(N_{\text{obs},i})$ associated with Poisson bins labeled by $i$, Eq.~\ref{eq:likelihoodratio-normalized} simplifies to
\begin{align}
-\frac{1}{2}u_\text{binned}(\mathcal{D}|\mu,\bn) &= -\mathcal{L}(\sigma(\mu,\bn) - \sigma(1,\bzero)) \nonumber\\&+ \sum_{i=1}^{N_\text{bins}}N_{\text{obs},i}\log\left(\frac{\sigma_i(\mu,\bn)}{\sigma_i(1,\bzero)}\right).\label{eq:likelihoodratio-normalized-binned}
\end{align}
The per-bin cross-section values $\sigma_i(1,\bzero)$ and $\sigma_i(\mu,\bn)$ can be approximately evaluated from simulated (training) data. The continuous interpolation as a function of $\bn$ is achieved by (log-)polynomial approximations~\cite{Cranmer:2014lly,CMS:2024onh}, and we will obtain it for FAIR-HUC in the following section. The total cross section is given as the sum over bins, $\sigma(\mu,\bn)=\sum_{i=1}^{N_{\text{bins}}}\sigma_i(\mu,\bn)$.

\subsection{Binned likelihoods}\label{sec:binned-LL}

In this section, we compute the necessary interpolations for the binned likelihood function defined in Eq.~\ref{eq:likelihoodratio-normalized-binned}, covering the case of an arbitrary number of Poisson-distributed observations indexed by \(i\) as commonly applied in traditional analyses. We extend the nuisance parameter dependence to higher-order polynomials in \bn and allow systematic effects on binned yields to be non-factorizable. Although the discussion is tailored to the seven parameters in FAIR-HUC, the presented methodology is general and provides a guardrail for the unbinned case. 

The binned case also illustrates conceptual similarities between the two approaches. Crucially, both binned and unbinned methods allow for iterative refinements without invalidating previous intermediate results. In the binned case, this is straightforward: additional subleading (background) contributions can be added to the Poisson yield as needed, and further systematic effects can be incorporated via nuisance parameters that rescale the affected predictions. It is of practical importance that this flexibility extends to the unbinned case.

We construct the model of binned Poisson yields by separating into contributions from the signal and the background processes as
\begin{align}
\sigma_i(\mu,\bn) &= \mu\,\sigma_{i,\PH}(\bn_{\text{calib}})+\sum_{p=\PZ,\ttbar,\VV}\sigma_{i,p}(\bn).\label{eq:binned-param}
\end{align}
The normalization uncertainties enter the per-bin per-process cross-sections as
\begin{subequations}
\label{eq:sigma_all}
\begin{align}
\sigma_{i,\PZ}(\bn)=(1+\alpha_\text{bkg})^{\nu_\text{bkg}}\sigma_{i,\PZ}(\bn_{\text{calib}}),\label{eq:sigma_i_PZ}\\
\sigma_{i,\ttbar}(\bn)=(1+\alpha_\text{bkg})^{\nu_\text{bkg}}(1+\alpha_{\ttbar})^{\nu_{\ttbar}}\sigma_{i,\ttbar}(\bn_{\text{calib}}),\\
\sigma_{i,\VV}(\bn)=(1+\alpha_\text{bkg})^{\nu_\text{bkg}}(1+\alpha_\VV)^{\nu_\VV}\sigma_{i,\VV}(\bn_{\text{calib}}),\label{eq:sigma_i_VV}
\end{align}
\end{subequations}
where we have introduced the abbreviation $\bn_{\text{calib}}=\{\nu_{\text{tes}},\nu_{\text{jes}},\nu_{\text{met}}\}$ which we use to parameterize the calibration-type systematic uncertainties dependence. 

We next parameterize $\sigma_{i,p}(\bn_\text{calib})$ for $p=\{\PH,\PZ,\ttbar,\VV\}$. In standard binned analyses as implemented with, e.g., the \textsc{combine} package~\cite{CMS:2024onh}, systematic up- and down variations corresponding to $\nu=\pm 1$ are used to obtain a continuous interpolation~\cite{Cranmer:2014lly} of the dependence of the simulated yields on the nuisance parameters. Typically, a log-polynomial expansion is used, and the nuisance parameter dependence is assumed to factorize. 
Here, we do not use this assumption and include cross-terms in a log-polynomial expansion, including the mixed terms, concretely
\begin{align}
    &\log \frac{\sigma_{i,p} (\bn_{\text{calib}})}{\sigma_{i,p}(\bzero)}=\nu_\text{tes}\,\hat\Delta_{i,p,\text{tes}}+\nu_\text{jes}\,\hat\Delta_{i,p,\text{jes}}+\nu_\text{met}\,\hat\Delta_{i,p,\text{met}}\nonumber\\    &\quad\quad+\nu_\text{tes}^2\,\hat\Delta_{i,p,\text{tes},\text{tes}}+\nu_\text{jes}^2\,\hat\Delta_{i,p,\text{jes},\text{jes}}+\nu_\text{met}^2\,\hat\Delta_{i,p,\text{met},\text{met}}\nonumber\\ &\quad\quad+\nu_\text{tes}\nu_\text{jes}\,\hat\Delta_{i,p,\text{tes},\text{jes}}+\nu_\text{jes}\nu_\text{met}\,\hat\Delta_{i,p,\text{jes},\text{met}}\nonumber\\
    &\quad\quad+\nu_\text{tes}\nu_\text{met}\,\hat\Delta_{i,p,\text{tes},\text{met}}=\nu_A\hat\Delta_{i,p,A}.\label{eq:poly-exp}
\end{align}
The first three coefficients correspond to the log-linear dependence, the three terms in the second line to the log-quadratic dependence and the remainder constitute the mixed terms. For each bin $i$ and process $p$, we thus need to determine 9 constants $\hat\Delta_{i,p,A}$ that approximate the l.h.s. 
To simplify the notation, we label the 9 linear, quadratic, and mixed coefficients as
\begin{align}
    A=\{&\{\text{tes}\},\{\text{jes}\},\{\text{met}\},\{\text{tes},\text{tes}\},\{\text{jes},\text{jes}\},\{\text{met},\text{met}\}, \nonumber\\
    &\{\text{tes},\text{jes}\},\{\text{jes},\text{met}\},\{\text{tes},\text{met}\}\},\label{eq:index-def}
\end{align}
which defines the last equality in Eq.~\ref{eq:poly-exp} where we use the Einstein sum convention for the new index $A$. Note that $\nu_A$ also contains terms linear and quadratic in $\bn_{\text{calib}}$, explicitly 
\begin{align}
    \nu_A=\{&\nu_\text{tes},\nu_\text{jes},\nu_\text{met},\nu_\text{tes}^2,\nu_\text{jes}^2,\nu_\text{met}^2,\nonumber\\&\nu_\text{tes}\nu_\text{jes},\nu_\text{jes}\nu_\text{met},\nu_\text{tes}\nu_\text{met}\}.\label{eq:nu-def2}
\end{align}
Note also that we are free to drop individual terms from Eq.~\ref{eq:index-def} if it simplifies the analysis. Finally, Eq.~\ref{eq:index-def} extends to an arbitrary number of nuisance parameters and polynomial order straightforwardly so that the following discussion and the ML strategy are generally applicable. 

We determine the constant coefficients $\hat\Delta_{i,p,A}$ from simulated yields $\sigma_{i,p}(\bn_\text{calib})$, where $\bn_\text{calib}=\bzero$, as usual, corresponds to the nominal calibration of the training data. The three-dimensional parameter points $\bn_\text{calib}$ come from a training set $\mathcal{V}$, which we must choose such that all coefficients $\hat\Delta_{i,p,A}$ can be determined. Given $\mathcal{V}$, we obtain the constants by minimizing
\begin{align}
    \chi^2 = \sum_{\bn\in\mathcal{V}}\left(\log \frac{\sigma_{i,p} (\bn)}{\sigma_{i,p}(\bzero)} - \nu_A\hat\Delta_{i,p,A}\right)^2
\end{align}
independently for each bin $i$ and process $p$. Here, we omit the subscript ``calib'' for readability. The solution is obtained analytically by differentiation~\cite{Schofbeck:2024zjo} as
\begin{align}
    \hat\Delta_{i,p,A} = \left[\;\sum_{\bn\in\mathcal{V}}\bn\bn^{\textrm{T}}\right]_{AB}^{-1}
    \left[\sum_{\;\bn\in\mathcal{V}}\bn\log\frac{\sigma_{i,p}(\bn)}{\sigma_{i,p}(\bzero)}\right]_B.\label{eq:icp}
\end{align}

From the first factor in Eq.~\ref{eq:icp}, it follows that the condition on $\mathcal{V}$ is that the $9\times9$ matrix $\left[\sum_{\bn\in\mathcal{V}}\bn\bn^{\textrm{T}}\right]_{AB}$ is invertible. As before, \(A\) and \(B\) index the nine basis terms used in the polynomial expansion. It is always possible to find a suitable $\mathcal{V}$~\cite{Schofbeck:2024zjo}; specifically, we select $\nu = -3,-2,-1,0,1,2,3$ for \texttt{tes} and \texttt{jes}, and $\nu_\text{met} = 0,1,2,3$. From all combinations of these integer values, we retain the 44 parameter vectors also satisfying $|\nu_\text{tes}| + |\nu_\text{jes}| + |\nu_\text{met}| \leq 3$. The resulting $\mathcal{V}$ is sufficient to evaluate Eq.~\ref{eq:icp}.

There is considerable flexibility in this procedure. Note, for example, that we can estimate $\hat\Delta_{i,p,A}$ separately for groups of nuisance parameters, provided that we do not include mixed terms that spoil the factorization of the effects of these groups in Eq.~\ref{eq:index-def}. 
In fully realistic settings, this allows the estimation of systematic effects independently of each other.

This completes the construction of the binned likelihood function, summarized as follows. The background normalization uncertainties are implemented using Eqs.~\ref{eq:sigma_i_PZ}--\ref{eq:sigma_i_VV}, where $\sigma_{i,p}(\bn_\text{calib})$ is computed as
\begin{align}
\sigma_{i,p}(\bn_\text{calib}) = \exp(\nu_A\hat\Delta_{i,p,A})\sigma_{i,p}(1,\bzero).\label{eq:sigma-param}
\end{align}
The three-dimensional $\bn_\text{calib}$ leads to nine coefficients $\hat\Delta_{i,p,A}$ up to quadratic order per bin and process, each labeled by $A$ as given in Eq.~\ref{eq:index-def}.
The $\hat\Delta_{i,p,A}$ are determined from Eq.~\ref{eq:icp}, using systematically varied yields evaluated at an adequate set of parameter points $\mathcal{V}$. With these quantities, the test statistic in Eq.~\ref{eq:test-stat} and the binned likelihood ratio in Eq.~\ref{eq:likelihoodratio-normalized-binned} can be evaluated. 

Table~\ref{tab:systematic_dependence} summarizes the parametrization of systematic dependencies in the cross-sections of the five regions defined in Table~\ref{tab:selections} up to log-quadratic order and each region treated as a single bin. It shows that $\nu_\text{tes}$ and $\nu_\text{jes}$ have impacts of $10^{-3}$--$10^{-2}$ dominated by the log-linear term, while $\nu_\text{met}$ is smaller overall but with a non-negligible log-quadratic term. Mixed contributions are generally small.

The unbinned estimates will follow the same procedure except that the constants $\hat\Delta_{i,p,A}$ become $\bx$-dependent functions which we implement with neural networks and train by minimizing a cross-entropy~(CE) loss function instead of a $\chi^2$.

\begin{table*}
    \centering
    \caption{Systematic dependence of the four processes (\(\PH\to\tau\tau\), \(\PZ\to\tau\tau\), \ttbar, and \VV) in the five regions. Each row represents the coefficients of a quadratic polynomial describing the impact of the nuisance parameters \(\nu_{\text{tes}}\), \(\nu_{\text{jes}}\), and \(\nu_{\text{met}}\).}
    \label{tab:systematic_dependence}
    \renewcommand{\arraystretch}{1.3}
    \begin{ruledtabular}
    \begin{tabular}{lrrrrrrrrrr}
 &  
$\nu_{\text{tes}}\cdot10^{-3}$&  
$\nu_{\text{jes}}\cdot10^{-3}$&  
$\nu_{\text{met}}\cdot10^{-5}$&  
$\nu_{\text{tes}}^2\cdot10^{-5}$ &  
$\nu_{\text{jes}}^2\cdot10^{-5}$ &  
$\nu_{\text{met}}^2\cdot10^{-5}$ &  
$\nu_{\text{tes}} \nu_{\text{jes}}\cdot10^{-5}$ &  
$\nu_{\text{jes}} \nu_{\text{met}}\cdot10^{-5}$ &  
$\nu_{\text{tes}} \nu_{\text{met}}\cdot10^{-5}$  \\ \hline
    \multicolumn{10}{c}{\texttt{lowMT-VBFJet}}\\ \hline
    \(\PH\to\tau\tau\)  & 6.03  & 13.0   & -3.40  & -9.91   & -16.5   & -9.15   & 4.63   & 5.28   & -0.451 \\
    \(\PZ\to\tau\tau\)  & 9.89  & 22.0   & 2.57   & -12.8   & -16.0   & -6.87   & 7.25   & 2.03   & -1.78 \\
    \(\ttbar\)          & 8.78  & 7.15   & -6.36  & -8.89   & -18.1   & -13.7   & 2.38   & 0.466  & -3.84 \\
    \(\VV\)             & 11.0  & 20.4   & 11.3   & -6.43   & -4.85   & -6.97   & 2.60   & -59.9  & 14.5  \\
        \hline
    \multicolumn{10}{c}{\texttt{highMT-VBFJet}}\\ \hline
    \(\PH\to\tau\tau\)  & -1.95 & 8.85   & 29.3   & 23.3    & 67.7    & 148     & -59.3  & -10.3  & 2.60  \\
    \(\PZ\to\tau\tau\)  & 8.32  & 3.60   & 22.5   & 34.8    & 149     & 333     & -119   & -11.7  & 29.5  \\
    \(\ttbar\)          & 6.25  & 6.50   & -0.402 & -6.40   & -8.06   & 14.6    & -2.05  & -1.05  & -0.172\\
    \(\VV\)             & 6.57  & 16.5   & -24.6  & -4.91   & -18.7   & 13.4    & 4.71   & 3.45   & -10.9 \\
 \hline
    \multicolumn{10}{c}{\texttt{highMT-noVBFJet}} \\\hline
    \(\PH\to\tau\tau\)  & -10.2 & -3.83  & -18.4  & 8.18    & 27.9    & 351     & -18.6  & 2.41   & -6.94 \\
    \(\PZ\to\tau\tau\)  & -7.65 & -2.56  & 46.7   & 2.35    & 46.4    & 930     & -37.5  & 5.82   & 20.2  \\
    \(\ttbar\)          & 5.32  & -24.0  & -23.9  & -5.14   & 9.06    & 7.79    & -3.07  & -9.88  & 3.33 \\
    \(\VV\)             & 6.99  & -3.16  & 7.56   & -11.7   & -2.86   & 5.61    & -6.29  & 8.78   & 5.92  \\
        \hline
    \multicolumn{10}{c}{\texttt{lowMT-noVBFJet-ptH100}} \\\hline
    \(\PH\to\tau\tau\)  & 5.38  & 2.49   & 5.27   & -7.84   & -11.9   & 24.4    & 9.03   & -4.20  & 1.61 \\
    \(\PZ\to\tau\tau\)  & 7.06  & 12.9   & -9.71  & -8.88   & -19.9   & 51.6    & 9.48   & -5.45  & -1.04 \\
    \(\ttbar\)          & 7.50  & -2.96  & 82.6   & -4.46   & -1.11   & 4.63    & 12.1   & 26.5   & 5.17  \\
    \(\VV\)             & 10.2  & 9.49   & 341    & 27.6    & -19.0   & -72.5   & 9.88   & 66.8   & 33.3  \\
    \hline
    \multicolumn{10}{c}{\texttt{lowMT-noVBFJet-ptH0to100
}} \\\hline
    \(\PH\to\tau\tau\)  & 6.42  & -4.83  & -0.726 & -11.1   & -0.15   & -16.2   & -0.20  & 0.205  & -0.44 \\
    \(\PZ\to\tau\tau\)  & 12.5  & -1.34  & -2.28  & -22.4   & -0.644  & -4.67   & 0.137  & -0.055& -1.57 \\
    \(\ttbar\)          & 8.40  & -28.4  & 15.0   & -10.9   & 2.93    & -11.3   & -3.67  & 6.35   & -8.47 \\
    \(\VV\)             & 14.2  & -4.07  & -4.42  & -10.8   & -0.172  & -8.79   & 5.44   & 0.067 & 1.59  \\
    \end{tabular}
    \end{ruledtabular}
\end{table*}

\subsection{Unbinned likelihoods}
\label{sec:unbinned}
Comparing Eq.~\ref{eq:likelihoodratio-normalized} to Eq.~\ref{eq:likelihoodratio-normalized-binned}, the unbinned likelihood can be viewed as the continuum limit of the binned Poisson likelihood\footnote{A first-principles derivation is provided in, e.g., Ref.~\cite{Schofbeck:2024zjo}.} where the granularity of the cross-section ratio in the logarithm becomes arbitrarily fine~\cite{GomezAmbrosio:2022mpm}. In analogy to the binned case, we therefore define an additive fully differential model
\begin{align}
\ddd\sigma(\bx|\mu,\bn) &= \mu\,\ddd\sigma_{\PH}(\bx|\bn_{\text{calib}})+\sum_{p=\PZ,\ttbar,\VV}\ddd\sigma_{p}(\bx|\bn).\label{eq:unbinned-additive-model}
\end{align}
The relation to the binned Poisson yields is given by 
\begin{align}
\sigma_{i,p}(\bn)=\int_{\Delta\bx_i}\frac{\ddd\sigma_{p}(\bx|\bn)}{\ddd\bx}\ddd\bx
\end{align}
with $\Delta\bx_i$ the phase-space region associated with bin~$i$.
It is important that Eq.~\ref{eq:unbinned-additive-model} is available via event generators only in the ``forward'' mode; we may specify $\bn$ and then obtain event samples, but we cannot use those to evaluate, e.g., the DCR in Eq.~\ref{eq:likelihoodratio-normalized} for newly observed data. This was, in fact, true already in the binned case: To obtain a parametrization that is continuous in \bn, we used the log-polynomial ansatz in Eq.~\ref{eq:poly-exp} and had to solve a system of equations for the per-bin constants $\hat\Delta_{i,p,A}$ based on training data at certain values of \bn. A similar interpolation is implemented in \textsc{combine}~\cite{CMS:2024onh}.

Our next goal is to define a fully unbinned model. We start with specifying the analytic dependence of the normalization-type systematic uncertainties in complete analogy with the binned case as
\begin{subequations}
\label{eq:nuisance-dep-unbinned}
\begin{align}
\ddd\sigma_{\PZ}(\bx|\bn)&=(1+\alpha_\text{bkg})^{\nu_\text{bkg}}\ddd\sigma_{\PZ}(\bx|\bn_{\text{calib}}),\\
\ddd\sigma_{\ttbar}(\bx|\bn)&=(1+\alpha_\text{bkg})^{\nu_\text{bkg}}(1+\alpha_{\ttbar})^{\nu_{\ttbar}}\ddd\sigma_{\ttbar}(\bx|\bn_{\text{calib}}),\\
\ddd\sigma_{\VV}(\bx|\bn)&=(1+\alpha_\text{bkg})^{\nu_\text{bkg}}(1+\alpha_\VV)^{\nu_\VV}\ddd\sigma_{\VV}(\bx|\bn_{\text{calib}}).
\end{align}
\end{subequations}
We note that Eq.~\ref{eq:likelihoodratio-normalized} only asks for the DCR, which we can write as 
\begin{align}
&\frac{\ddd\sigma(\bx|\mu,\bn)}{\ddd\sigma(\bx|1,\bzero)\,}=\mu\frac{\ddd\sigma_\PH(\bx|\bn_\text{calib})}{\ddd\sigma(\bx|1,\bzero)}\nonumber\\
&\qquad+(1+\alpha_\text{bkg})^{\nu_\text{bkg}}\Bigg(\frac{\ddd\sigma_\PZ(\bx|\bn_\text{calib})}{\ddd\sigma(\bx|1,\bzero)}\nonumber\\
&\qquad+(1+\alpha_{\ttbar})^{\nu_{\ttbar}}\,\frac{\ddd\sigma_{\ttbar}(\bx|\bn_\text{calib})}{\ddd\sigma(\bx|1,\bzero)}\nonumber\\
&\qquad+(1+\alpha_\VV)^{\nu_\VV}\,\frac{\ddd\sigma_{\VV}(\bx|\bn_\text{calib})}{\ddd\sigma(\bx|1,\bzero)}\Bigg).\label{eq:xsec-ratio}
\end{align}
We expand the DCRs for $p=$\PH, \PZ, \ttbar, \VV on the r.h.s. as
\begin{align}
\frac{\ddd\sigma_p(\bx|\bn_\text{calib})}{\ddd\sigma(\bx|1,\bzero)}&=\frac{\ddd\sigma_p(\bx|\bn_\text{calib})}{\ddd\sigma_p(\bx|\bzero)}\frac{\ddd\sigma_p(\bx|\bzero)}{\ddd\sigma(\bx|1,\bzero)}\nonumber\\
&\simeq \hat S_p(\bx|\bn_\text{calib})\hat g_p(\bx)
\end{align}
where the second line uses the ML approximations 
\begin{align}
    \hat g_p(\bx)&\simeq\frac{\ddd\sigma_p(\bx|\bzero)}{\ddd\sigma(\bx|1,\bzero)}=\frac{\ddd\sigma_p(\bx|\bzero)}{\sum_q\ddd\sigma_q(\bx|\bzero)}\;\text{and}\label{eq:ML-gp}\\
    \hat S_p(\bx|\bn_\text{calib})&\simeq\frac{\ddd\sigma_p(\bx|\bn_\text{calib})}{\ddd\sigma_p(\bx|\bzero)}.\label{eq:ML-S}
\end{align}
The sum over $q$ extends over \PH, \PZ, \ttbar, and \VV.
Note that none of these quantities depend on $\mu$ or the normalization-type nuisance parameters.
Because the denominator in Eq.~\ref{eq:ML-gp} is just the total nominal differential cross-section, the quantity $\hat g_p(\bx)$ predicts the \bx-dependent DCR of process $p$ to the total differential cross-section. Similarly, the parametric estimator $\hat S_p(\bx|\bn_\text{calib})$ predicts the relative dependence of the differential cross-section of a process $p$ on the nuisance parameters $\bn_\text{calib}$ and \bx. 

In summary, our ML surrogate model of the DCR is
\begin{align}
&\frac{\ddd\sigma(\bx|\mu,\bn)}{\ddd\sigma(\bx|1,\bzero)\,}\simeq\hat R(\bx|\mu,\bn)=\mu\hat g_\PH(\bx)\hat S_\PH(\bx|\bn_\text{calib})\nonumber\\
&\qquad+(1+\alpha_\text{bkg})^{\nu_\text{bkg}}\Big(\hat g_\PZ(\bx)\hat S_\PZ(\bx|\bn_\text{calib})\nonumber\\
&\qquad+(1+\alpha_{\ttbar})^{\nu_{\ttbar}}\,\hat g_{\ttbar}(\bx)\hat S_{\ttbar}(\bx|\bn_\text{calib})\nonumber\\
&\qquad+(1+\alpha_\VV)^{\nu_\VV}\,\hat g_{\VV}(\bx)\hat S_{\VV}(\bx|\bn_\text{calib})\Big).\label{eq:xsec-ratio-ML}
\end{align}
We devise training strategies for $\hat g_p(\bx)$ and $\hat S_p(\bx|\bn_\text{calib})$ in~Sec.~\ref{sec:ML}.

\newpage
While our DCR surrogate captures the full dependence on all seven model parameters, practical applications typically require incremental refinement. Binned analysis development starts with the dominant backgrounds and leading uncertainties and then adds sub‐leading components and systematic effects to the Poisson yields. Our approach extends this staged workflow to unbinned modeling, allowing the gradual inclusion of new effects, in analogy to the binned case.

For instance, unmodeled systematics can be incorporated into Eq.~\ref{eq:xsec-ratio-ML} by multiplying with additional factors of \(\hat S\), using per‐process approximations as in Eq.~\ref{eq:ML-S}. If mixed terms between the new and nominal systematics are negligible, this single factor suffices, and the existing ML surrogates remain valid.

Similarly, unaccounted background contributions can be included. For example, we can refine the existing differential cross-section by adding a new component $\ddd\sigma_\text{new}$,
\begin{align}
\ddd\sigma'(\bx|\mu,\bn)=\ddd\sigma(\bx|\mu,\bn)+\ddd\sigma_\text{new}(\bx|\bn).
\end{align}
We can then rewrite the updated DCR in terms of the original model as
\begin{align}
&\frac{\ddd\sigma'(\bx|\mu,\bn)}{\ddd\sigma'(\bx|1,\bzero)\,}=\frac{\ddd\sigma(\bx|\mu,\bn)+\ddd\sigma_\text{new}(\bx|\bn)}{\ddd\sigma(\bx|1,\bzero)+\ddd\sigma_\text{new}(\bx|\bzero)}\nonumber\\
&\qquad=\frac{\frac{\ddd\sigma(\bx|\mu,\bn)}{\ddd\sigma(\bx|1,\bzero)\,}+\frac{\ddd\sigma_\text{new}(\bx|\bn)}{\ddd\sigma_\text{new}(\bx|\bzero)}\frac{\ddd\sigma_\text{new}(\bx|\bzero)}{\ddd\sigma(\bx|1,\bzero)}}{1+\frac{\ddd\sigma_\text{new}(\bx|\bzero)}{\ddd\sigma(\bx|1,\bzero)}}\nonumber\\
&\qquad\simeq\frac{\hat R(\bx|\mu,\bn)+\hat S_\text{new}(\bx|\bn)\hat g_\text{new}(\bx)}{1+\hat g_\text{new}(\bx)}\label{eq:DCR-refined}
\end{align}
where the only additional surrogates needed are
\begin{align}
    \hat g_\text{new}(\bx)&\simeq\frac{\ddd\sigma_\text{new}(\bx|\bzero)}{\ddd\sigma(\bx|1,\bzero)},\;\text{and}\nonumber\\
    \hat S_\text{new}(\bx|\bn)&\simeq\frac{\ddd\sigma_\text{new}(\bx|\bn)}{\ddd\sigma_\text{new}(\bx|\bzero)}.
\end{align}
Expressing the refined DCR in Eq.~\ref{eq:DCR-refined} entirely in terms of the original DCR and only two surrogates for the new component mirrors the flexibility of the binned approach and facilitates the refinable modeling for unbinned analyses.

\subsection{Inclusive cross-section parametrization}\label{sec:CSI}

If Eq.~\ref{eq:ML-gp} and Eq.~\ref{eq:ML-S} are available, we can efficiently evaluate the inclusive cross-section difference, which is the final ingredient required in Eq.~\ref{eq:likelihoodratio-normalized}. Using a nominal simulation at $\mu=1$ and $\bn=0$, i.e., an event sample $\mathcal{D}_{1,\bzero}\sim\ddd\sigma(\bx|1,\bzero)$, we obtain
\begin{align}
    &\mathcal{L}(\sigma(\mu,\bn) - \sigma(1,\bzero))\nonumber\\
    &\quad\quad=\mathcal{L}\int \left(\frac{\ddd\sigma(\bx|\mu,\bn)}{\ddd\sigma(\bx|1,\bzero)\,}-1\right)\frac{\ddd\sigma(\bx|1,\bzero)}{\ddd\bx}\ddd\bx\nonumber\\
    &\quad\quad\approx\sum_{\{w_i,\bx_i\}\in\mathcal{D}_{1,\bzero}} w_i\left(\hat R(\bx_i|\mu,\bn)-1\right).\label{eq:inc-xsec}
\end{align}
The event weights $w_i$ implement the cross-section normalization such that 
\begin{align}
    \sum_{D_{1,\bzero}^{(p)}}\,w_i=\mathcal{L}\sigma_p(\bzero),
\end{align}
where $D_{1,\bzero}^{(p)}$ is the sample with nominal simulation of process $p$ in Table~1 of Ref.~\cite{Bhimji:2024bcd}. Because $\sum_p \hat g_p(\bx)=1$ by construction, it follows from Eq.~\ref{eq:xsec-ratio-ML} that $\hat R(\bx|1,\bzero)=1$, which expresses the trivial fact that the DCR of (1,\bzero) to the reference hypothesis, which is characterized by the same parameters, is one. 

This completes the construction of the unbinned surrogate model, which is, in principle, ready for inference once $\hat g_p(\bx)$ and $\hat\Delta_{p,A}(\bx)$ are available.

In practice, Eq.~\ref{eq:inc-xsec} must be evaluated using the training dataset. During the profiling of nuisance parameters when computing $q_\mu$, many evaluations are required. Although the surrogate functions $\hat g_p(\bx)$ and $\hat \Delta_{p,A}(\bx)$ are independent of the model parameters and can thus be precomputed and stored even for large samples, the computational cost of repeatedly evaluating Eq.~\ref{eq:inc-xsec} can still become prohibitive for large samples. Additionally, the expression involves products of several terms that are close to unity near the nominal parameter point, $(\mu,\bn)=(1,\bzero)$, potentially causing numerical instabilities. To mitigate the computational cost and to improve numerical stability, we separate the evaluation into a numerically stable, analytically calculable part and interpolate the residual dependence on $\bn_\text{calib}$. Specifically, we rearrange Eq.~\ref{eq:inc-xsec} as
\begin{align}
&\sum_{\{w_i,\bx_i\}\in\mathcal{D}_{1,\bn}} \left(\hat R(\bx_i|\mu,\bn)-1\right)w_i\nonumber\\
&= \mu\, \textrm{CSI}_\PH(\bn_\text{calib})\nonumber\\
&\quad+(1+\alpha_\text{bkg})^{\nu_\text{bkg}}\, \textrm{CSI}_\PZ(\bn_\text{calib})\nonumber\\
&\quad+(1+\alpha_\text{bkg})^{\nu_\text{bkg}}(1+\alpha_{\ttbar})^{\nu_{\ttbar}}\, \textrm{CSI}_{\ttbar}(\bn_\text{calib})\nonumber\\
&\quad+(1+\alpha_\text{bkg})^{\nu_\text{bkg}}(1+\alpha_{\VV})^{\nu_{\VV}}\, \textrm{CSI}_{\VV}(\bn_\text{calib})\nonumber\\
&\quad+(\mu-1)\,\textrm{CSIC}_\PH\nonumber\\
&\quad+((1+\alpha_\text{bkg})^{\nu_\text{bkg}}-1)\, \textrm{CSIC}_\PZ\nonumber\\
&\quad+((1+\alpha_\text{bkg})^{\nu_\text{bkg}}\, (1+\alpha_{\ttbar})^{\nu_{\ttbar}}-1)\textrm{CSIC}_{\ttbar}\nonumber\\
&\quad+((1+\alpha_\text{bkg})^{\nu_\text{bkg}}\, (1+\alpha_{\VV})^{\nu_{\VV}}-1)\textrm{CSIC}_{\VV}\label{eq:CSI-master}
\end{align}
with
\begin{subequations}
\label{eq:CSI-terms}
\begin{align}
    \textrm{CSI}_p(\bn_\text{calib})&=\sum_{\{w_i,\bx_i\}\in\mathcal{D}_{1,\bzero}} w_i\, \hat g_p(\bx) (\hat S_p(\bx|\bn_\text{calib})-1),\label{eq:CSI-gp}\\
    \textrm{CSIC}_p&=\sum_{\{w_i,\bx_i\}\in\mathcal{D}_{1,\bzero}} w_i\,\hat g_p(\bx).\label{eq:CSI-conts}
\end{align}
\end{subequations}

Equation~\ref{eq:CSI-master} is a cross-section interpolation~(CSI) that reflects the known analytic dependence on $\mu$ and the normalization-type uncertainties. It is written such that training-data sums over many small terms remain numerically stable: the parenthesis in the last four terms in Eq.~\ref{eq:CSI-master} and on the r.h.s of Eq.~\ref{eq:CSI-gp} can be accurately determined by the \texttt{expm1} method in \textsc{numpy}~\cite{harris2020array}. The dependence on $\bn_\text{calib}$ is encoded in Eq.~\ref{eq:CSI-gp}, which we evaluate on a fine grid within three standard deviations for the three nuisance parameters in $\bn_\text{calib}$ and a coarser grid that fills the entire allowed parameter space. In total 6561 grid points are used and a three-dimensional quintic spline, implemented in the \textsc{scipy}-package~\cite{2020SciPy-NMeth}, provides a fast interpolation separately for each process. The constant terms $\textrm{CSIC}_p$ need only be evaluated once. We emphasize that this interpolation removes the main bottleneck in CPU consumption during the profiling and is numerically stable even for large simulated datasets.

If we sum the unbinned log-likelihood over the $N_r=3$ regions marked with UB in Table~\ref{tab:selections}, we arrive at
\begin{align}
&-\frac{1}{2}u_{\text{UB}}(\mathcal{D}|\mu,\bn) = \sum_{r=1}^{N_r}\Bigg[-\mathcal{L}(\sigma_r(\mu,\bn) - \sigma_r(1,\bzero)) \nonumber\\
&\qquad+ \sum_{\bx_i\in\mathcal{D}}\log\left(\frac{\ddd\sigma_r(\bx_i|\mu,\bn)}{\ddd\sigma_r(\bx_i|1,\bzero)}\right)\Bigg]\nonumber\\
&\quad\approx\sum_{r=1}^{N_r}\Bigg[-\sum_{\substack{\{w_i,\bx_i\}\in\\\mathcal{D}_{1,\bzero}\cap r}}w_i\left(\hat R_r(\bx_i|\mu,\bn)-1\right)\nonumber\\
&\qquad+\sum_{\bx_i\in\mathcal{D}}\log \hat R_r(\bx_i|\mu,\bn)\Bigg]\label{eq:u-unbinned}
\end{align}
where we denote by $\mathcal{D}_{1,\bzero}\cap r$ a set of \emph{simulated} events at $(\mu,\bn)=(1,\bzero)$ that satisfy the selection requirements of region $r$. The ratio in the logarithm in the first term pertains to region $r$, i.e., it is the same as in Eq.~\ref{eq:xsec-ratio}. In the second line, the DCR surrogate $\hat R_r(\bx|\mu,\bn)$ from Eq.~\ref{eq:xsec-ratio-ML} is the same in both terms with multiclassifiers and neural network parameterizations learned separately for each region $r$. The observed dataset $\mathcal{D}$ only enters in the second term in both lines. 

\section{Machine learning}\label{sec:ML}
\subsection{Calibrated multiclassifier}\label{sec:calib-classifier}
For estimating the DCR with $\hat g_p(\bx)$ in Eq.~\ref{eq:ML-gp}, we could use the CE loss function and train a multi-classifier with simulated samples $\mathcal{D}_p\sim\ddd\sigma_p(\bx|\bzero)$ at nominal values for the nuisance parameters. In Table~\ref{tab:selections}, however, we observe relative normalizations spanning three orders of magnitude. Therefore, a cross-section-weighted training is highly imbalanced. 

To stabilize the training, we equalize the normalization by dividing each process by its total cross-section. The resulting CE multi-classifier loss function is
\begin{align}
L_{\textrm{CE}}[\hat f_p] &= -\sum_{p=1}^{N_p}\int \frac{\ddd\sigma_p(\bx)}{\sigma_p}\,\log\Bigl(\hat f_p(\bx)\Bigr)\label{eq:ce}
\end{align}
where $\hat f_p(\bx)$ is a neural network with 28 input nodes and $N_p=4$ outputs. A \texttt{softmax} output layer guarantees 
\begin{align}
    \sum_{p=1}^{N_p}\hat f_p(\bx) &= 1.\label{eq:constraint}
\end{align}
The empirical version, suitable for implementation in computer code, is
\begin{align}
L_{\textrm{CE}}[\hat f_p] &\approx -\sum_{p=1}^{N_p}\sum_{\{\bx_i,w_i\}\in\mathcal{D}_p} \frac{w_i}{\mathcal{L}\sigma_p}\log\Bigl(\hat f_p(\bx_i)\Bigr).\label{eq:ce-empirical}
\end{align}

Functional minimization of Eq.~\ref{eq:ce} and treating Eq.~\ref{eq:constraint} with a Lagrange multiplier shows that the minimum is attained for
\begin{align}
\hat f_p^\ast(\bx) &\simeq \frac{\ddd\sigma_p(\bx)/\sigma_p}{\sum_{q=1}^{N_p}\ddd\sigma_q(\bx)/\sigma_q},\label{eq:learned-LL-f}
\end{align}
which is nothing but the likelihood ratio, learned by a conventional multi-classifier. 

To convert to the DCR, we invert the normalization that was included in the loss and arrive at
\begin{align}
    \hat g^\ast_p(\bx)=\frac{\hat f^\ast_p(\bx)\sigma_p}{\sum_q \hat f^\ast_q(\bx)\sigma_q}\simeq\frac{\ddd\sigma_p(\bx|\bzero)}{\sum_q\ddd\sigma_q(\bx|\bzero)}
\end{align}
which now exactly corresponds to Eq.~\ref{eq:ML-gp}.

For FAIR-HUC, we train one multiclassifier for each of the three regions marked with UB in Table~\ref{tab:selections} using \texttt{tensorflow 2.11}~\cite{tensorflow2015-whitepaper}. All neural networks have a straightforward architecture with 28 input nodes, corresponding to the features listed in Table~\ref{tab:event_features}, and four \texttt{softmax} output nodes. 
We selected the three‑layer, 64‑unit architecture and its regularization parameters by first running a coarse grid search over hidden‑layer sizes (32, 64, 128) and depths (2–4 layers) on a held‑out validation split, then refining dropout (0.1–0.5) and penalties (\texttt{L1}=0.1, \texttt{L2}=0.05) through random search. All models were trained for up to 1000 epochs with early stopping (patience 10 epochs) monitored on validation loss. 

\begin{figure*}
    \centering
    \includegraphics[width=0.48\linewidth]{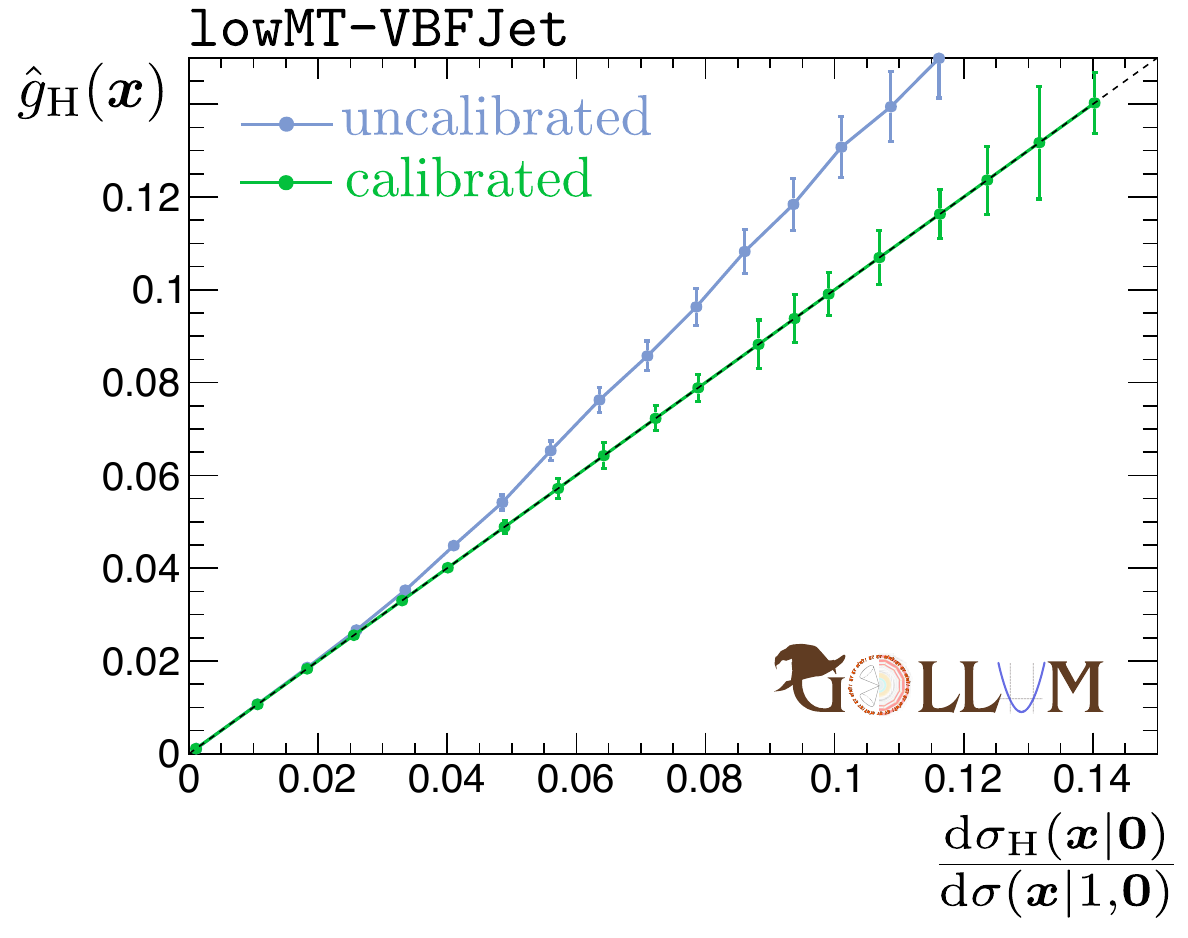}\hfill
    \includegraphics[width=0.48\linewidth]{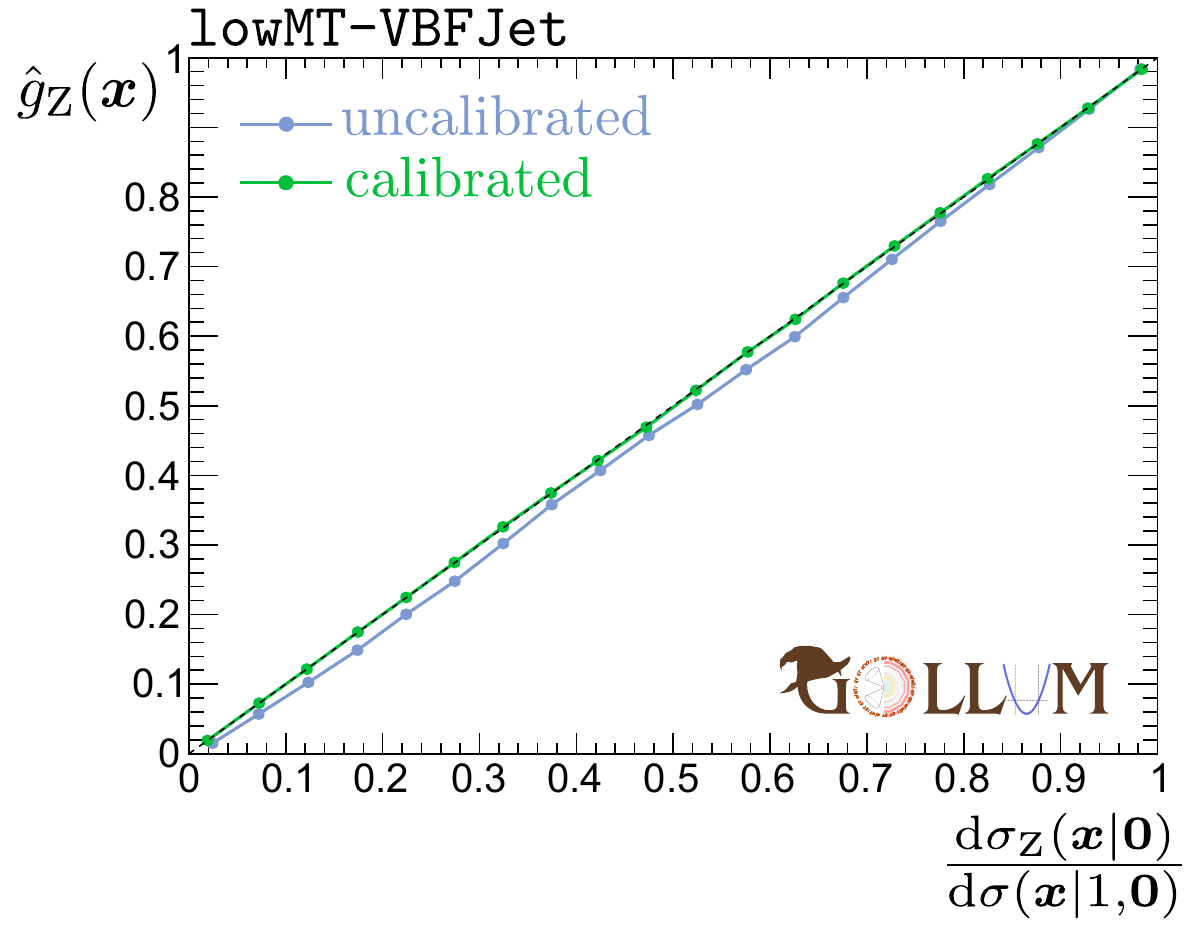}\\
    \includegraphics[width=0.48\linewidth]{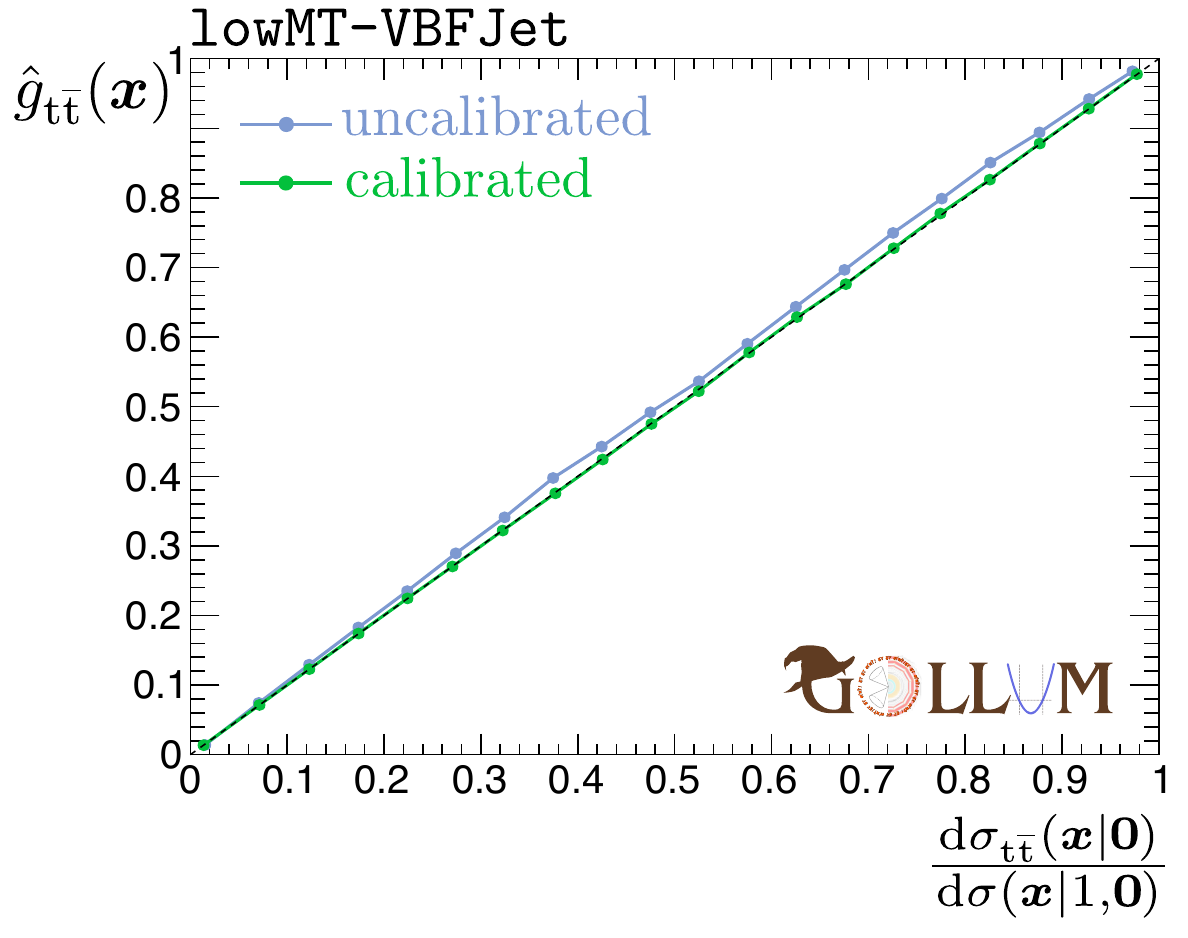}\hfill
    \includegraphics[width=0.48\linewidth]{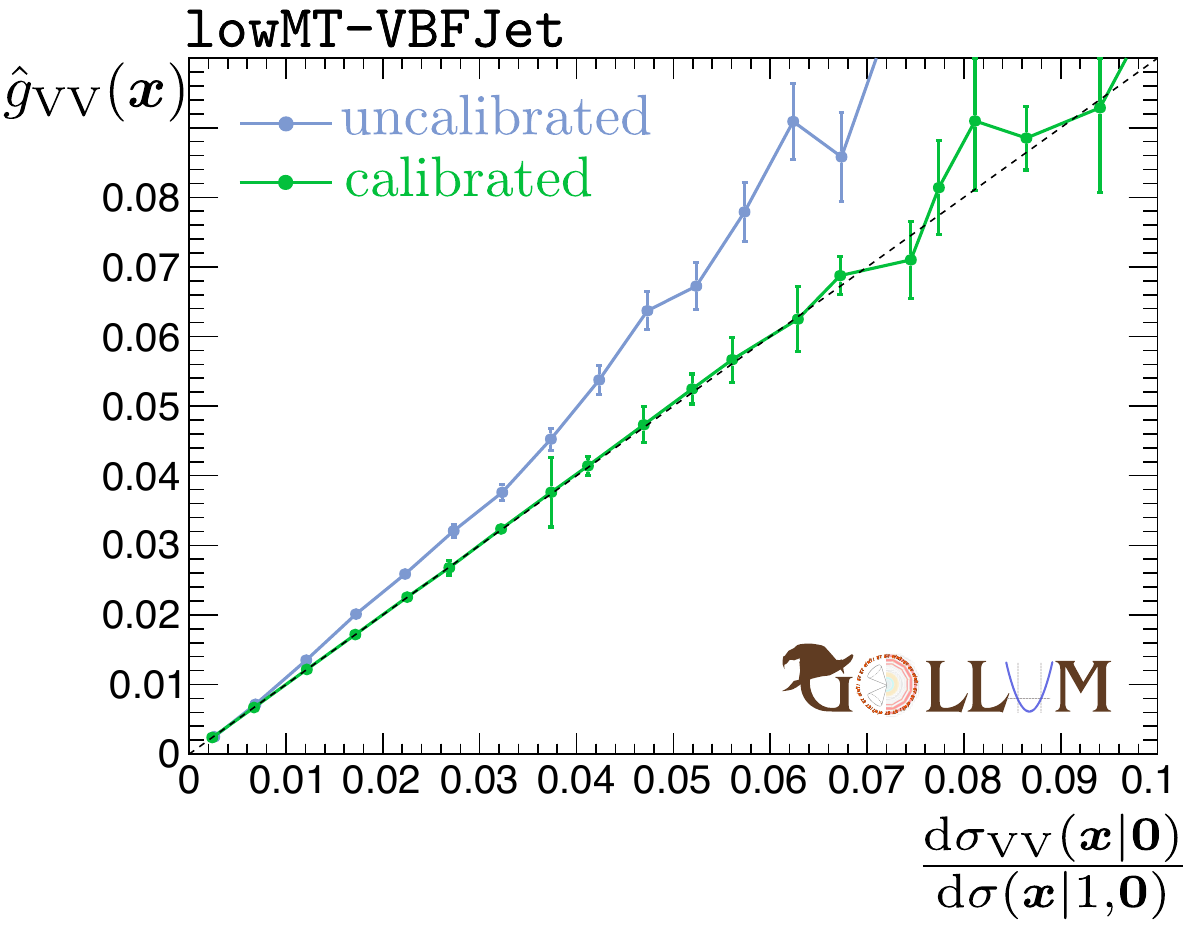}
    \caption{\label{fig:calibration} Calibration of the DCR in the \texttt{lowMT-VBFJet} region of the four processes in FAIR-HUC: $\PH\to\tau\tau$~(top left), $\PZ\to\tau\tau$~(top right), \ttbar~(bottom left), and \VV~(bottom right).}
\end{figure*}

An important step is the calibration of the multi-classifier $\hat f_p(\bx)$ and the corresponding DCR estimate $\hat g_p(\bx)$. A classifier is calibrated if its output $\hat f_p(\bx)$ accurately reflects the true fraction of events in the vicinity of $\bx$ belonging to class $p$. Consequently, a value $\hat g_p(\bx)\approx 0.1$ implies that approximately 10\% of cross-section-weighted events near $\bx$ originate from process $p$. Modern classifiers are typically not well calibrated~\cite{2017arXiv170604599G,Cranmer:2015bka}, as illustrated in Fig.~\ref{fig:calibration}. For instance, the predicted signal DCR tends to exceed the true expected DCR.

Binary classifiers can be recalibrated after training by applying a monotonic function --- such as isotonic regression~(IREG) --- to adjust the classifier outputs without altering decision boundaries~\cite{Cranmer:2015bka}. Among several extensions of the isotonic regression to the multi-class case, we chose the following: first, we independently calibrate each of the four $\hat g_p(\bx)$ outputs using one-vs-rest binary isotonic regression, implemented via {\tt sklearn}~\cite{scikit-learn}. Then, we retain the calibrated output for the signal-vs-background class ($p=\PH$) and rescale the calibrated outputs of the three background classes so that the four DCR outputs sum to unity:
\begin{align}
\hat g_p(\bx) = \begin{cases}
\textrm{IREG}(g_\PH^\ast(\bx)), & \text{if } p=\PH \\[5pt]
\frac{\textrm{IREG}(g_p^\ast(\bx))(1-\textrm{IREG}(g_\PH^\ast(\bx)))}{\sum_{q=\PZ,\ttbar,\VV} \textrm{IREG}(g_{q}^\ast(\bx))}, & \text{otherwise}.
\end{cases}
\end{align}
Prioritizing the calibration of the signal class ensures accurate estimation of the signal DCR, reducing biases from the dominant backgrounds and thereby improving precision in the extraction of~$\mu$. 

Figure~\ref{fig:calibration} illustrates the effect of calibration on the DCR distributions. The corrections for the $\PH\to\tau\tau$ signal and the \VV background are substantial. The corrections to the $\hat g_\PZ$ are at most 2\%. However,  the large cross-section of this background entails a significant degradation of the measurement of $\mu$ if the miscalibration is not corrected. 

\begin{figure*}
    \centering
    \hspace{-.3cm}\includegraphics[width=0.48\linewidth]{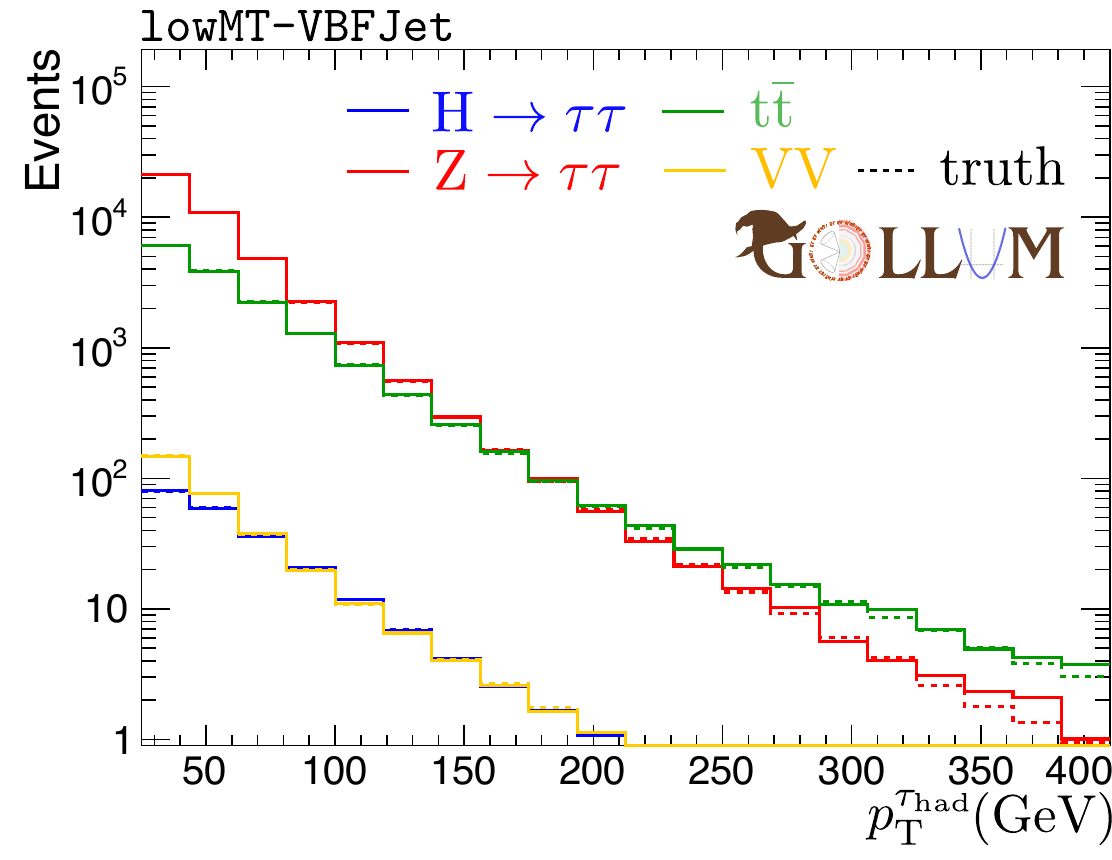}\hfill
    \includegraphics[width=0.48\linewidth]{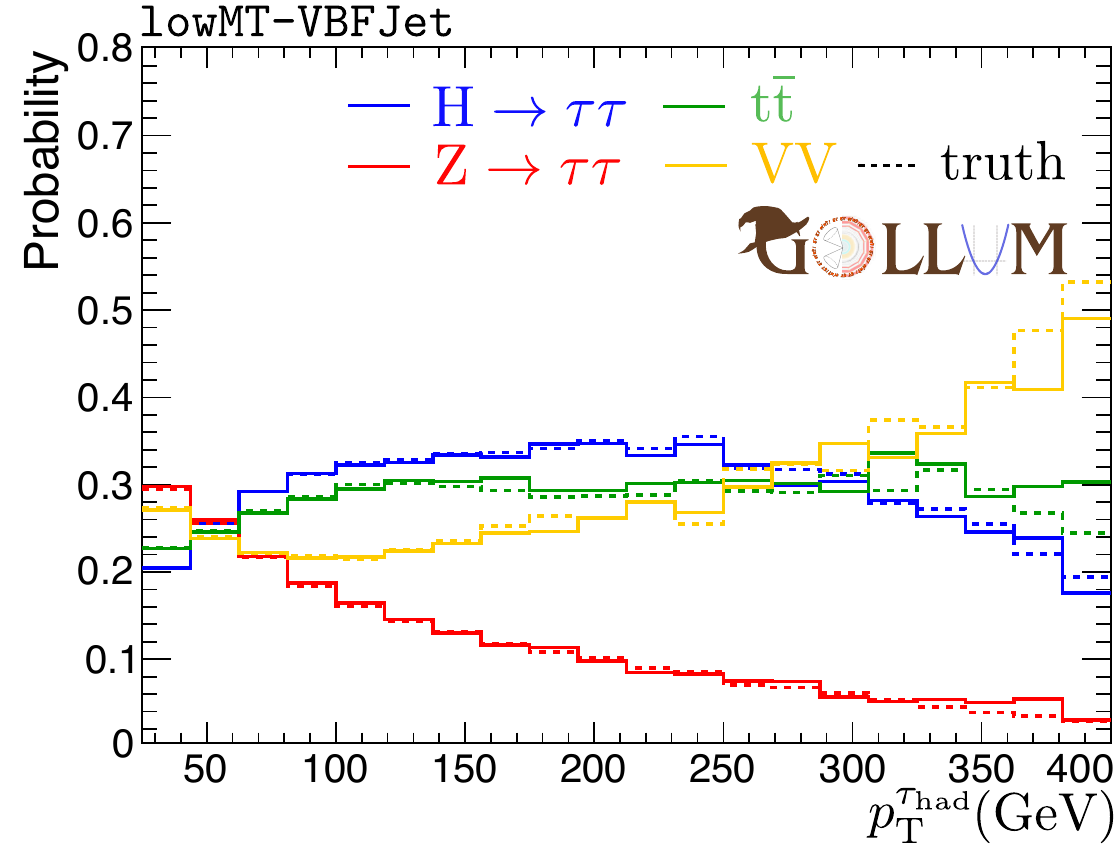}
    \hspace{.15cm}
    \caption{\label{fig:closure-tfmc-1} Comparison of the true spectra of $\pt^{\tau_\text{had}}$~(dashed lines) for the four processes in the \texttt{lowMT-VBFJet} region: $\PH\to\tau\tau$~(blue),  $\PZ\to\tau\tau$~(red), \ttbar~(green), and \VV~(yellow) with the prediction from the ML surrogates~(solid lines). The left column shows the spectra normalized to the expected number of events, while the right column shows a comparison of the per-bin fractions of the area-normalized shapes of each process.}
\end{figure*}

\begin{figure*}
    \centering
    \hspace{-.3cm}\includegraphics[width=0.48\linewidth]{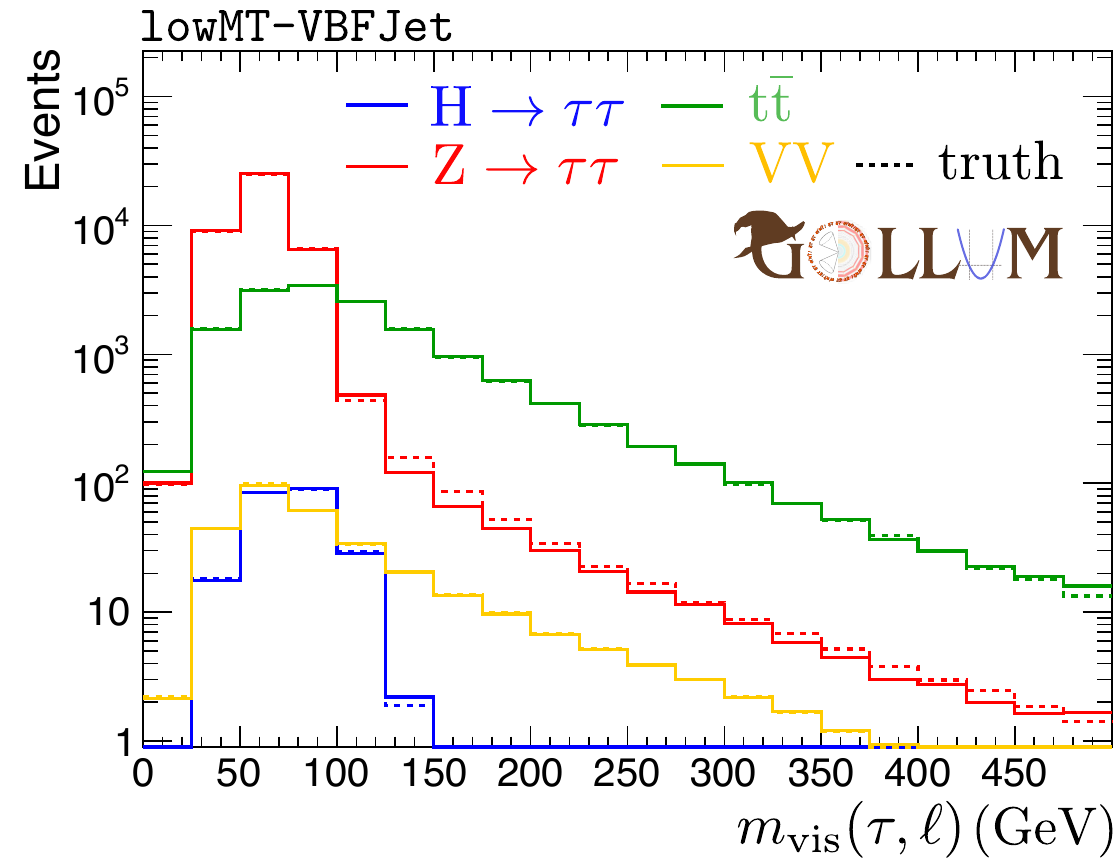}\hfill
    \includegraphics[width=0.48\linewidth]{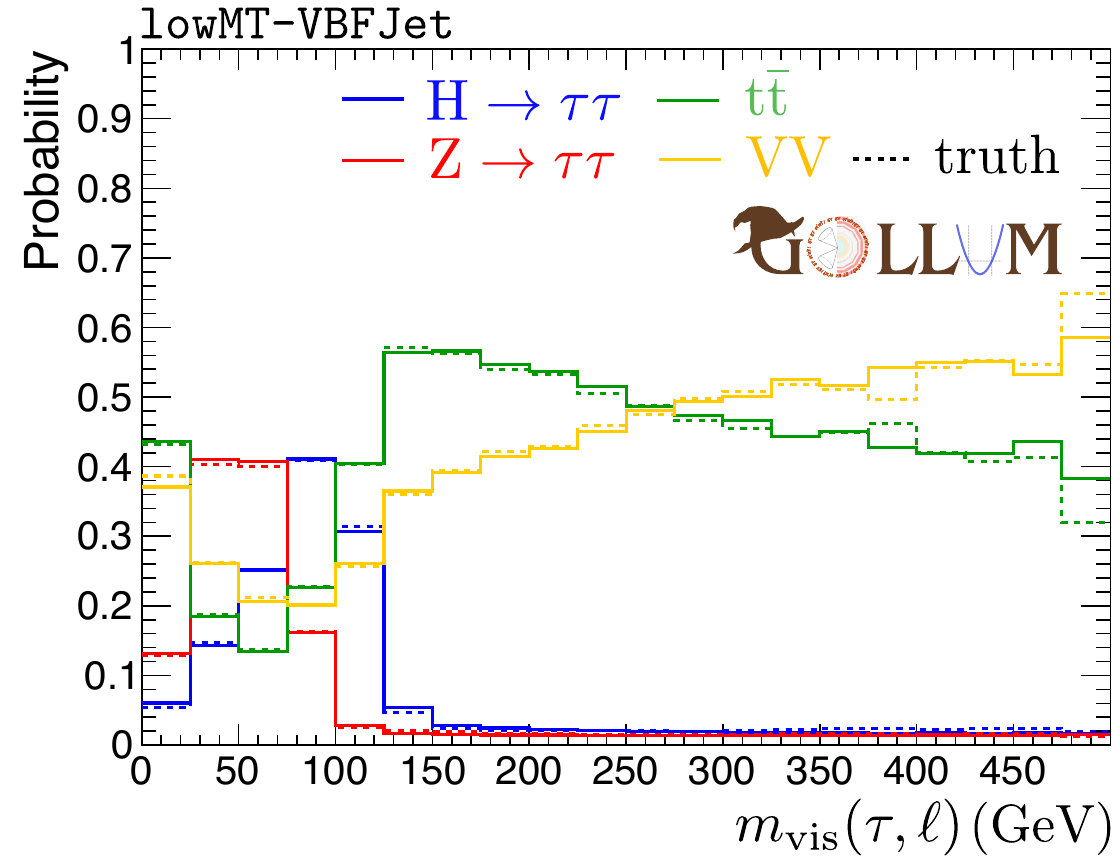}
    \hspace{.15cm}
    \caption{\label{fig:closure-tfmc-2} Same as Fig.~\ref{fig:closure-tfmc-1}, but for $m_\text{vis}(\tau,\ell)$.}
\end{figure*}

\begin{figure*}
    \centering
    \hspace{-.3cm}\includegraphics[width=0.48\linewidth]{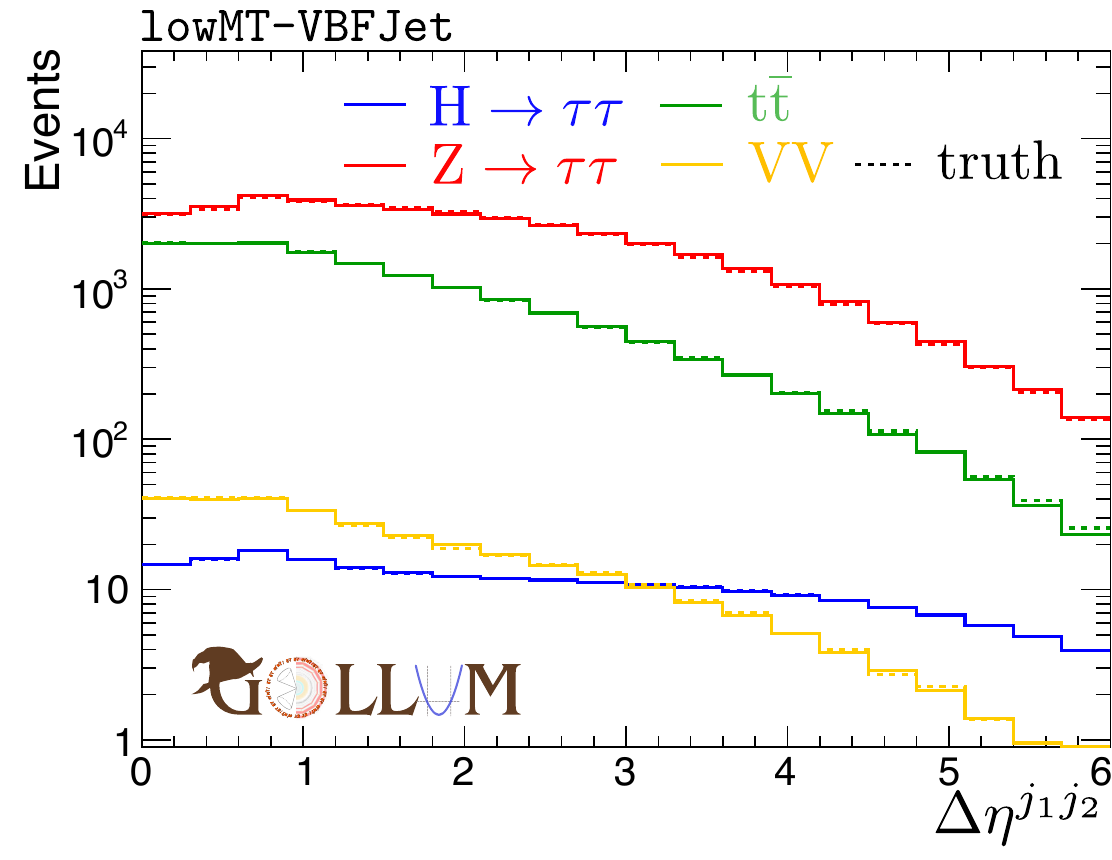}\hfill
    \includegraphics[width=0.48\linewidth]{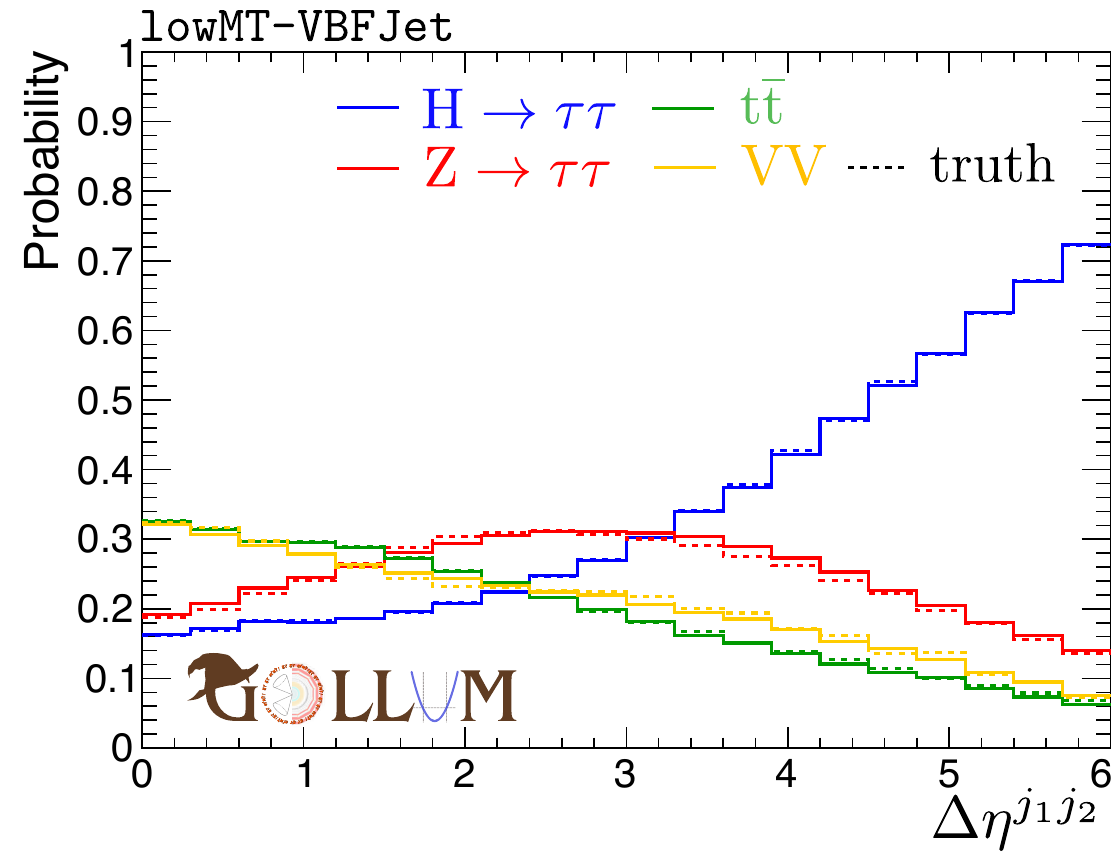}
    \hspace{.15cm}
    \caption{\label{fig:closure-tfmc-3} Same as Fig.~\ref{fig:closure-tfmc-1}, but for $\Delta\eta^{j_1j_2}$.}
\end{figure*}

Figures~\ref{fig:closure-tfmc-1}--\ref{fig:closure-tfmc-3} demonstrate the closure of the calibrated surrogate for several distributions in the \texttt{lowMT-VBFJet} region. The simulated (true) distributions of $\pt^{\tau_\text{had}}$, $m_\text{vis}(\tau,\ell)$, and $\Delta\eta^{j_1j_2}$ (dashed lines) for the four processes $\PH\to\tau\tau$, $\PZ\to\tau\tau$, \ttbar, and \VV are compared to the predictions obtained from the surrogate $\hat g_p(\bx)$ (solid lines). Specifically, the surrogate predictions shown correspond to binned yields computed as $\sum_{\bx_i\in \text{bin}} w_i\,\hat g_p(\bx_i)$. The left column shows spectra normalized to the expected event counts, while the right column compares their shapes. After calibration, the agreement between surrogate predictions and simulated spectra is excellent, spanning several orders of magnitude.

\begin{figure*}
    \centering
    \hspace{-.5cm}\includegraphics[width=0.48\linewidth]{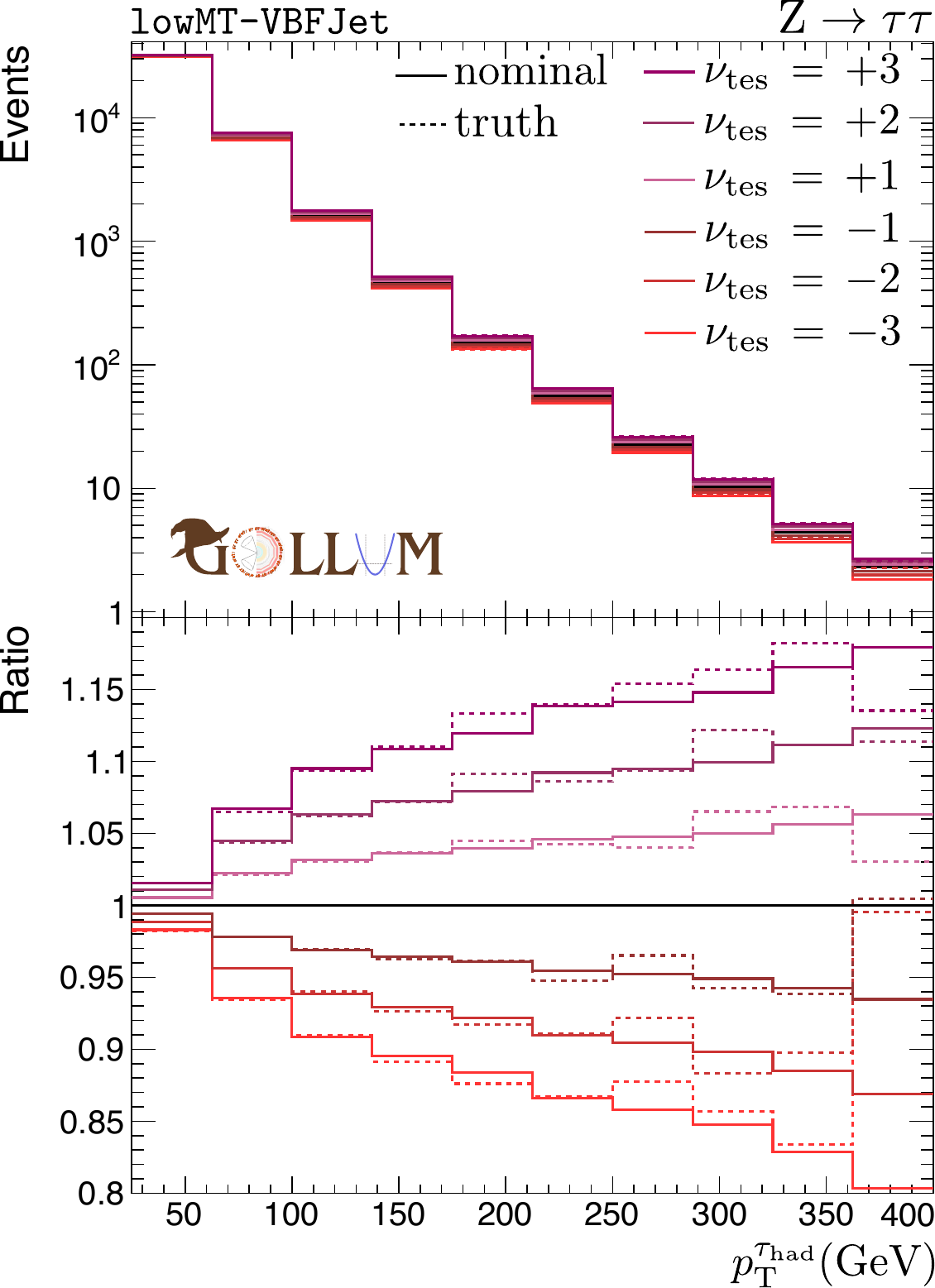}\hfill
    \includegraphics[width=0.48\linewidth]{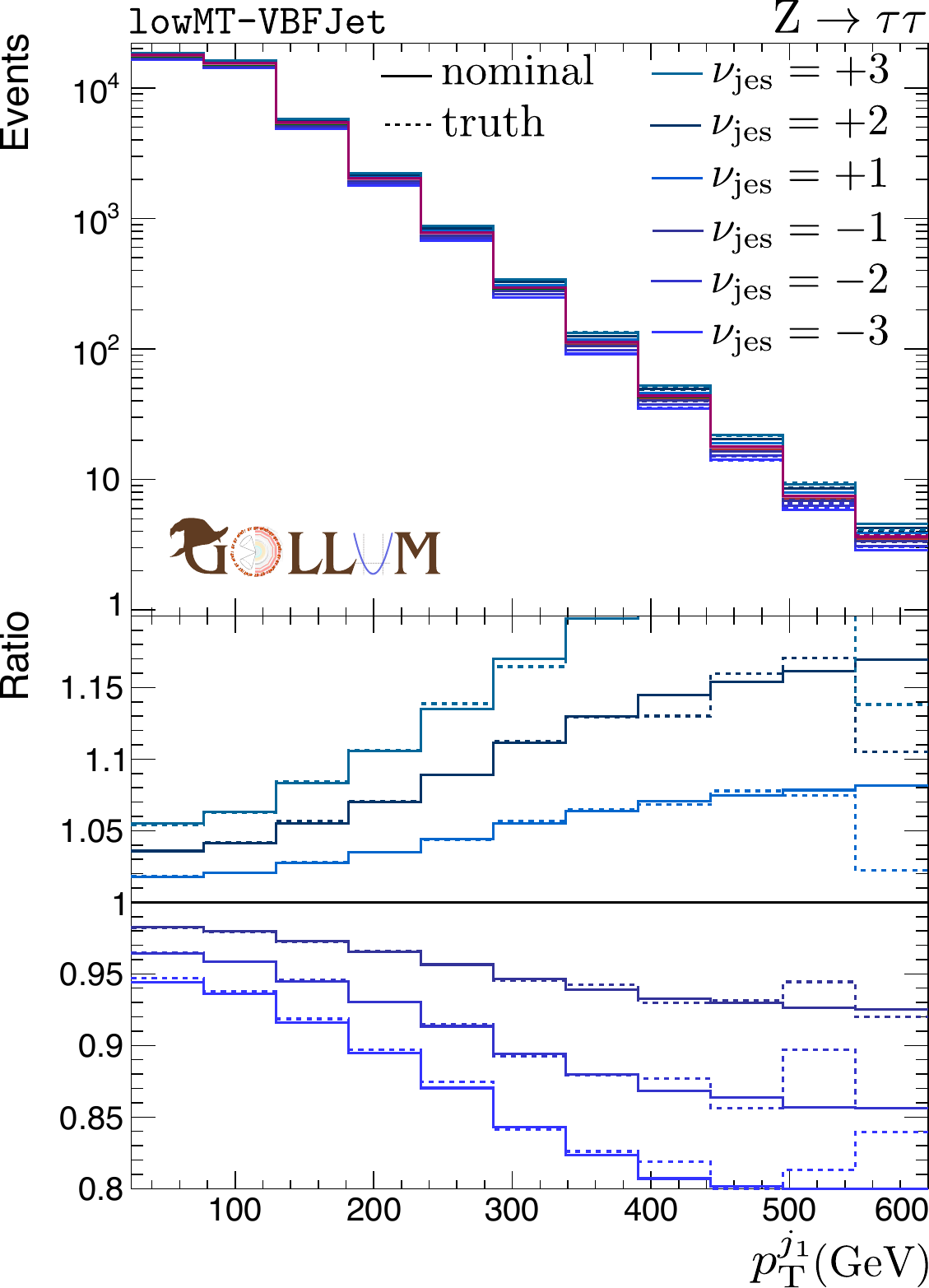}
    \hspace{.1cm}
    \caption{\label{fig:PNN-closure} Comparison of the $\PZ\to\tau\tau$ training data spectra in the $\texttt{lowMT-VBFJet}$ region with the prediction from the parametric neural network.  We show $\pt^{\tau_\text{had}}$ with \texttt{tes} variations~(left) and $\pt^{j_1}$ with \texttt{jes} variations~(right).}
\end{figure*}

\subsection{Neural network parametrization}\label{sec:pnn}
Next, we train a parametric neural network that estimates the nuisance parameter dependence of the DCR in Eq.~\ref{eq:ML-S}, following in broad terms Ref.~\cite{Schofbeck:2024zjo} but using neural networks instead of a boosted tree algorithm. In this section, we remove notational clutter and solve the generic training task
\begin{align}
    \hat S(\bx|\bn)&\simeq\frac{\ddd\sigma(\bx|\bn)}{\ddd\sigma(\bx|\bzero)}.
\end{align}
The first step is to choose a sufficiently large set of nuisance parameter vectors $\bn\in\mathcal{V}$, where the same considerations apply as in Sec.~\ref{sec:binned-LL}, and thus we keep the choice discussed there. For every $\bn\in\mathcal{V}$, we obtain a systematically varied training sample $\mathcal{D}_\bn$.  The nominal sample $\mathcal{D}_\bzero$ is removed from $\mathcal{V}$ and plays a special role. 

If we train a binary classifier between a specific $\mathcal{D}_\bn$ and the nominal with the CE loss\footnote{The FAIR-HUC datasets provide event weights that are normalized to the event yield $\mathcal{L}\sigma$, technically including an unspecified luminosity factor. Therefore, in principle, we only have access to $L_\text{CE}[\hat f]=\mathcal{L}\int\ddd\sigma\cdots$. However, we can still drop this distracting constant in the notation because it does not affect the minimum. }
\begin{align}
    L_{\textrm{CE}}[\hat f]&=-\int\ddd\sigma(\bx|\bzero)\log\hat f({\bx})-\int\ddd\sigma(\bx|\bn)\log(1-\hat f(\bx)),\nonumber
\end{align}
the network $\hat f(\bx)$ attains its minimum at
\begin{align}
    f^{\ast}(\boldsymbol{\bx})=&\left(1+\frac{\ddd\sigma(\bx|\bn)}{\ddd\sigma(\bx|\bzero)}\right)^{-1}.\label{eq:CE-analytic-sol}
\end{align}
While $f^{\ast}(\boldsymbol{\bx})$ is a monotonous function of the required DCR, it is not parametric in \bn. To arrive at a fully parametric estimate, we combine the analytic structure in  Eq.~\ref{eq:CE-analytic-sol} with an ansatz for the \bn-dependence chosen as $\exp(\nu_A\hat\Delta_{A}(\bx))$ in analogy to Eq.~\ref{eq:sigma-param}. Taken together, the ansatz is
\begin{align}
\hat f_\bn(\bx)&=\frac{1}{1+\exp(\nu_A\hat\Delta_{A}(\bx))},\label{eq:parametric-ansatz}
\end{align} 
where the $\hat\Delta_A(\bx)$ are polynomial coefficient functions. 
These provide the unbinned generalization, replacing the per-bin constants in Eq.~\ref{eq:icp}. They are implemented as neural networks and designed to learn the systematic effects continuously as a function of \bx. The ensuing parameterization is also continuous in $\bn$ and given by the combination $\nu_A\hat\Delta_{A}(\bx)$. The  $\nu_A$ are polynomial in $\bn$ and given by Eq.~\ref{eq:index-def} as in the binned case.

Similar to Eq.~\ref{eq:poly-exp}, this expansion can encode arbitrary polynomial orders. For the FAIR-HUC, the nine output nodes of the neural network correspond to the nine components of $\hat\Delta_A(\bx)$ in the expansion
\begin{align}
&\nu_A\hat\Delta_A(\bx)=\nu_\text{tes}\,\hat\Delta_{\text{tes}}(\bx)+\nu_\text{jes}\,\hat\Delta_{\text{jes}}(\bx)+\nu_\text{met}\,\hat\Delta_{\text{met}}(\bx)\nonumber\\    &\quad+\nu_\text{tes}^2\,\hat\Delta_{\text{tes},\text{tes}}(\bx)+\nu_\text{jes}^2\,\hat\Delta_{\text{jes},\text{jes}}(\bx)+\nu_\text{met}^2\,\hat\Delta_{\text{met},\text{met}}(\bx)\nonumber\\ &\quad+\nu_\text{tes}\nu_\text{jes}\,\hat\Delta_{\text{tes},\text{jes}}(\bx)+\nu_\text{jes}\nu_\text{met}\,\hat\Delta_{\text{jes},\text{met}}(\bx)\nonumber\\
&\quad+\nu_\text{tes}\nu_\text{met}\,\hat\Delta_{\text{tes},\text{met}}(\bx).\label{eq:poly-exp-unbinned}
\end{align}
Each network has 28 inputs, in accordance with the multiclassifier case. To train all coefficient functions simultaneously, we sum the CE loss over $\bn\in\mathcal{V}$. A short calculation reveals
\begin{subequations}
\begin{align}
&L[\hat \Delta_A]=\sum_{\bn\in\mathcal{V}}\Bigg[\int\ddd\sigma(\bx|\bzero)\,\textrm{Soft}^+(\nu_A\hat\Delta_A(\bx))\nonumber\\
&\qquad\quad\;\;+\int\ddd\sigma(\bx|\bn)\,\textrm{Soft}^+(-\nu_A\hat\Delta_A(\bx))\Bigg]\label{eq:loss-analytic-SoftMax}\\
&\quad\approx\sum_{\bn\in\mathcal{V}}\Bigg[\sum_{\{\bx_i,w_i\}\in\mathcal{D}_\bzero}w_i\,\textrm{Soft}^+(\nu_A\hat\Delta_A(\bx_i))\nonumber\\
&\qquad\quad\;\;+\sum_{\{\bx_i,w_i\}\in\mathcal{D}_\bn}w_i\,\textrm{Soft}^+(-\nu_A\hat\Delta_A(\bx_i))\Bigg],\label{eq:loss-empirical-SoftMax}
\end{align}
\end{subequations}
where $\textrm{Soft}^+(x)=\log(1+\exp(x))$ and the last approximation defines the empirical version of the loss function. We emphasize that for each term in the sum over $\mathcal{V}$, the sample $\mathcal{D}_0$ in the first term is the same. The functional minimum of Eq.~\ref{eq:loss-analytic-SoftMax} is given by
\begin{align}
 \hat S^\ast(\bx|\bn)=\exp\left(\nu_A\hat\Delta_A(\bx)\right)&\simeq\frac{\ddd\sigma(\bx|\bn)}{\ddd\sigma(\bx|\bzero)},\label{eq:S-estimate}
\end{align}
which is the DCR we need for Eq.~\ref{eq:ML-S}. 

For FAIR-HUC, we train neural network parametrizations separately for each region and process, which amounts to a total of 12 networks. Each consists of two hidden layers with 128 nodes each and nine output nodes. Otherwise, the training closely follows the multi-classifier setup, using the \texttt{Adam} optimizer with a learning rate of $10^{-3}$. 
We stress that the nuisance parameters do not enter the neural networks which solely predict $\hat\Delta_A(\bx)$. Instead, $\bn$ enters  Eq.~\ref{eq:S-estimate}.

Figure~\ref{fig:PNN-closure} shows examples of the parametric dependence learned for the $\PZ\rightarrow\tau\tau$ process in the \texttt{lowMT-VBFJet} selection. The approximations for \texttt{tes} and \texttt{jes} as functions of $\pt^{\tau_\text{had}}$ and $\pt^{j_1}$ are stable across the full phase space. At high tails, where stochastic fluctuations in the training data variations (``truth'') become large, the model interpolates smoothly. We further assess the quality of the model through extensive toy-data studies in Sec.~\ref{sec:toy-results}.

\section{Limit setting}\label{sec:limit-setting}
\subsection{Binned control regions and test statistics}\label{sec:CR-binning}

To exploit the constraining power of the control regions, we include the $N_\textrm{CR}=3$ regions marked as CR in Table~\ref{tab:selections}. The \texttt{lowMT-noVBFJet0pt0to100} region is by far the largest. Tests showed that treating this region unbinned does not improve the measurement of $\mu$, but as an inclusive control region, it mildly helps constrain because $(\mathcal{L}\sigma(1,\bzero))^{-1/2}\approx10^{-3}\approx\alpha_\text{bkg}$.

The $\ttbar$ and \VV processes have larger uncertainties, motivating the inclusion of CRs based on \texttt{highMT-noVBFJet}. 
In this selection, we adopt the learned likelihood ratio $\hat f_{\ttbar}$ and $\hat f_{\VV}$ from Eq.~\ref{eq:learned-LL-f} to define CRs enriched in \ttbar and \VV. An additional requirement of $\hat f_{\ttbar}>0.4$ defines \texttt{highMT-noVBFJet-tt} with a purity in \ttbar of approx. 90\%. Similarly, the requirements $\hat f_{\ttbar}\leq0.4$ and $\hat f_{\VV}>0.5$ select the \VV contribution with a purity of 60\%. We denote this region by \texttt{highMT-noVBFJet-VV}. 
Given $(\mathcal{L}\sigma_{\ttbar}(1,\bzero))^{-1/2}\approx0.015<\alpha_{\ttbar}=0.02$ and $(\mathcal{L}\sigma_{\VV}(1,\bzero))^{-1/2}\approx0.037<\alpha_{\VV}=0.25$, we expect significant constraints on $\nu_{\ttbar}$ and $\nu_{\VV}$. 
The two regions are shown in Table~\ref{tab:selections}.

In summary, the log-likelihood ratio from the three CRs defined in this section is
\begin{align}
-\frac{1}{2}u_\text{CR}(\mathcal{D}|\mu,\bn) =& \sum_{i=1}^{N_{\text{CR}}}\Bigg[-\sum_{p}\mathcal{L}\left(\sigma_{i,p}(\mu,\bn) - \sigma_{i,p}(1,\bzero)\right) \nonumber\\&+ N_{i,\text{obs}}\log\left(\frac{\sum_p\sigma_{i,p}(\mu,\bn)}{\sum_p\sigma_{i,p}(1,\bzero)}\right)\Bigg],
\end{align}
with $\sigma_{i,p}(\mu,\bn)$ again defined by Eq.~\ref{eq:binned-param}, using components computed from Eq.~\ref{eq:sigma_all}, Eq.~\ref{eq:icp}, and Eq.~\ref{eq:sigma-param} and the sums over the processes $p$ run over \PH, \PZ, \ttbar, and \VV.

Along the same lines, we also define a binned likelihood in regions that are otherwise treated as unbinned and marked with UB in Table~\ref{tab:selections}. Doing so provides a fully binned reference likelihood for performance comparisons. For each UB region, we construct the distribution of the signal node of the multi-classifier~($\hat f_\PH$) using nominal simulation. Bins in this observable are defined as follows.

\begin{figure}
    \centering
    \includegraphics[width=0.99\linewidth]{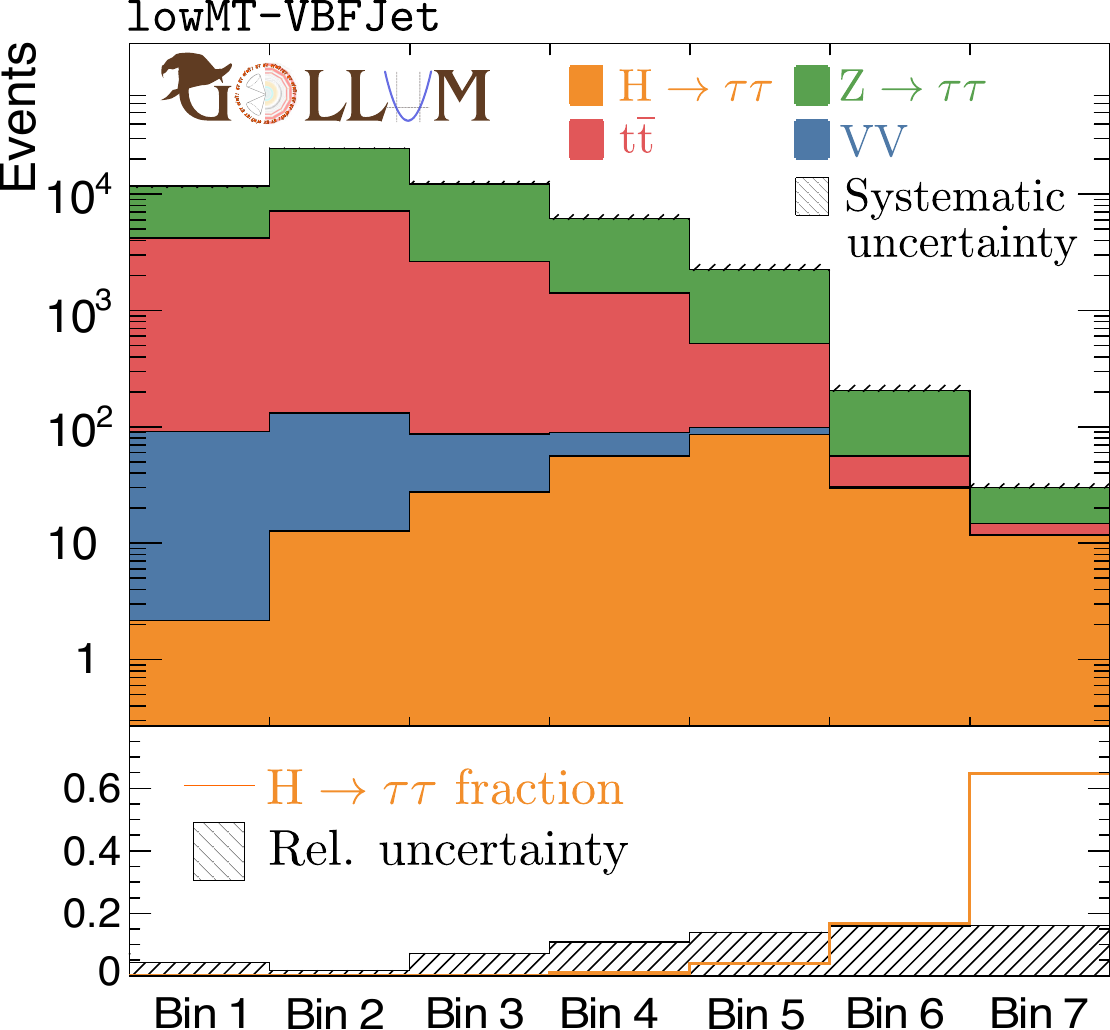}
    \caption{The binned distribution of $\hat f_\PH(\bx)$.}
    \label{fig:binned}
\end{figure}

The weighted distribution of $\hat f_\PH$ is integrated from a lower threshold up to its maximum value of one. We first set the lower integration bound to one, resulting in zero integral, and then gradually lower it, computing the $\PH \to \tau \tau$ signal yield ($S$) and cumulative background yield ($B$) at each step. These yields define a simple proxy for the signal significance in the interval, $s = S/\sqrt{S+B}$. We continue lowering the threshold until $s=2$ or until $s$ reaches a maximum. At that point, a bin is formed from the integration interval, the corresponding phase space is removed, and the procedure is repeated until all events are binned. The resulting distribution in the \texttt{lowMT-VBFJet} region is shown in  Fig.~\ref{fig:binned}.
This binning procedure roughly approximates a traditional analysis, with the classifier output used in a binned template fit carried out simultaneously in all UB regions. We tested finer variants of the binning procedure that led to similar results, as reported below.
The corresponding binned reference log-likelihood is given by
\begin{widetext}
\begin{align}
-\frac{1}{2}u_\text{B}(\mathcal{D}|\mu,\bn) =
\sum_{r=1}^{N_{\text{UB}}}\sum_{i=1}^{N_{\text{bins,r}}}\Bigg[-\sum_{p}\mathcal{L}\left(\sigma_{r,i,p}(\mu,\bn) - \sigma_{r,i,p}(1,\bzero)\right) + N_{r,i,\text{obs}}\log\left(\frac{\sum_p\sigma_{r,i,p}(\mu,\bn)}{\sum_p\sigma_{r,i,p}(1,\bzero)}\right)\Bigg].\label{eq:u-binned}
\end{align}
\end{widetext}

Finally, the likelihood of the hypothetical auxiliary measurements that provide the FAIR-HUC nuisance parameter constraints is 
\begin{align}
u_\text{P}=\nu_\text{bkg}^2+\nu_{\ttbar}^2+\nu_{\VV}^2+\nu_\text{tes}^2+\nu_\text{jes}^2+\nu_\text{met}^2\label{eq:penalty}
\end{align}
and acts as a penalty during the profiling. We do not allow nuisance parameter values outside the boundaries discussed in Sec.~\ref{sec:sys}.

With these ingredients, we define the unbinned and binned log-likelihoods as
\begin{subequations}\label{eq:binned-and-unbinned}
\begin{align}
    u_\text{unbinned} &= u_\text{UB}+u_\text{CR}+u_\text{P} 
   \label{eq:u-all-unbinned},\\
    u_\text{binned}   &= u_\text{B}\;\;+u_\text{CR}+u_\text{P},
\end{align}
\end{subequations}
and use those in the profiling in Eq.~\ref{eq:test-stat}. We emphasize that both unbinned and binned likelihoods include a binned component from the CRs~($u_\text{CR}$) and the penalty~($u_\text{P}$).

\subsection{Asimov results}\label{sec:asimov-results}

We use the Asimov dataset~\cite{Cowan:2010js} to compute expected confidence intervals for all test statistics. The unbinned Asimov dataset is introduced in Ref.~\cite{GomezAmbrosio:2022mpm} and is adapted to our notation in Appendix~\ref{sec:app-asimov}.
\begin{figure}
    \centering
    \includegraphics[width=0.98\linewidth]{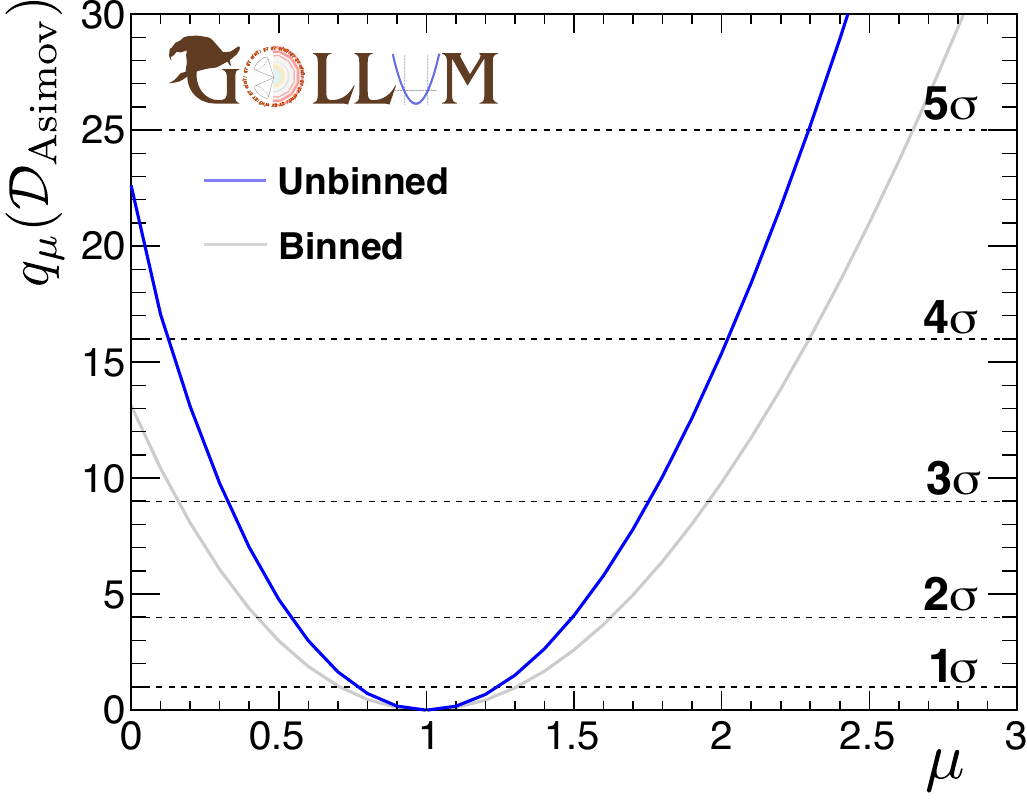}
    \caption{\label{fig:profiled-likleihood} The profiled likelihood test statistic $q_\mu(\mathcal{D}_\text{Asimov})$ for the binned and unbinned case with Asimov data. }
\end{figure}
The profiling is performed with the \texttt{MINUIT} package~\cite{James:1975dr}. 
Figure~\ref{fig:profiled-likleihood} shows the Asimov expected likelihood test statistic for the (assumed) true values $\mu^\text{true}=1$ and $\bn^\text{true}=\bzero$. We observe no noticeable bias in the location of the minimum: the maximum likelihood estimate~(MLE) is found at $(\bar\mu,\bar\bn)=(1,\bzero)$ with an accuracy of $10^{-3}$. Repeating the Asimov fit for different values of $\mu^\text{true}$ in the range $0.1 \leq \mu \leq 3$ confirms this conclusion. Any residual imperfection in the ML surrogates thus has minimal impact on the result. However, their overall quality remains critical: For example, removing the isotonic regression calibration from Sec.~\ref{sec:calib-classifier} induces a bias on $\bar\mu$ in the range 0.3--0.4, with strong dependence on $\bn^\text{true}$. This highlights the necessity of calibrating the classifier output using this or similar SBI methods~\cite{Cranmer:2015bka}.

The binned profiled likelihood ratio in Fig.~\ref{fig:profiled-likleihood} exceeds the $1\sigma$ threshold outside the interval $\mu = 1.00 \pm 0.21$. For the unbinned case, the interval narrows to 0.17, yielding a 20\% improvement in the FAIR-HUC setup. The hypothesis $\mu=0$ is excluded at $3.6\sigma$ for the binned case and $4.8\sigma$ for the unbinned case, marking a significant gain.

\begin{figure*}
    \centering
    \includegraphics[width=0.48\linewidth]{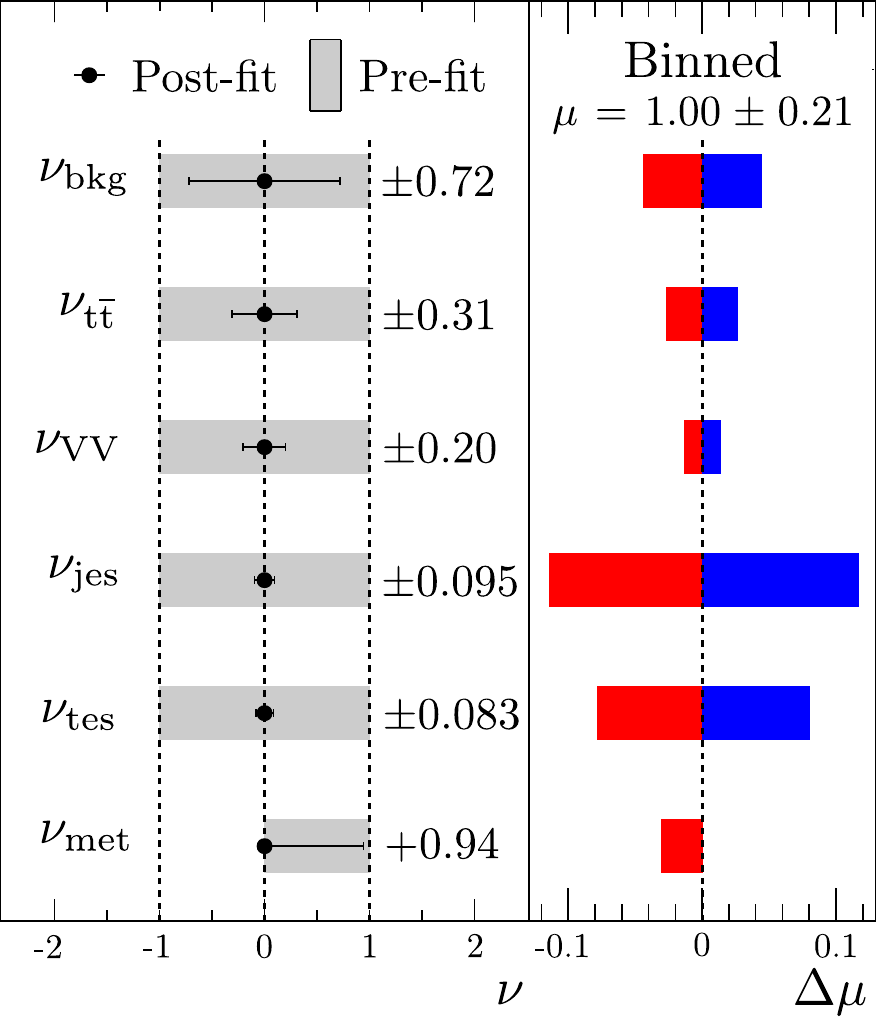}\hfill
    \includegraphics[width=0.48\linewidth]{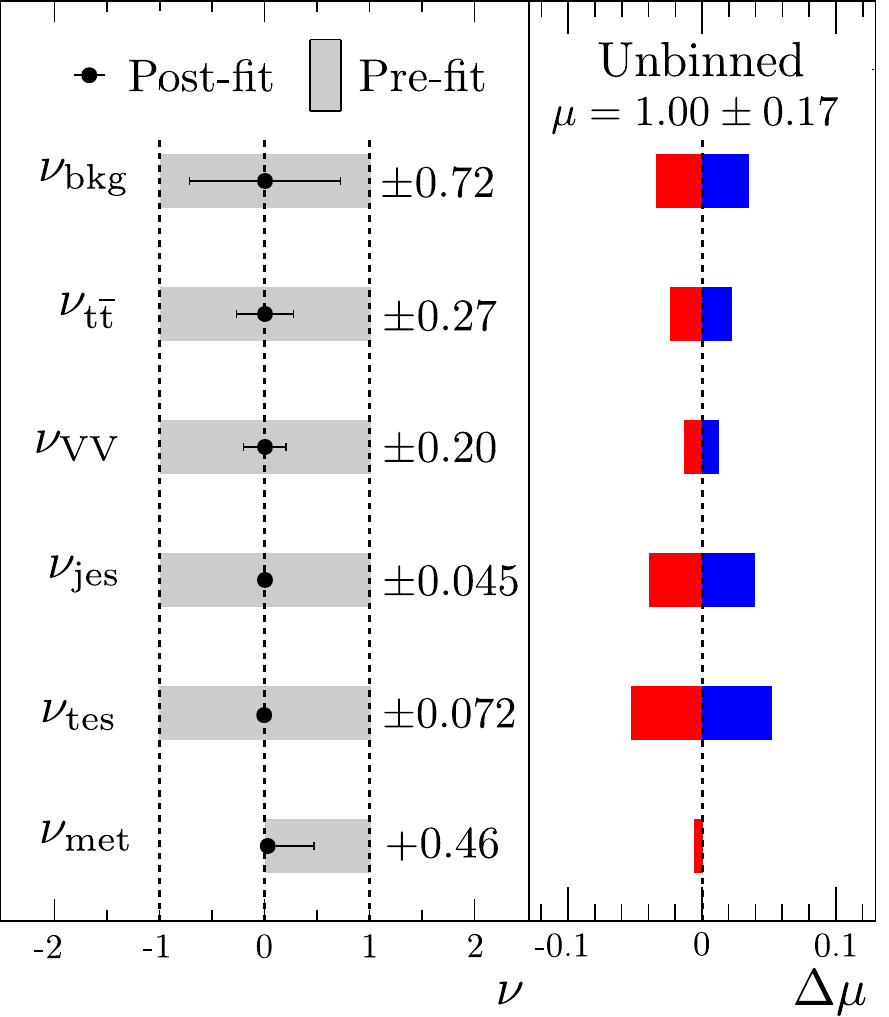} 
    \caption{\label{fig:impacts}  Pre-fit and post-fit nuisance parameter values for the binned analysis~(left) and the unbinned analysis~(right) in the Asimov fit to the nominal simulation. In each figure, the impacts on the measured signal strength are also shown. The red color corresponds to shifts of $\mu$ associated with upward variations of the nuisance parameters, and the blue color corresponds to downward variations. The \texttt{met} uncertainty is one-sided.}
\end{figure*}

The impacts of the nuisance parameters on the extracted signal strength, shown in Fig.~\ref{fig:impacts}, highlight differences between the binned and unbinned approaches. We show pre-fit and post-fit nuisance parameter values for both analyses based on an Asimov fit to the nominal simulation. Starting from the MLE, each nuisance parameter is varied until the likelihood ratio test statistic increases by one. This defines the post-fit uncertainty. Re-fitting the signal strength at this configuration yields the corresponding impact.

In both the binned and unbinned fits, we observe significant post-fit constraints on the normalization uncertainties. This is expected for FAIR-HUC, where the auxiliary likelihood is less constraining than in realistic applications with high-quality calibrations that reduce these impacts already prior to the fit.

Notably, the unbinned model provides slightly stronger constraints on the \ttbar background normalization, while other normalization uncertainties remain similar between the two approaches—indicating that the common binned control regions predominantly determine these parameters. The \texttt{jes} and \texttt{tes} nuisance parameters are constrained to 9.5\% and 8.3\% in the binned case, respectively, while the unbinned fit yields much tighter constraints of 4.5\% and 5.2\%. This suggests that the high-dimensional unbinned likelihood captures subtle shape information more effectively. Moreover, the largest impact in the binned fit arises from the \texttt{jes} nuisance (11.5\%), which is reduced to 4\% and becomes subleading in the unbinned model. The impact of the \texttt{tes} nuisance also improves, decreasing from 8.3\% to 5.2\%, a 37\% reduction. 
Additionally, only the unbinned model exhibits a meaningful constraint on the \texttt{met} nuisance, which remains largely unconstrained in the binned case. The strong reduction in the impact of \texttt{met} in the unbinned analysis indicates a dominantly non-linear influence, which the high-dimensional likelihood captures efficiently. Overall, the results demonstrate that the unbinned likelihood efficiently exploits the information distributed across the feature space. We conjecture that realistic measurements with typically more than 100 nuisance parameters will benefit even more.

Finally, we note that all positive variations of the nuisance parameters around the MLE (shown in red) reduce the signal strength, which is consistent with the training data where positive values of the nuisance parameters increase the background yield.

\begin{figure}
    \centering
    \includegraphics[width=0.98\linewidth]{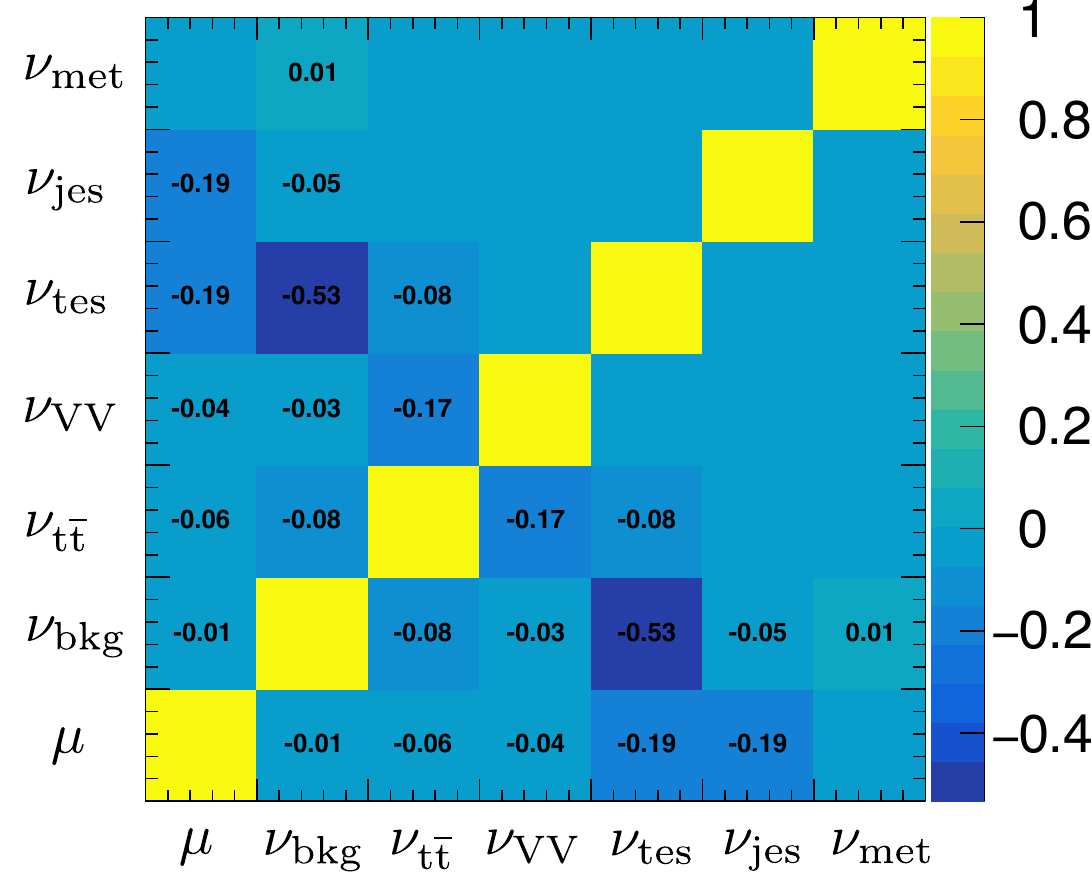}
    \caption{\label{fig:asimov-correlation} The correlation of the seven model parameters at the best-fit point in the unbinned Asimov dataset. Correlations below an absolute value of 1\% and along the diagonal are not shown.}
\end{figure}

We show the correlation matrix of the model parameters in Fig.~\ref{fig:asimov-correlation} for the unbinned models. Notably, the parameters $\nu_\text{tes}$ and $\nu_\text{bkg}$ exhibit a correlation coefficient of $-0.53$. This anti-correlation arises from a normalization change induced by the \texttt{tes} variation across all processes, which results from a shift that moves low-energetic $\tau_\text{had}$ candidates above the selection threshold. All other correlation coefficients have a magnitude of less than 0.2.



\subsection{Fisher information matrix}\label{sec:fisher}
The learned likelihood ratio can be used to assess differences in the approaches for parameter points close to the true values.
We can use the Fisher information matrix to express the likelihood in the vicinity of the fixed but unknown true parameter value  $\bt^\text{true}$ compactly as
\begin{align}
L(\mathcal{D}|\bt) \approx L(\mathcal{D}|\bt^\text{true}) \exp\!\left[-\frac{1}{2} \Delta\bt^T I(\bt^\text{true}) \Delta\bt\right],
\end{align}
where we arrange the seven model parameters as $\bt=(\mu,\bn)$ and $\Delta\bt = \bt-\bt^\text{true}$.
The Fisher information matrix $I(\bt)$, for general \bt, is given by
\begin{align}
I_{ab}(\bt)=\left\langle\frac{\partial}{\partial \theta_a}\log p(\mathcal{D}|\boldsymbol{\theta})\frac{\partial}{\partial \theta_b}\log p(\mathcal{D}|\boldsymbol{\theta})\right\rangle_{\bt},\label{eq:fisher}
\end{align} 
where $a,b$ label the model parameters. The probability density function $p(\mathcal{D}|\boldsymbol{\theta})$ here is the same as the extended likelihood in Eq.~\ref{eq:extended-likelihood} but is interpreted as a function of the dataset.
The Fisher information matrix has many useful applications in estimation problems. For example, the Cram\'er–Rao bound~\cite{RaoCR,CrammerH} states that the covariance matrix of an unbiased estimator, such as $\bar{\bt}$, is bounded from below by the inverse of the Fisher information matrix, i.e.,
\begin{align}
\operatorname{Cov}(\bar\bt) \ge I^{-1}(\bt^\text{true}).
\end{align}
This inequality, understood in the sense that $\operatorname{Cov}(\bar\bt) - I^{-1}(\bt^\text{true})$ is positive semidefinite, establishes a fundamental lower bound on the achievable variance and covariance for any unbiased estimator of the parameters. We evaluate the Fisher information in the Asimov data set by differentiating our parametric model using binned and unbinned components exactly as in Eq.~\ref{eq:binned-and-unbinned}. The details are provided in App.~\ref{sec:FI-entries}.

\begin{figure}
    \centering
    \includegraphics[width=0.99\linewidth]{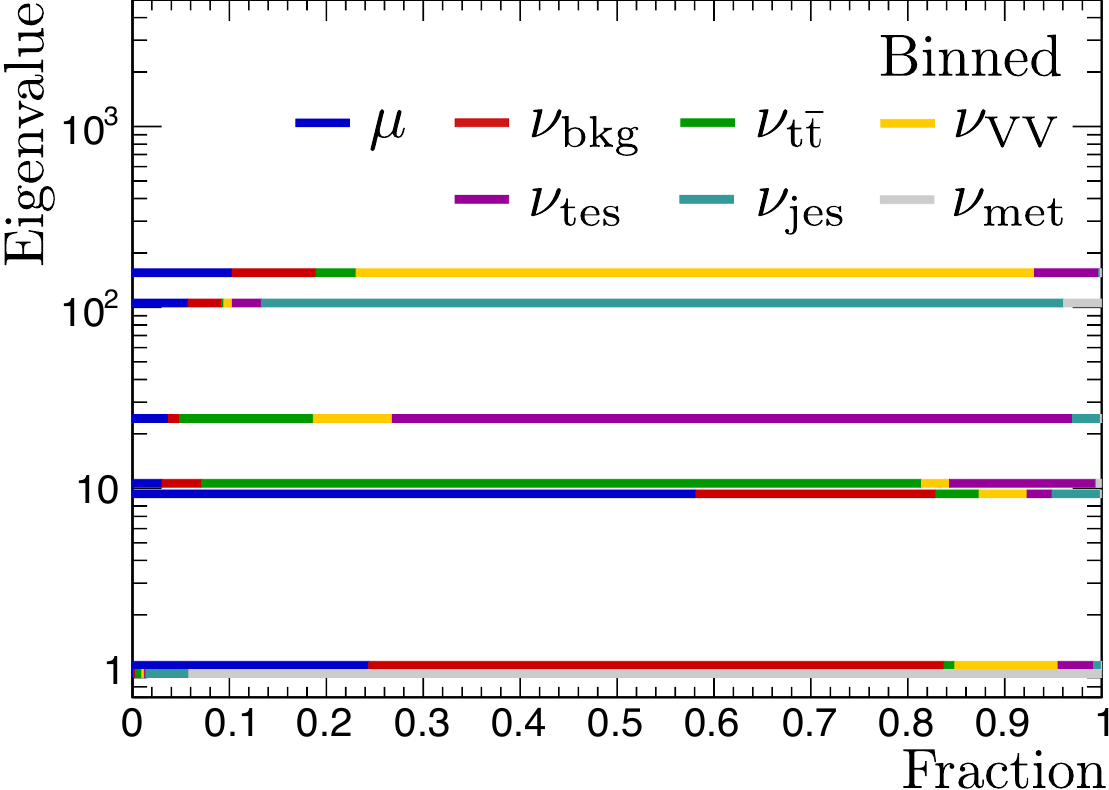}\\\hspace{.5cm}
    \includegraphics[width=0.99\linewidth]{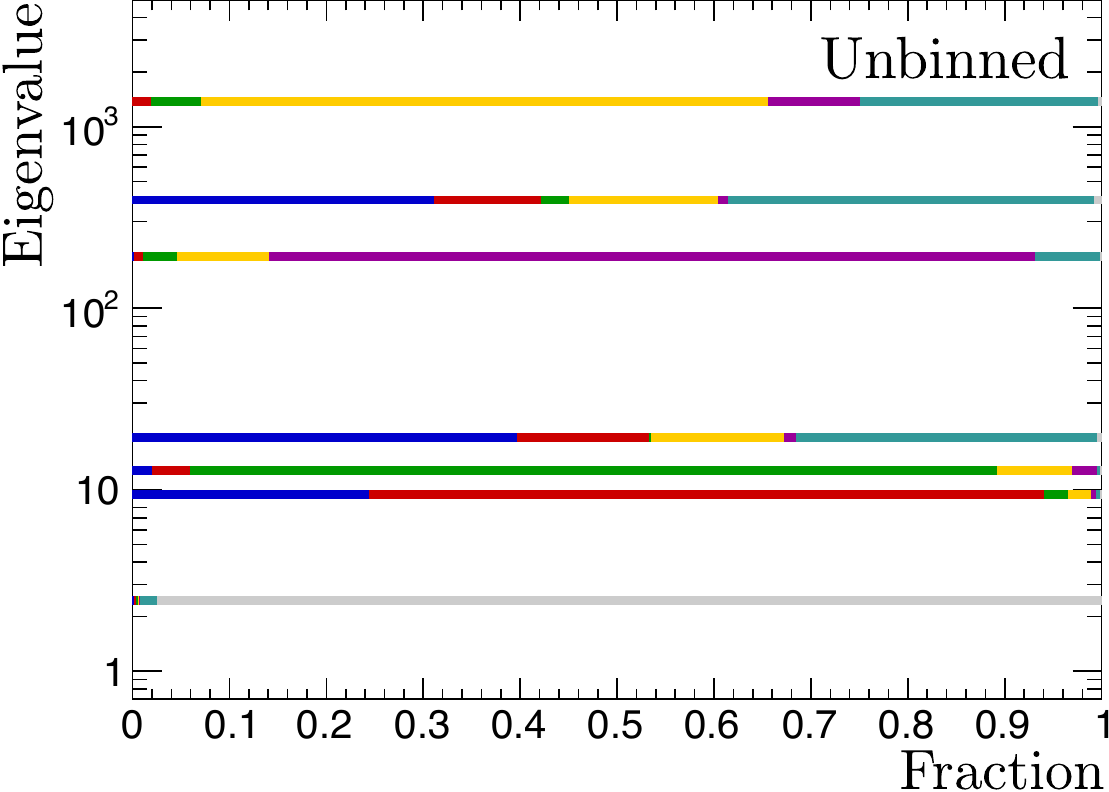} 
    \caption{\label{fig:Fisher-eigensystem} A visual representation of the eigensystems of the Fisher information in the binned~(top) and unbinned case~(bottom). The eigenvalue gives the position on the y-axis, and the color-coded fractions in the horizontal direction represent the absolute values of the coefficients of the corresponding eigenvector.}
\end{figure}

We provide a visual representation of the eigensystems of the Fisher information in Fig.~\ref{fig:Fisher-eigensystem} for both the binned and unbinned cases. In these plots, the eigenvalues are shown on the vertical axis, indicating the sensitivity of the model to the combination of model parameters encoded as eigenvectors.  On the horizontal axis, color-coded bars represent the absolute values of the corresponding eigenvector coefficients, illustrating the fractional contributions of each parameter. The uncertainty in a given eigendirection with eigenvalue EV is $\textrm{EV}^{-1/2}$ such that higher values are desirable. Notably, in the unbinned case, the largest eigenvalue is associated with an eigenvector that corresponds to a tight constraint on a combination of nuisance parameters --- a feature that does not appear in the binned Fisher eigensystem. Moreover, the second largest eigenvalue in the unbinned case is linked to an eigenvector with a signal fraction of 30\%, whereas the binned analysis yields two eigenvectors with a similar signal contribution but a lower combined fraction. This observation supports our understanding that the unbinned analysis is more effective at disentangling nuisance parameters. Finally, the $\nu_\text{met}$ direction appears poorly constrained, reflecting its small linear impact, which dominates the Fisher information analysis.

\subsection{Toy studies}\label{sec:toy-results}
To further assess the quality of the approach, we probe the unbinned model in Eq.~\ref{eq:u-all-unbinned} with toy data samples. 

\begin{figure}
    \centering
    \includegraphics[width=0.98\linewidth]{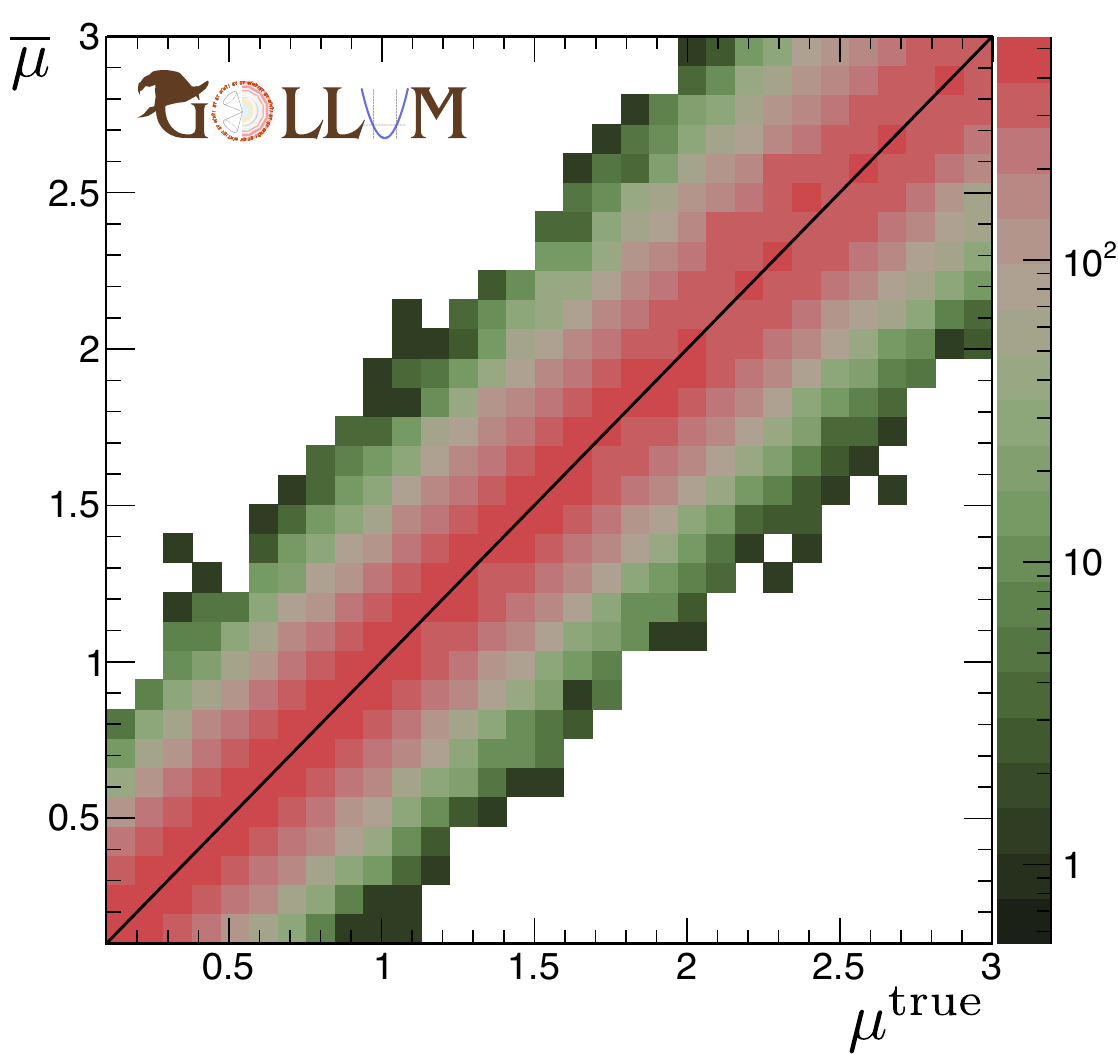}
    \caption{\label{fig:toy-scatter-mu} Scatter plot of the true value of the $\PH\to\tau\tau$ signal strength parameter $\mu$ and the MLE $\bar\mu$ for $5\cdot10^4$ toys. }\label{fig:mu-corr}
\end{figure}

We sample $5\cdot10^4$ model parameter configurations, with $\mu^\text{true} \sim U(0.1,3)$ uniformly distributed, $\nu^\text{true}_\text{met} \sim \text{lnN}(0,1)$ log-normally distributed, and the remaining five model parameters drawn from normal distributions. Samples with parameters outside the boundaries described in Sec.~\ref{sec:sys} are removed.  This reproduces the probability density function used to assess model performance in FAIR-HUC~\cite{Bhimji:2024bcd}. 

\begin{figure}
    \centering
    \includegraphics[width=0.9\linewidth]{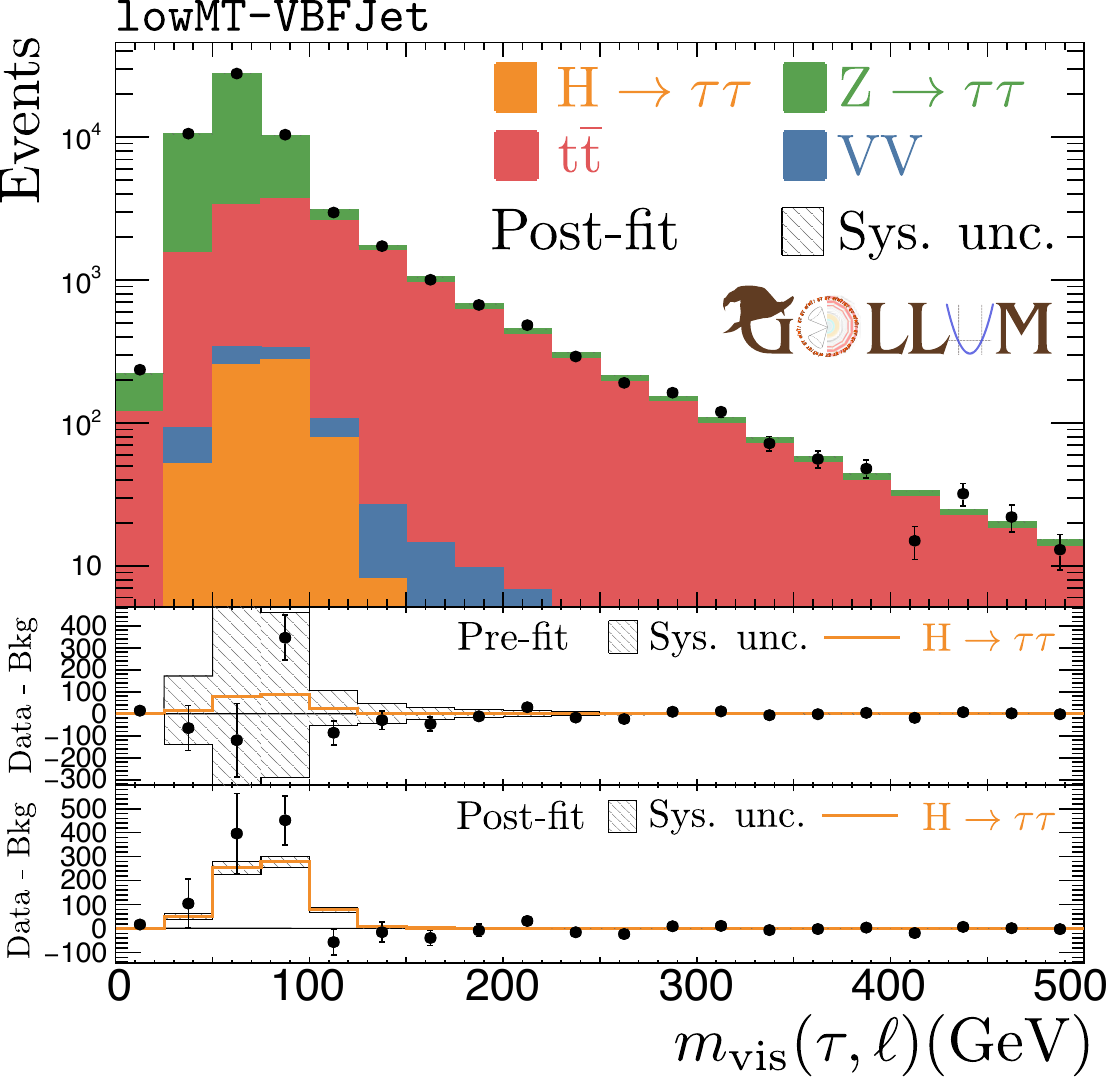}\\
    \hspace{-.2cm}\includegraphics[width=0.875\linewidth]{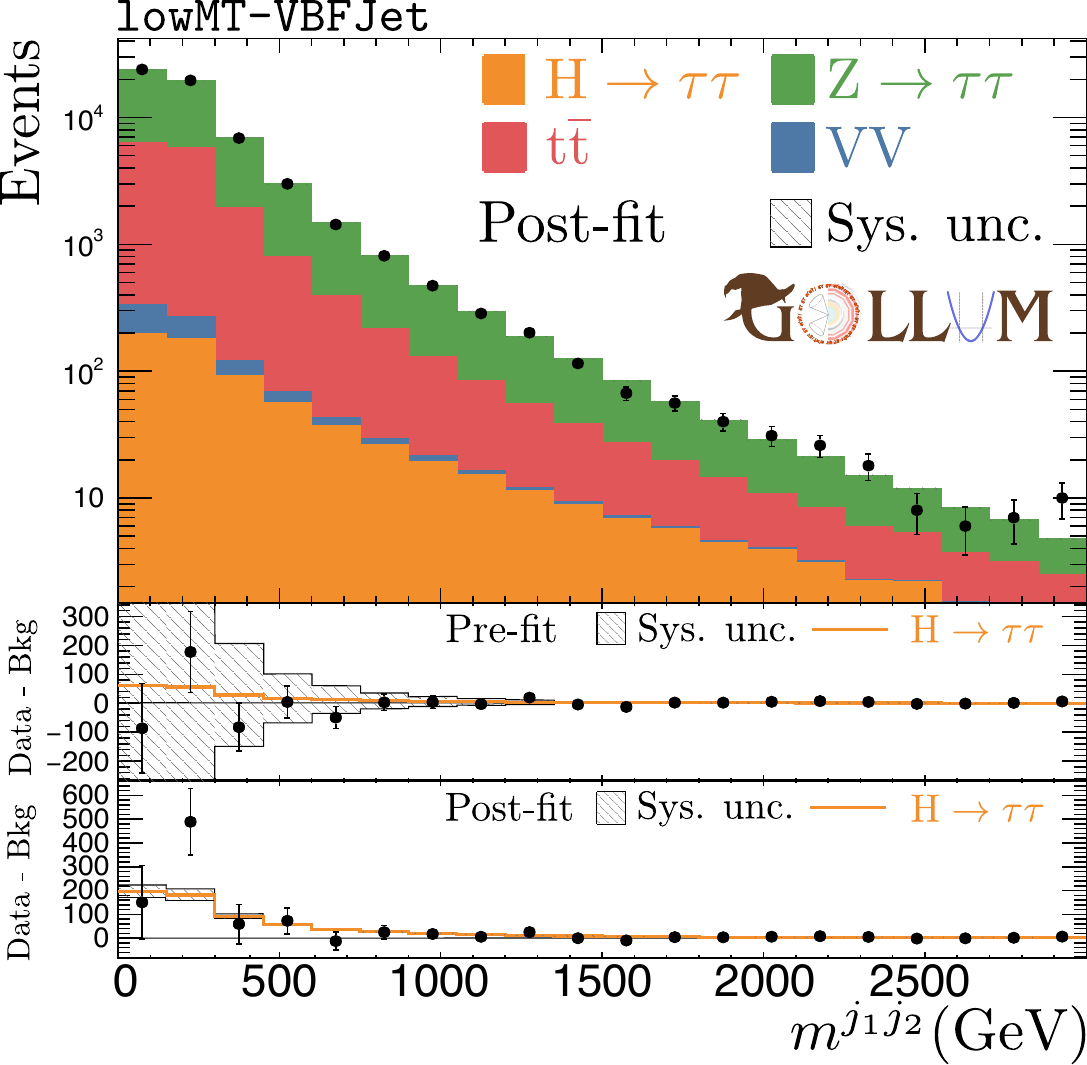}\\
    \includegraphics[width=0.89\linewidth]{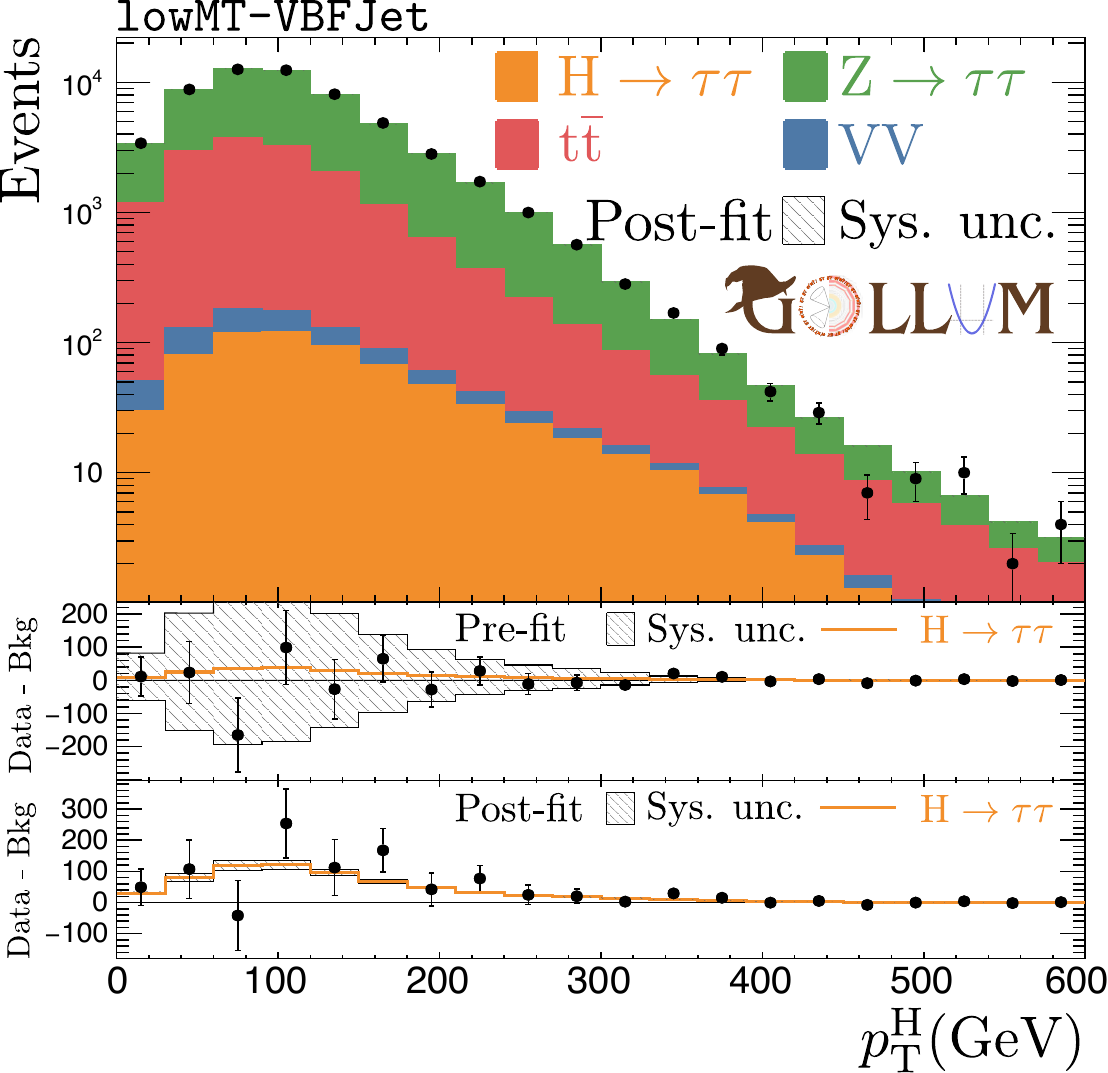}
    \caption{\label{fig:toy-postfit} Pre- and post-fit distributions in the \texttt{lowMT-VBFJet} region for the observables $m_\text{vis}(\eta,\ell)$~(top), $m^{j_1j_2}$~(middle), and $\pt^\PH$~(bottom) for a randomly chosen toy dataset with $\mu^\text{true}=3$. The MLE is $\bar\mu=3.24\pm0.25.$}
\end{figure}

For each configuration, we sample from test data that was not used during surrogate training. The values of the signal strength and normalization uncertainties modify the expected event yields. The effects of the calibration-type uncertainties \texttt{tes}, \texttt{jes}, and \texttt{met} are applied using the code snippets from Ref.~\cite{Bhimji:2024bcd}, which adjust both primary and derived features. These modifications are computed on the fly for each toy dataset. The resulting datasets are used to compute the test statistic $q_\mu(\mathcal{D})$, the maximum likelihood estimate $(\bar\mu,\bar\bn)$, the 68\% confidence interval boundaries, $\mu_{16}$ and $\mu_{84}$, and the 95\% confidence interval boundaries, $\mu_{2.5}$ and $\mu_{97.5}$.

\begin{figure*}
    \centering
    \includegraphics[width=0.32\linewidth]{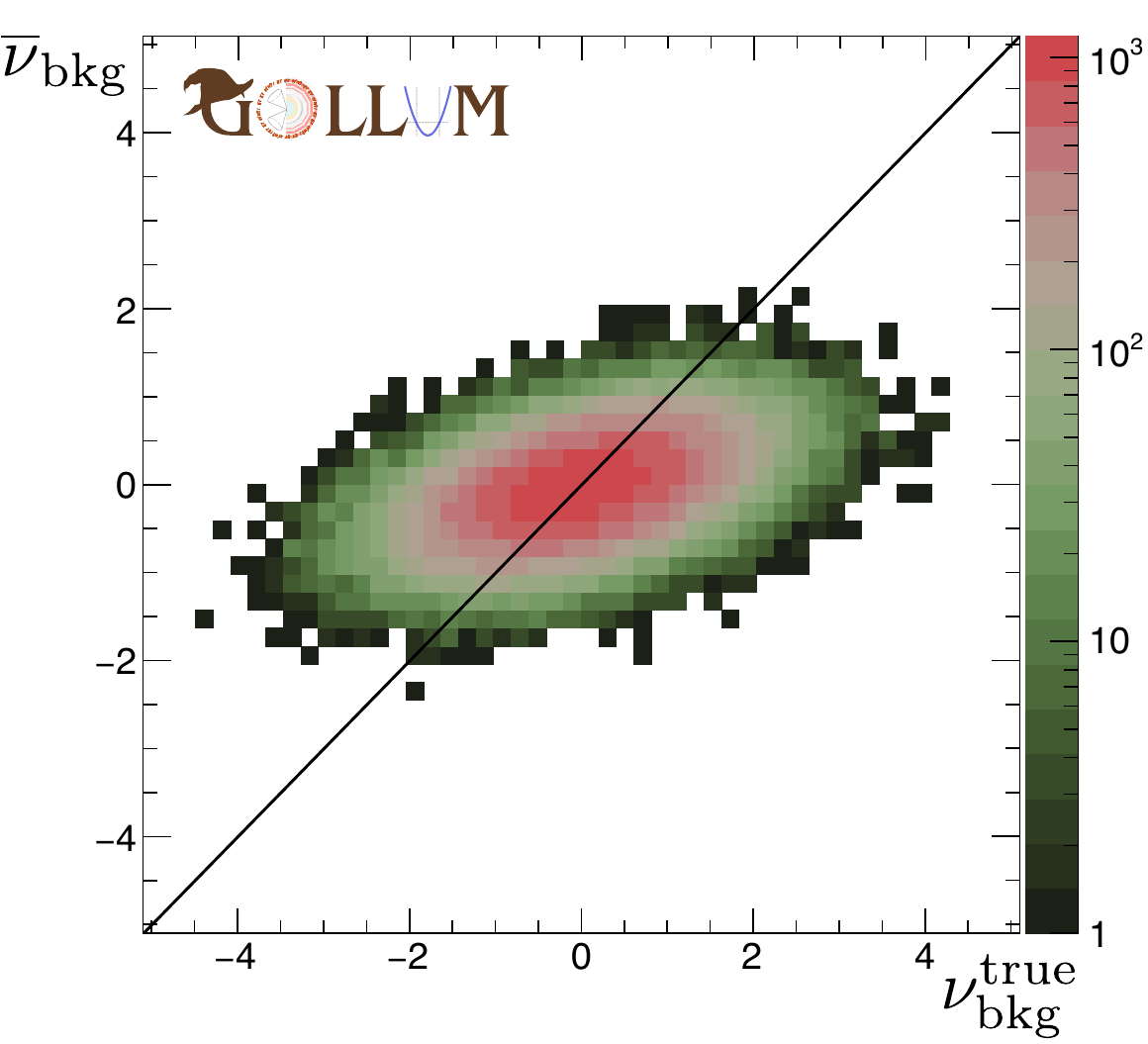}
    \vspace{-.1cm}\includegraphics[width=0.32\linewidth]{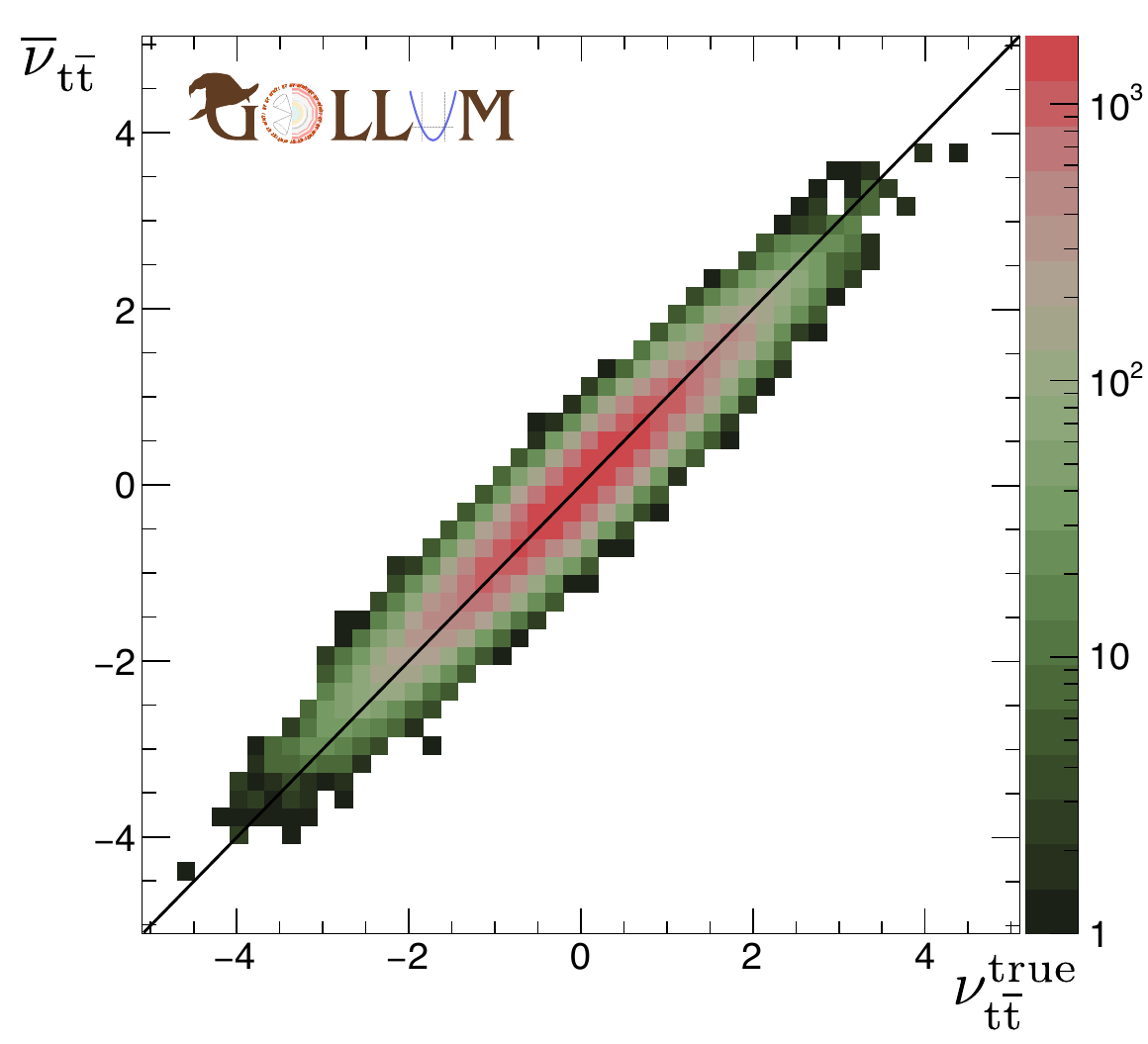}
    \includegraphics[width=0.32\linewidth]{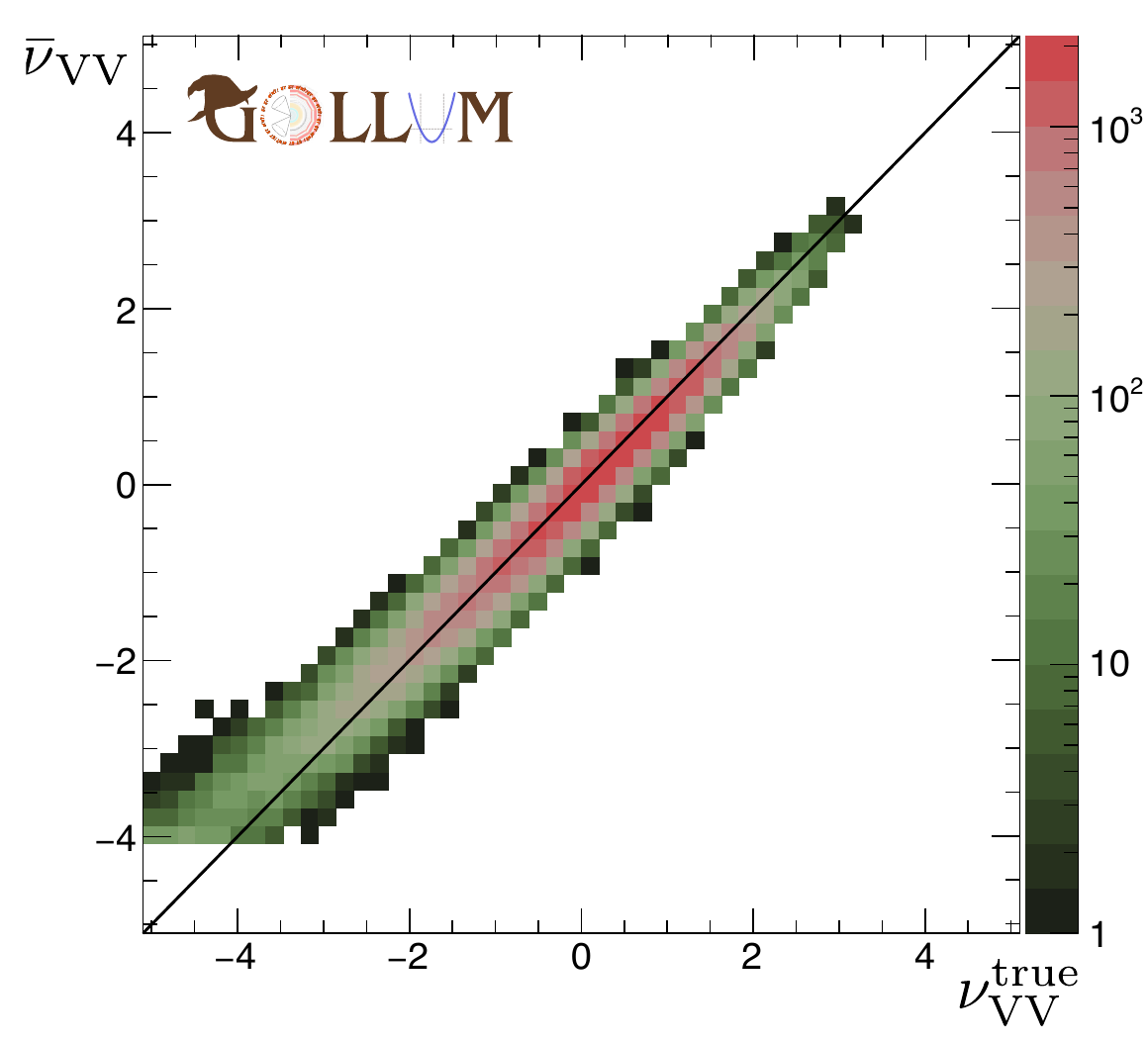}\hfill
    \caption{\label{fig:toy-scatter-nuisances-2} Same as Fig.~\ref{fig:mu-corr}, but for $\nu_\text{bkg}$~(left), $\nu_{\ttbar}$~(middle), and $\nu_{\VV}$~(right).}\label{fig:scatter-nuis-1}
\end{figure*}

\begin{figure*}
    \centering
      \includegraphics[width=0.32\linewidth]{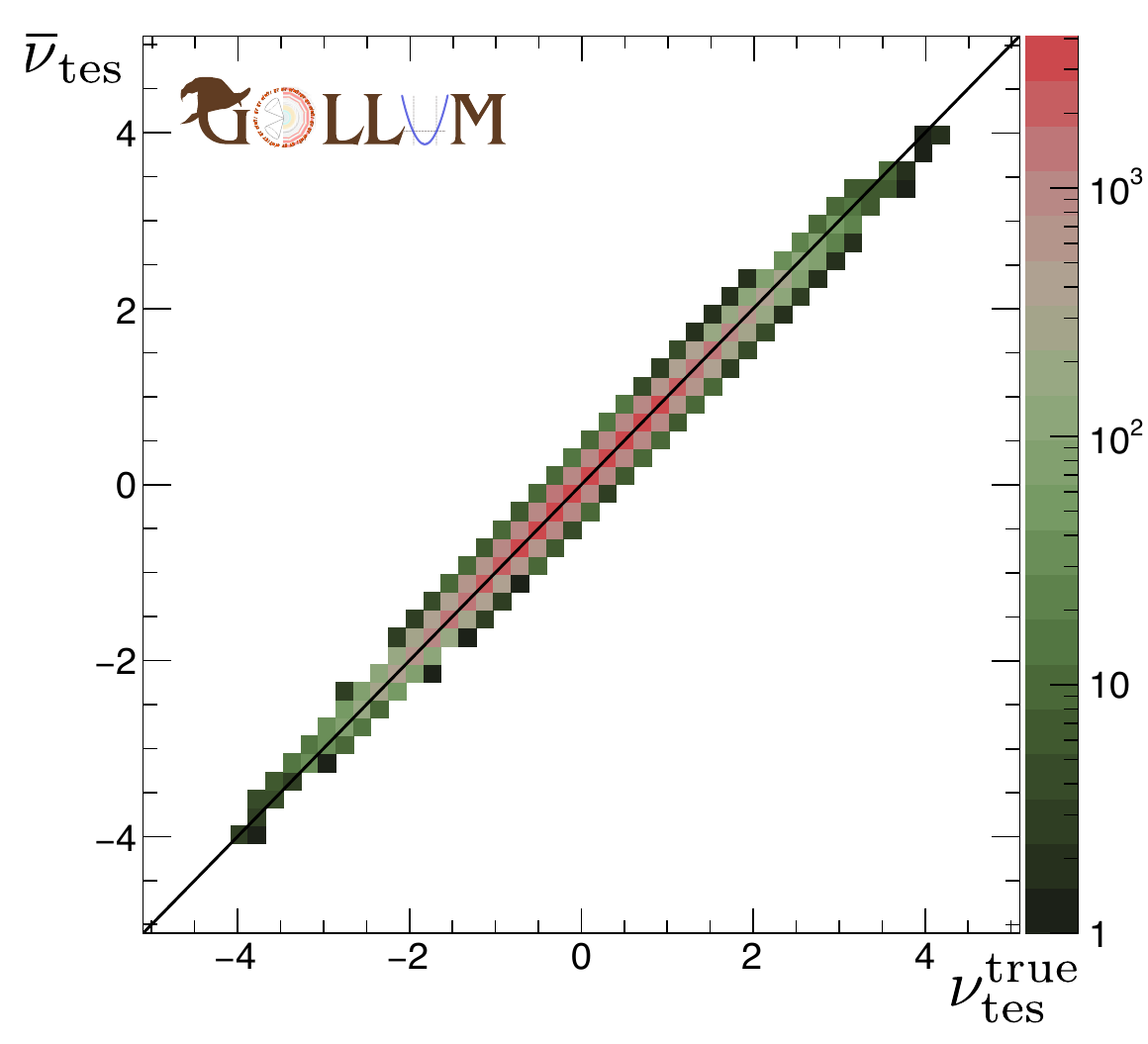}
    \includegraphics[width=0.32\linewidth]{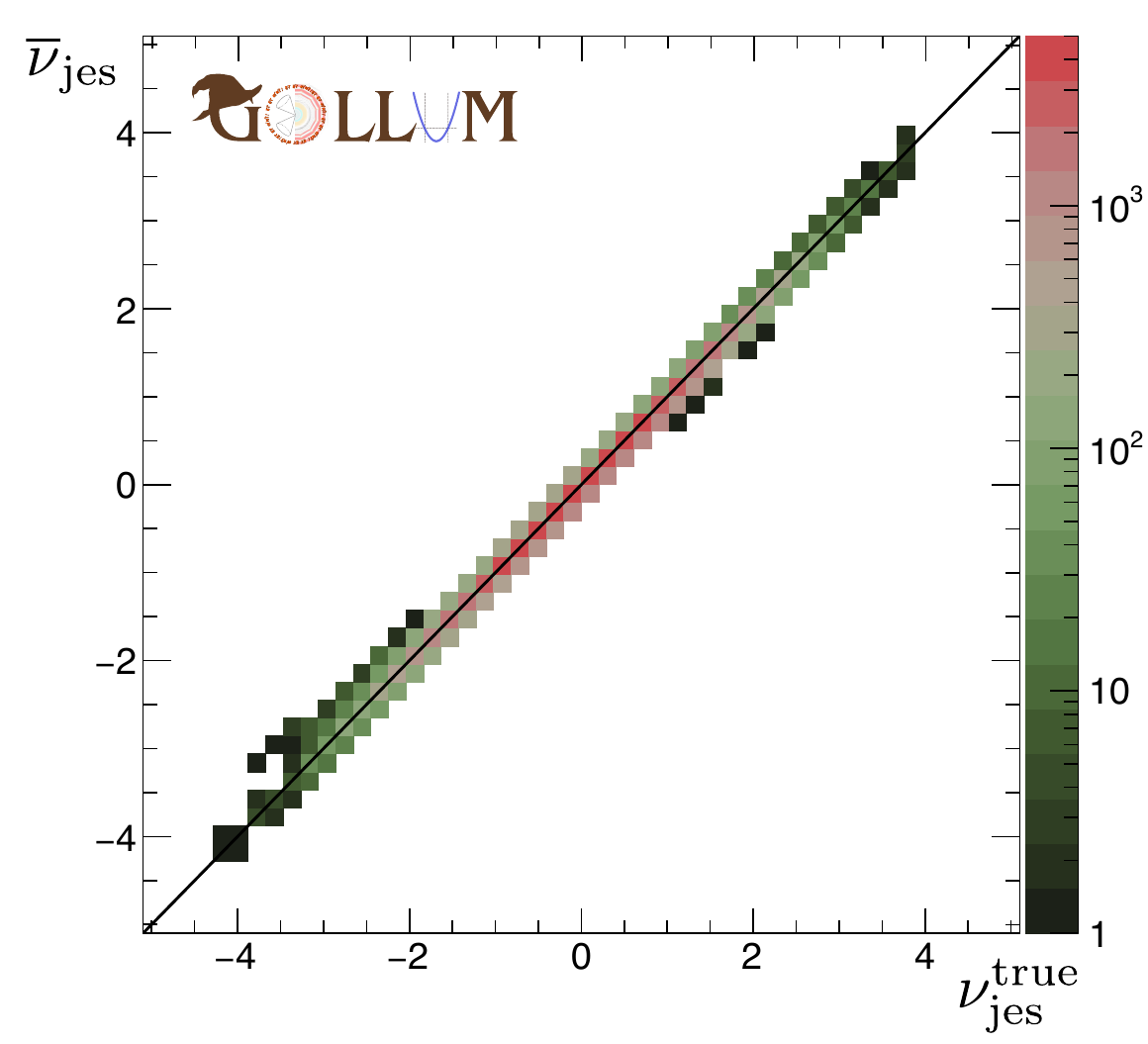}
    \includegraphics[width=0.32\linewidth]{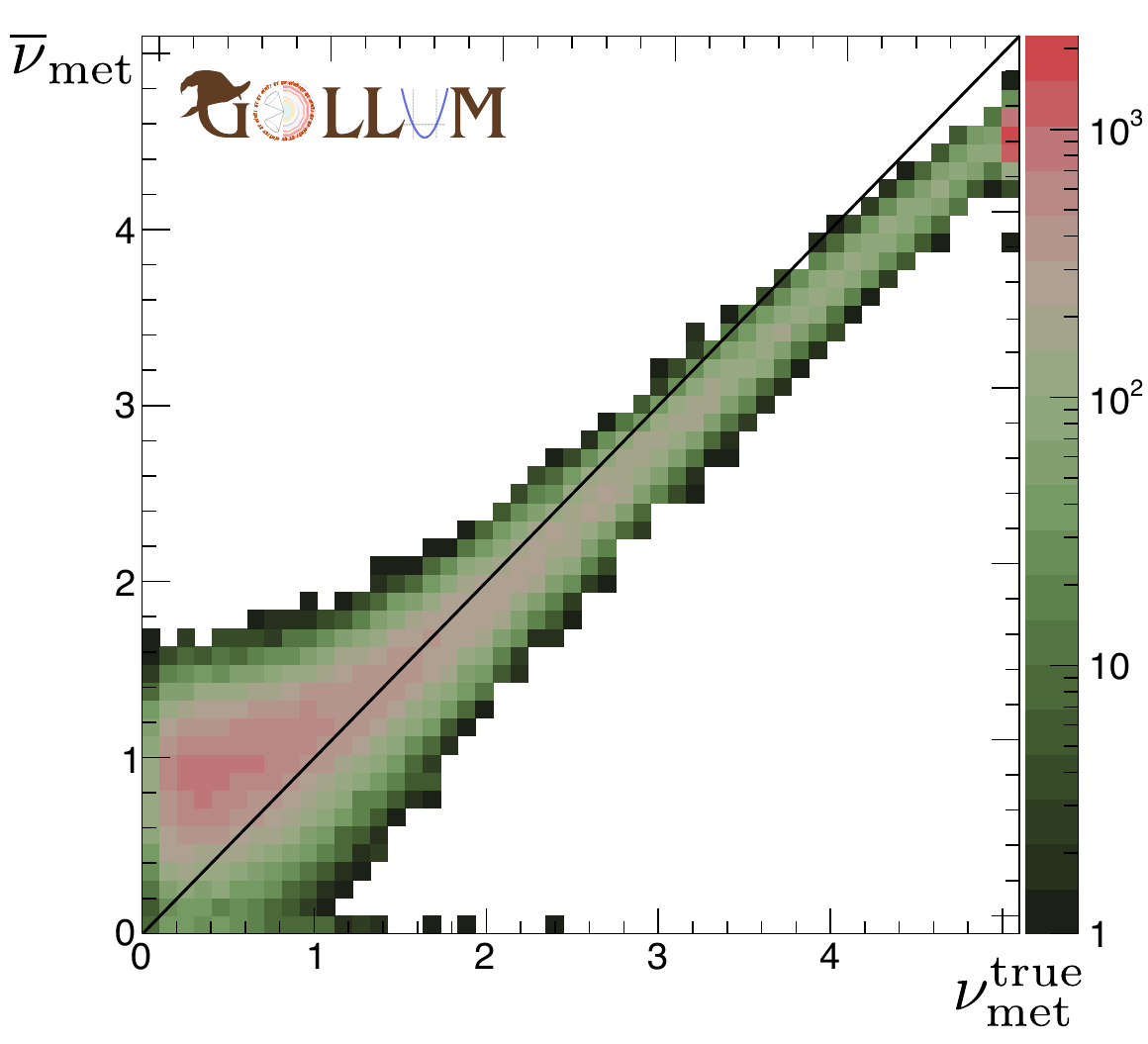}
    \caption{\label{fig:toy-scatter-nuisances-1} Same as Fig.~\ref{fig:mu-corr}, but for $\nu_\text{tes}$~(left), $\nu_\text{jes}$~(middle), and $\nu_\text{met}$~(right).}\label{fig:scatter-nuis-2}
\end{figure*}

\begin{figure*}
    \centering
    \includegraphics[width=0.48\linewidth]{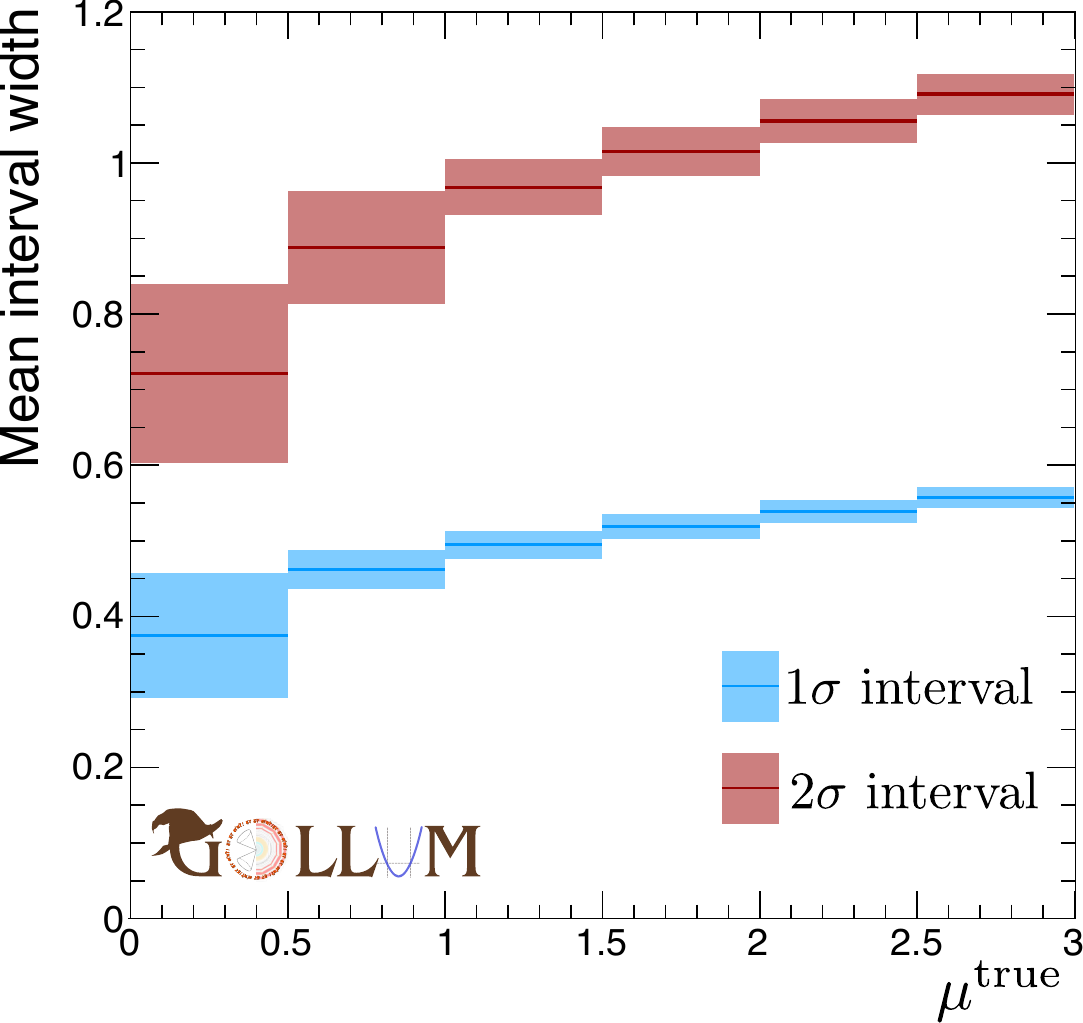}\hfill
    \includegraphics[width=0.48\linewidth]{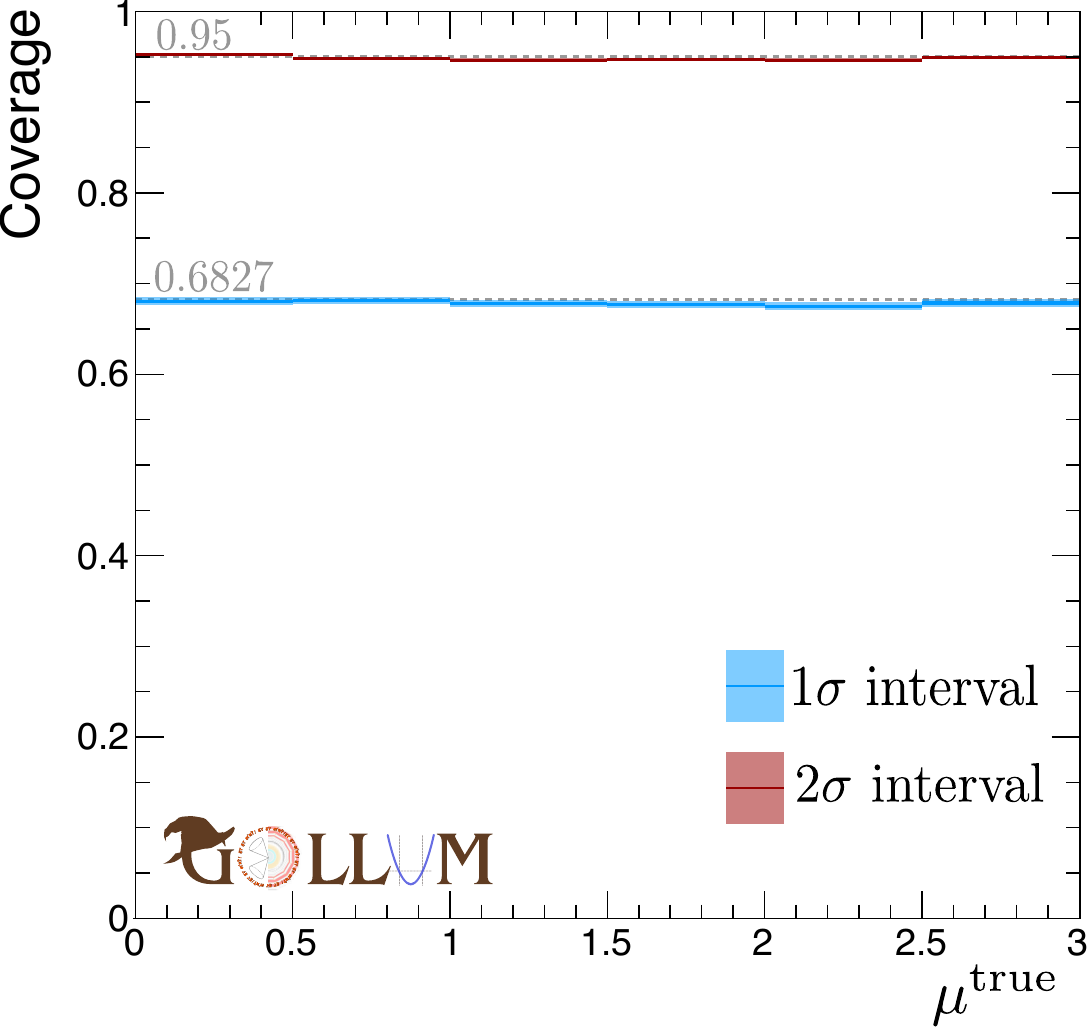}\\
    \caption{\label{fig:toy-coverage} Mean interval width of the signal strength parameter $\mu$~(left) and coverage~(right) as a function of $\mu^\text{true}$ obtained from $5\cdot 10^4$ toy datasets.}
\end{figure*}

In Fig.~\ref{fig:mu-corr}, we show the scatter plot of $\mu^\text{true}$ versus $\bar\mu$. The MLE points lie along the diagonal with no visible bias. 
There is no population of outliers, indicating excellent stability of the \texttt{MINUIT} minimization and the unbinned surrogate model. 

In Fig.~\ref{fig:toy-postfit}, we show the post-fit distribution of a randomly chosen toy fit for the observables $m_\text{vis}(\eta,\ell)$, $m^{j_1j_2}$, and $\pt^\PH$. In the bottom panels we compare the pre-fit and post-fit uncertainties. The much narrower post-fit uncertainty bands visually demonstrate how the high-dimensional approach constraints the systematic uncertainties and extracts sensitivity. For this particular toy, we use $\mu^\text{true}=3$ and found $\bar\mu=3.24\pm0.25$.

Figures~\ref{fig:scatter-nuis-1} and \ref{fig:scatter-nuis-2} show analogous scatter plots for the nuisance parameters. As discussed in Sec.~\ref{sec:asimov-results}, $\nu_\text{bkg}$ is only weakly constrained, and the MLE shows little correlation with the true value. In contrast, $\nu_{\ttbar}$ and $\nu_{\VV}$ cluster tightly along the diagonal, although very small values of $\nu_{\VV}$ are harder to reconstruct accurately. The values of $\nu_\text{tes}$ and $\nu_\text{jes}$ are precisely recovered, consistent with their tight post-fit constraints. For $\nu_\text{met}$, we observe a slight bias at both ends of the parameter range, but this has negligible impact given the small post-fit impact of $\nu_\text{met}$. 
Overall, we observe very few outliers, indicating stable reconstruction across all parameters.

In Fig.~\ref{fig:toy-coverage}~(left), we show the mean interval widths at the $1\sigma$ and $2\sigma$ confidence levels, given by $\mu_{84}-\mu_{16}$ and $\mu_{97.5}-\mu_{2.5}$, respectively. Both increase slightly with $\mu^\text{true}$ across the probed range.

The final evaluation, which also tests the quality of the learned parametrization, is the coverage test shown in Fig.~\ref{fig:toy-coverage}~(right). As a function of $\mu^\text{true}$, we show the fraction of toy fits for which $\mu^\text{true}$ lies within the reconstructed $1\sigma$ and $2\sigma$ intervals. We observe excellent agreement, indicating that the unbinned model provides reliable coverage.

Our code ``Guaranteed Optimal LikeLihood-based Unbinned Method'' (\textsc{GOLLUM}) for ML and inference is publicly available at Ref.~\cite{GOLLUM}.

\section{Conclusions}\label{sec:outlook}

We have presented a novel unbinned methodology for inclusive cross-section measurements that systematically incorporates machine-learned approximations of systematic uncertainties. 
By combining a calibrated multi-class classifier with a parametric neural network surrogate for the differential cross-section ratio, we reconstruct a high-dimensional likelihood ratio test statistic that captures both known analytic and unknown data-driven dependencies on nuisance parameters. 
Using the FAIR-HUC benchmark dataset, we demonstrate significant improvements over traditional binned analyses. In Asimov data and large-scale toy studies, the unbinned model outperforms the binned reference by reducing the uncertainty in the signal strength by 20\% and achieving higher significance for signal exclusion. 
The Fisher information matrix further confirms that the unbinned likelihood extracts more information from the nuisance parameters. The method is fully refinable and compatible with established workflows, enabling stage-wise improvements without retraining the entire analysis. Our implementation is provided in the public \textsc{GOLLUM} codebase~\cite{GOLLUM}, offering a flexible and robust toolkit for unbinned likelihood-based inference at the LHC and beyond.

\vspace{-.3cm}
\section*{Acknowledgments}
We thank the organizers of the FAIR-HUC challenge for the organization and technical support. C.G. and M.S. received support from the Austrian Science fund~(FWF) under grant I5687. D.S. received support under grant P33771. The computational results were obtained using the Vienna Bio Center and the CLIP computing cluster of the Austrian Academy of Sciences  at~\url{https://www.clip.science/}.

\section*{Data availability}
The data that support the findings of this article are openly available~\cite{benato_2025_15322773,GOLLUM}. The underlying FAIR-HUC data~\cite{bhimji_2025_15131565} is described in~\cite{Bhimji:2024bcd}.

\appendix

\section{Binned and unbinned Asimov data}\label{sec:app-asimov}

Asimov data~\cite{Cowan:2010js} can be used to compute sampling-free expected results within continuous parametric models.  
In the well-known binned case, the Asimov expectation amounts to replacing the integer-count observation in, e.g., Eq.~\ref{eq:u-binned}, by the expected yield under the (assumed) true hypothesis, which we denote by $\mu'\,\bn'$, that is
\begin{align}
    \langle N_{\textrm{obs}}\rangle_{\mu',\bn'}\rightarrow \mathcal{L}\sigma(\mu'\bn')
\end{align} 
for every bin. 

The unbinned generalization, first obtained in Ref~\cite{GomezAmbrosio:2022mpm}, is the continuum limit of this procedure and amounts to replacing the sum over observed events in the last term in Eq.~\ref{eq:u-unbinned} by a weighted sum over simulated events, concretely 
\begin{align}
\left\langle\sum_{i=1}^{N_\text{obs}}(\cdots)\right\rangle_{\mu',\bn'}\;&\rightarrow \sum_{\{w_i,\bx_i\}\in\mathcal{D}_{\mu',\bn'}}w_i\times(\cdots).
\end{align}
For obtaining Asimov limits with arbitrary $\mu'$ it thus suffices to compute the log-term in  the unbinned test statistic as a weighted sum and scale the signal weights by $\mu'$.

We can also use the learned model instead, that is,
\begin{align}
\left\langle\sum_{i=1}^{N_\text{obs}}(\cdots)\right\rangle_{\mu',\bn'}\;&\rightarrow \sum_{\{w_i,\bx_i\}\in\mathcal{D}_{1,\bzero}}\;w_i \hat R(\bx_i|\mu'\bn')\times (\cdots).\label{eq:asimov-exp-learned}
\end{align}
The approximate expectation of the log-likelihood ratio then becomes
\begin{align}
&\left\langle-\frac{1}{2}u_{\text{UB}}(\mathcal{D}|\mu,\bn)\right\rangle_{\mu',\bn'}\approx\sum_{r=1}^{N_r}\sum_{\substack{\{w_i,\bx_i\}\in\\\mathcal{D}_{1,\bzero}\cap r}}w_i\Big[1-\hat R_r(\bx_i|\mu,\bn)\nonumber\\
&\qquad\quad+\hat R_r(\bx_i|\mu',\bn')\log \hat R_r(\bx_i|\mu,\bn)\Big],\label{eq:asimov-exp}
\end{align}
which we can evaluate parametrically. 

Because we defined $u$ in Eq.~\ref{eq:test-stat} with respect to $(\mu,\bn)=(1,\bzero)$, we need not subtract the reference log-likelihood to re-obtain the general case. The Asimov expectation for generic $(\mu',\bn')$ is 
\begin{align}
-\frac{1}{2}\Lambda&\equiv\left\langle\frac{L(\mathcal{D}|\mu,\bn)\;\;}{L(\mathcal{D}|\mu',\bn')}\right\rangle_{\mu',\bn'}\nonumber\\
&=-\frac{1}{2}\langle u_\text{UB}(\mathcal{D}|\mu,\nu)-u_\text{UB}(\mathcal{D}|\mu',\nu')\rangle_{\mu',\bn'}\nonumber\\
&=\sum_{r=1}^{N_r}\sum_{\substack{\{w_i,\bx_i\}\in\\\mathcal{D}_{1,\bzero}\cap r}} w_i\;\Bigg[\hat R_r(\bx_i|\mu',\bn')-\hat R_r(\bx_i|\mu,\bn)\nonumber\\
&\quad+\hat R_r(\bx_i|\mu',\bn')\log\frac{\hat R_r(\bx_i|\mu,\bn)\;\;\;}{\hat R_r(\bx_i|\mu',\bn')}\Bigg],
\end{align}
which has the same form as Eq.~B.15 in Ref.~\cite{GomezAmbrosio:2022mpm}. The number $\Lambda$ on the l.h.s. is the non-centrality parameter of a non-central $\chi^2$ distribution which, according to Wald's theorem~\cite{Wald}, is the asymptotical limit of the distribution of the test-statistic. Using the learned model in Eq.~\ref{eq:asimov-exp}, we have $\Lambda$ parametrically as a function of $\mu'$ and $\bn'$.

\section{Entries of the Fisher information matrix}\label{sec:FI-entries}

In this section, we compute the entries of the seven-parameter Fisher information matrix for binned and unbinned datasets.
Combining the components of the model for binned yields in Sec.~\ref{sec:binned-LL} and abbreviating $\bt=(\mu,\bn)$, we have
\begin{align}
    \lambda(\bt)&=\mathcal{L}\Bigg[\mu\,\sigma_{\PH}\hat S_\PH(\bn)+(1+\alpha_\text{bkg})^{\nu_\text{bkg}}\times\nonumber\\
    &\qquad\Big(\sigma_{\PZ}\hat S_\PZ(\bn)+(1+\alpha_{\ttbar})^{\nu_{\ttbar}}\sigma_{\ttbar}\hat S_{\ttbar}(\bn)\nonumber\\
    &\quad+(1+\alpha_{\VV})^{\nu_{\VV}}\sigma_{\VV}\hat S_{\VV}(\bn)\Big)\Bigg]
\end{align}
where $\hat S_p=\exp(\nu_A\hat\Delta_{A,p})$. 

The Fisher information is additive, so if $r$ labels all bins contributing to the likelihood, we find from Eq.~\ref{eq:fisher}
\begin{align}
    &I^{(\text{binned})}_{ab}(\bt)\nonumber\\
    &\quad=\sum_r\left\langle\frac{\partial}{\partial \theta_a}\log\text{P}(N_{r,\text{obs}}|\lambda_r(\boldsymbol{\theta}))\frac{\partial}{\partial \theta_b}\log\text{P}(N_{r,\text{obs}}|\lambda_r(\boldsymbol{\theta}))\right\rangle_{\bt}\nonumber\\
    &\quad=\sum_r\frac{1}{\lambda_r(\bt)}\frac{\partial\lambda_r(\bt)}{\partial\theta_a}\frac{\partial\lambda_r(\bt)}{\partial\theta_b}\label{eq:FI-binned}
\end{align}
where we label the seven parameters in \bt with indices $a,b$.
The derivatives, evaluated at $\bt=(1,\bzero)$, are
\begin{subequations}\label{eq:FI-binned-der}
    \begin{align}
    \frac{\partial}{\partial\mu}\lambda(\bt)\Big|_{1,\bzero}&=\mathcal{L}\sigma_\PH\\
    \frac{\partial}{\partial\nu_\text{bkg}}\lambda(\bt)\Big|_{1,\bzero}&=\mathcal{L}(\sigma_\PZ+\sigma_{\ttbar}+\sigma_{\VV})\log(1+\alpha_\text{bkg})\\
    \frac{\partial}{\partial\nu_{\ttbar}}\lambda(\bt)\Big|_{1,\bzero}&=\mathcal{L}\sigma_{\ttbar}\log(1+\alpha_{\ttbar})\\
    \frac{\partial}{\partial\nu_{\VV}}\lambda(\bt)\Big|_{1,\bzero}&=\mathcal{L}\sigma_{\VV}\log(1+\alpha_{\VV})\\
    \frac{\partial}{\partial\nu_{\text{tes}}}\lambda(\bt)\Big|_{1,\bzero}&=\mathcal{L}(\sigma_\PH\hat\Delta_{\text{tes},\PH}+\sigma_\PZ\hat\Delta_{\text{tes},\PZ}\nonumber\\
    &\quad+\sigma_{\ttbar}\hat\Delta_{\text{jes},\ttbar}+\sigma_{\VV}\hat\Delta_{\text{jes},\VV})\\
        \frac{\partial}{\partial\nu_{\text{jes}}}\lambda(\bt)\Big|_{1,\bzero}&=\mathcal{L}(\sigma_\PH\hat\Delta_{\text{jes},\PH}+\sigma_\PZ\hat\Delta_{\text{jes},\PZ}\nonumber\\
    &\quad+\sigma_{\ttbar}\hat\Delta_{\text{jes},\ttbar}+\sigma_{\VV}\hat\Delta_{\text{jes},\VV})\\
    \frac{\partial}{\partial\nu_{\text{met}}}\lambda(\bt)\Big|_{1,\bzero}&=\mathcal{L}(\sigma_\PH\hat\Delta_{\text{met},\PH}+\sigma_\PZ\hat\Delta_{\text{met},\PZ}\nonumber\\
    &\quad+\sigma_{\ttbar}\hat\Delta_{\text{met},\ttbar}+\sigma_{\VV}\hat\Delta_{\text{met},\VV})
\end{align}
\end{subequations}
where we suppressed the index $r$ on all yields $\lambda$ on the l.h.s. and on all per-bin cross-sections $\sigma$ on the r.h.s. From these equations, we can compute $I^{(\text{binned})}$. While Eq.~\ref{eq:FI-binned-der} is not particularly illuminating, it implies that in the absence of nuisance parameters, i.e., for an idealized single-parameter measurement of $\mu$ in the presence of backgrounds, 
\begin{align}
I^{(\text{binned})}_{\mu\mu}(1,\bzero)=\frac{(\mathcal{L}\sigma_H)^2}{\mathcal{L}(\sigma_H+\sigma_{Z}+\sigma_{\ttbar}+\sigma_{\VV})}=\rho\,\mathcal{L}\,\sigma_H
\end{align}
where $\rho$ is the signal purity. The Cram\'er-Rao bound then states that
\begin{align}
    \sqrt{C_{\mu\mu}}\geq\frac{1}{\sqrt{\rho\,\mathcal{L}\,\sigma_H}}
\end{align}
i.e. the uncertainty of an unbiased estimator is bounded by the reciprocal square root of the expected signal yield, as usual, but it is also penalized by the purity in the bin.

Next, we compute the Fisher information for the unbinned extended likelihood where we drop the sum over the unbinned regions for simplicity. The general expression then is 
\begin{align}
    I_{ab}(\bt)&=\int\ddd\mathcal{D}\,p(\mathcal{D}|\bt)\,\partial_a \log p(\mathcal{D}|\bt)\,\partial_b\log p(\mathcal{D}|\bt)\nonumber\\
    &=-\int\ddd\mathcal{D}\,p(\mathcal{D}|\bt)\,\partial_a\partial_b \log p(\mathcal{D}|\bt),
\end{align}
where the last line follows from partial integration. The integration measure over the space of the observed variable-length datasets is $\ddd\mathcal{D}=\ddd N\prod_{i=1}^n\ddd\bx_i$ where $\ddd N$ is the counting measure on $\mathbb{N}_0$ and the $\ddd \bx_i$ are standard Lebesgue measures on $\mathbb{R}^{N_f}$ with, in our case, $N_f=28$. Inserting Eq.~\ref{eq:extended-likelihood}, we find after a few steps
\begin{align}
    I_{ab}^{(\text{unbinned})}(\bt)&=\int\ddd\bx\,\frac{\partial_a (\mathcal L\sigma(\bt)p(\bx|\bt))\,\partial_b (\mathcal L\sigma(\bt)p(\bx|\bt))}{\mathcal L\sigma(\bt)p(\bx|\bt)}\label{eq:FI-unbinned}
\end{align}
where we have used Eq.~\ref{eq:normalization} to convert $p(\bx|\bt)$ to DCRs. Eq.~\ref{eq:FI-unbinned} is the continuum limit of Eq.~\ref{eq:FI-binned} and we evaluate it for $\bt\equiv(\mu,\bn)=(1,\bzero)$. 
Using a simulated sample $\mathcal{D}_{1,\bzero}$, we get
\begin{align}
    I_{ab}^{(\text{unbinned})}(1,\bzero)\approx\sum_{\substack{\{w_i,\bx_i\}\\\in\mathcal{D}_{1,\bzero}}} w_i\, \partial_a\hat R(\bx_i|\mu,\bn)\Big|_{1,\bzero}\,\partial_b\hat R(\bx_i|\mu,\bn)\Big|_{1,\bzero}
\end{align}
with $\hat R(\bx|\bt)$ again taken from Eq.~\ref{eq:xsec-ratio-ML}.
\vfill\null 
The required derivatives  are
\begin{subequations}
    \begin{align}
    \frac{\partial}{\partial\mu}\hat R(\bx|\mu,\bn)\Big|_{1,\bzero}&=\hat g_\PH(\bx)\nonumber\\
    \frac{\partial}{\partial\nu_\text{bkg}}\hat R(\bx|\mu,\bn)\Big|_{1,\bzero}&=(1-\hat g_\PH(\bx))\log(1+\alpha_\text{bkg})\nonumber\\
    \frac{\partial}{\partial\nu_{\ttbar}}\hat R(\bx|\mu,\bn)\Big|_{1,\bzero}&=\hat g_{\ttbar}(\bx)\log(1+\alpha_{\ttbar})\nonumber\\
    \frac{\partial}{\partial\nu_{\VV}}\hat R(\bx|\mu,\bn)\Big|_{1,\bzero}&=\hat g_{\VV}(\bx)\log(1+\alpha_{\VV})\nonumber\\
    \frac{\partial}{\partial\nu_{\text{tes}}}\hat R(\bx|\mu,\bn)\Big|_{1,\bzero}&=(\hat g_\PH(\bx)\hat\Delta_{\text{tes},\PH}(\bx)+\hat g_\PZ(\bx)\hat\Delta_{\text{tes},\PZ}(\bx)\nonumber\\
    \qquad+\hat g_{\ttbar}(\bx)\hat\Delta_{\text{jes},\ttbar}(\bx)&+\hat g_{\VV}(\bx)\hat\Delta_{\text{jes},\VV}(\bx))\nonumber\\
    \frac{\partial}{\partial\nu_{\text{jes}}}\hat R(\bx|\mu,\bn)\Big|_{1,\bzero}&=(\hat g_\PH(\bx)\hat\Delta_{\text{jes},\PH}(\bx)+\hat g_\PZ(\bx)\hat\Delta_{\text{jes},\PZ}(\bx)\nonumber\\
    \qquad+\hat g_{\ttbar}(\bx)\hat\Delta_{\text{jes},\ttbar}(\bx)&+\hat g_{\VV}(\bx)\hat\Delta_{\text{jes},\VV}(\bx))\nonumber\\
    \frac{\partial}{\partial\nu_{\text{met}}}\hat R(\bx|\mu,\bn)\Big|_{1,\bzero}&=(\hat g_\PH(\bx)\hat\Delta_{\text{met},\PH}(\bx)+\hat g_\PZ(\bx)\hat\Delta_{\text{met},\PZ}(\bx)\nonumber\\
    \qquad+\hat g_{\ttbar}(\bx)\hat\Delta_{\text{met},\ttbar}(\bx)&+\hat g_{\VV}(\bx)\hat\Delta_{\text{met},\VV}(\bx))\nonumber.
\end{align}
\end{subequations}
For example, the systematics-free Fisher information of the signal strength parameter is 
\begin{align}
    I^{(\text{unbinned})}_{\mu\mu}(1,\bzero)=\sum_{\{w_i,\bx_i\}\in\mathcal{D}_{1,\bzero}} w_i\, \hat g_\PH^2(\bx_i),
\end{align}
a quantity that could be visualized in one- or two-dimensional distributions to identify regions that contribute more substantially to the measurement.

\bibliographystyle{unsrturl}  
\bibliography{references} 

\end{document}